%% file: main_v2.tex
\documentclass[a4paper,11pt]{article}
\pdfoutput=1 
\usepackage{jheppub} 
\usepackage[bottom]{footmisc}
\usepackage{amssymb}
\usepackage{amsmath}
\usepackage{amsthm}
\usepackage[usenames,dvipsnames]{xcolor}
\usepackage{epsfig}
\usepackage{dcolumn}
\usepackage{tikz}
\usetikzlibrary{shapes.geometric, arrows}
\usepackage{upgreek}
\usepackage{setspace}
\usepackage{subfig}
\usepackage{enumitem}
\usepackage{array,multirow,bigdelim,arydshln}
\usepackage{appendix}
\usepackage{xparse}
\usepackage{stmaryrd}
\usepackage[T1]{fontenc} 
\usepackage{mathtools}
\usepackage{physics} 
\usepackage{adjustbox}
\usepackage{multirow}
\usepackage{graphicx} 
\usepackage{float} 
\graphicspath{{./images/}}
\usepackage[nottoc]{tocbibind}
\usepackage{hyperref}
\usepackage[utf8]{inputenc}
\usepackage{CJK}
\hypersetup{
	colorlinks,
	urlcolor=Maroon,
	linkcolor=Maroon,
	citecolor=Maroon
	}
\usepackage[capitalize]{cleveref}  
  
  \crefname{section}{Sec.}{Secs.}
  \crefname{appendix}{App.}{Apps.}

\usetikzlibrary{arrows}
\newcommand{\midarrow}{\tikz \draw[-Stealth] (0,0) -- +(.1,0);}

\tikzset{
    vector/.style={
        decoration={snake, aspect=0.75, mirror, segment length=2mm},
        decorate
    },
    photon/.style={decorate, decoration={snake, amplitude=1pt, segment length=4pt}}
}
\NewDocumentCommand{\binomial}{omm}
 {%
  \genfrac(){0pt}{}{#2}{#3}%
  \IfValueT{#1}{_{\!#1}}%
 }
\NewDocumentCommand{\eulerian}{omm}
 {%
  \genfrac<>{0pt}{}{#2}{#3}%
  \IfValueT{#1}{_{\!#1}}%
 }

\usepackage{latexsym}
\usepackage{tikz}
\newcommand*\diff{\mathop{}\!\mathrm{d}}

\theoremstyle{plain}

\theoremstyle{definition}

\def\bea#1\eea{\begin{eqnarray}#1\end{eqnarray}}
\def\be#1\ee{\begin{equation}#1\end{equation}}
\def\ba#1\ea{\begin{align}#1\end{align}}

\def\AA{\mathcal{A}}
\def\W{\mathcal{W}}

\usepackage{amsmath}
\usepackage{multicol}
\usepackage{bbm}

\usepackage{amsthm}
\usepackage[scr]{rsfso}
\usepackage{upgreek}
\usepackage{amssymb}
\usepackage{mathrsfs}
\usepackage{bm}
\usepackage{setspace}
\usepackage{array,multirow,arydshln}
\usepackage{bigdelim}
\usepackage{scalerel}
\usepackage{diagbox}
\usepackage{tabularx}
\usepackage{tikz}

\usetikzlibrary{shapes.geometric,arrows,arrows.meta,decorations.pathmorphing,decorations.markings,patterns}

\usepackage[percent]{overpic}
\usepackage{multirow} 
\usepackage{slashed}
\usepackage{titlesec}
\usepackage{amsmath}

\titleformat{\part} {\centering\normalfont\Large\bfseries}
  {\partname~\thepart:}
  {0pt}
  {\Large}

\makeatletter
  \@ifundefined{c@lofdepth}{}{\let\c@lofdepth\relax}
  \@ifundefined{c@lotdepth}{}{\let\c@lotdepth\relax}
\makeatother

\usepackage{tocloft}

\newcommand*\listofparts{%
  \section*{List of Parts}%
  \addcontentsline{toc}{section}{List of Parts}%
  \@starttoc{prt}%
}

\usepackage{xpatch}
\makeatletter
  \xpretocmd\part{%
    \addcontentsline{prt}{part}{%
      \protect\numberline{\thepart}#1%
    }%
  }{}{\fail}
\makeatother

\title{Surface Gauge Invariance, Soft Limits and the Transmutation of Gluons into Scalars}

\author{Jeffrey V. Backus,}
\author{Carolina Figueiredo}

\affiliation{Jadwin Hall, Princeton University, Princeton, NJ 08540, USA}

\emailAdd{jvabackus@princeton.edu}
\emailAdd{cfigueiredo@princeton.edu}

\abstract{Over the past year, the ``scalar-scaffolding'' formalism has revealed a number of new features of gluon amplitudes. In this paper, we leverage these developments to study two distinct but related questions, linked by the scaffolding statement of gauge invariance. We start by revisiting the soft expansion of gluon amplitudes. The scaffolding picture allows for a precise definition of the soft limit and a canonical way to expand the amplitude. At tree-level, this reproduces the classic Weinberg soft theorem, and at one-loop, using surface kinematics, we derive an extension of this theorem valid at the level of the loop integrand. We then switch gears and describe a new relationship between gluon and scalar amplitudes. The expression of surface gauge invariance naturally suggests a certain differential operator acting on individual external gluons. Remarkably, we find that, both for the tree-level amplitude and the surface one-loop integrand, repeated applications of this operator \textit{transmutes} gluon amplitudes/integrands into those of Tr$(\phi^3)$ scalars. This is an interesting counterpart to the $\delta$-shift connection that lifts ``stringy'' Tr$(\phi^3)$ amplitudes to those of gluons.}

\begin{document}

\maketitle
\addtocontents{toc}{\protect\setcounter{tocdepth}{2}}

\input{Sections_v2/Intro_v2}
\input{Sections_v2/ScalarScaff_v2}

\input{Sections_v2/Soft_v2}

\input{Sections_v2/Transmutation_v2}
\input{Sections_v2/Outlook_v2}

\acknowledgments 
The authors thank Nima Arkani-Hamed, Clifford Cheung, Jin Dong and Alfredo Guevara for useful discussions. J.B. is supported by the NSF Graduate Research Fellowship under Grant No.~KB0013612. C.F. is supported  by FCT - Fundacao para a Ciencia e Tecnologia, I.P. (2023.01221.BD and DOI  https://doi.org/10.54499/2023.01221.BD)

\appendix

\input{Sections_v2/AppHigherOrderTree_v2}

\input{Sections_v2/AppHigherOrderLoop_v2}

\bibliographystyle{JHEP}\bibliography{Refs}

\end{document}

%% file: Sections_v2/Intro_v2.tex
\section{Introduction}
\label{sec:Intro}

In the past year, a new approach has been proposed for studying scattering amplitudes of massless colored particles, from Tr$(\phi^3)$ theory to pions and gluons \cite{Gluons,Zeros,NLSM}. This picture not only makes the standard features of amplitudes manifest --- such as factorization on poles --- but also brings to light previously-hidden properties of these objects \cite{Zeros,Cachazo_2022,Splits,Zhang:2024iun,Zhang:2024efe,Zhou:2024ddy,Li:2024bwq,Bartsch:2024amu,Rodina:2024yfc,Backus:2025hpn,Jones:2025rbv,Feng:2025ofq}.

In this formulation, gluon amplitudes are recast using different kinematic variables,
which come from thinking of each external gluon as being produced by a pair of scalars ---
the gluons are \textit{scalar-scaffolded} \cite{Gluons}. In this way, the gluon amplitude is written exclusively in terms of dot products of the scalar momenta. This simple change of kinematic variables
turns out to have important consequences, in particular allowing us to write the gluon
amplitude in a canonical way, completely free of gauge redundancies \cite{Gluons}. In addition, at loop level, this formulation naturally suggests an extension of kinematics --- \textit{surface kinematics} --- that lets us write down well-defined gluon integrands that correctly match all cuts and are also gauge invariant, before loop integration \cite{YMIntegrand}.

Given this progress, it is natural to ask how we can recast well-known features of gluon amplitudes in this new language, as well as if we can uncover any new properties that are exposed by this formalism. We will pursue both of these goals in this paper, which is divided into two parts, linked by the scalar-scaffolding description of gauge invariance. In \cref{part:1}, we use scalar-scaffolding to define an on-shell soft limit and study the soft expansion of gluon amplitudes both at tree-level and one-loop. In \cref{part:2}, we explain how the expression of gauge invariance suggests the introduction of simple differential operators acting on individual external gluons. These operators turn out to have a surprising property: successively applying them to all but two gluons converts Yang-Mills (YM) amplitudes to amplitudes for Tr$(\phi^3)$ theory, both at tree-level and one-loop.

\subsubsection*{\cref{part:1}: Soft Limits at Tree-level and One-loop}

The study of soft limits in particle scattering is an ancient subject that has revealed many deep aspects of fundamental physics, starting from Weinberg's pioneering analysis of the consistency conditions on theories of massless particles with spin~\cite{Weinberg2,Weinberg}.

More recently, soft limits have been understood as part of a much larger narrative in the modern revival of our understanding of scattering amplitudes.  From the on-shell perspective, soft limits are intimately connected with the physics of factorization. For example, in color-ordered YM amplitudes, a gluon goes soft by becoming collinear with its two adjacent gluons. Factorization near poles has played a major role in recursive methods to compute amplitudes~\cite{BCFW, BCF}, and similarly, some advantage has been obtained from behavior of amplitudes under soft limits~\cite{Cheung:2014dqa,Cheung:2015ota,Cheung:2016drk,Zhou:2022orv,Zhou:2024qjh,Hu:2023lso,Cachazo:2016njl}. In another vein, leading and subleading soft limits can be interpreted as the appearance of asymptotic symmetries on the celestial sphere~\cite{Strominger:2013jfa,Strominger:2013lka,Strominger:2017zoo,Lysov:2014csa,Cachazo:2014fwa}. Subleading and higher orders in the soft expansion have also been extensively studied in the literature (at tree-level and beyond) for their own sake; see $e.g.$ Refs.~\cite{Li:2018gnc,Laddha:2017ygw,Sen:2017nim,Bern:2014oka,Klose:2015xoa,Hamada:2018vrw}.

We start by studying the soft expansion for scalar-scaffolded gluons at tree-level. The upshot of scalar-scaffolding is that the gluon amplitude has a completely \textit{locked} form, free of gauge redundancies. This means that it has a completely well-defined Laurent expansion in the kinematic variables, and therefore takes a similarly well-defined Laurent expansion around soft limits. In addition, since the polarizations and momenta of the gluons are traded for the momenta of $2n$ scalars, we can implement the soft expansion directly at the level of the momentum invariants of the $2n$ scalars, allowing us to take the soft limit while manifesting momentum conservation.
Undertaking this limit, we derive leading and subleading soft terms in this language and understand why there are no universal soft terms at higher orders in the expansion. For the leading contribution, we find a simple result that matches Weinberg's soft theorem:
\begin{equation*}
    \mathcal{A}_n^{\text{Gluon}} \to \left( \frac{X_{2n-3,2n}}{X_{1,2n-3}} + \frac{X_{3,2n}}{X_{3,2n-1}} - 1\right) \times   \mathcal{A}_{n-1}^{\text{Gluon}}, 
\end{equation*}
where $X_{i,j} =(p_i+p_{i+1}+\cdots p_{j-1})^2$ are the planar Mandelstam invariants for the momenta of the $2n$ scalars that scaffold the $n$ gluons.

In \cref{sec:one-loop-soft}, we extend this study to the one-loop YM integrand defined in terms of \textit{surface kinematics} \cite{YMIntegrand, Salvatori:2018aha, Arkani-Hamed:2019vag}. Here, since this extension of kinematics gives the integrand a well-defined notion of gauge invariance as well as factorization on cuts prior to loop integration, we may expect to find a statement similar to the one found for the tree-level answer. Indeed, in the context of NLSM amplitudes \cite{circles}, we have seen that this particular kinematic extension can be used to define a pion integrand that makes manifest the Adler zero directly at the level of the loop integrand (rather than post-loop integration) as well as admits predictions for its behavior under multi-soft limits \cite{Splits}. Therefore, it is natural to ask what this extension gives in the context of the YM integrand soft expansion. 

Quite remarkably, we find that, at leading order in the soft expansion, we obtain precisely the same soft Weinberg term as in the tree-level case plus corrections:
\begin{equation}
\begin{aligned}
    \mathcal{I}_n =& \left( \frac{X_{2n,3}}{X_{2n-1,3}} +   \frac{X_{2n-3,2n}}{X_{2n-3,1}} - 1\right)\mathcal{I}_{n-1}(s,2,3,\cdots,2n-2)\vert_{\mathcal{S}} \\
    &+X_{s,s}\left[\frac{X_{2n,3} - X_{2n,2}}{X_{2n-1,3}} \frac{\partial \mathcal{I}_{n-1}}{\partial X_{s,2}}\bigg \vert_{\mathcal{S}} + \frac{X_{2n-3,2n} - X_{2n-2,2n}}{X_{2n-3,1}} \frac{\partial \mathcal{I}_{n-1}}{\partial X_{2n-2,s}} \bigg \vert_{\mathcal{S}}\right]  + \mathcal{O}(\delta^0).
\end{aligned}
\end{equation}
The correction terms are proportional to the tadpole variables $X_{i,i}$ and therefore translate into scaleless contributions that vanish upon loop integration. It would be interesting to understand how this correction interacts with the IR divergences that appear post-loop integration, a topic we leave to future work. 

It is not especially surprising to be able to translate known statements about the soft expansion into scalar-scaffolding language at tree-level. It is somewhat more remarkable that the extension of kinematics which makes the one-loop integrand well-defined also plays nicely with the soft expansion, yielding the expected Weinberg term plus the simple correction factors described above.

\subsubsection*{\cref{part:2}: Transmutation of Gluons into Scalars}

As emphasized in \cref{part:1}, one striking aspect of the scalar-scaffolding formalism is the fact that it eliminates gauge redundancies, turning gauge invariance and linearity in each polarization vector into unified statements on the form of the amplitude~\cite{Gluons}. In addition to its utility in defining a soft expansion, this representation of the amplitude suggests that we study a simple operator $\W_{2i}$ that acts on the $i^{\rm{th}}$ external gluon:
\begin{equation*}
    \W_{2i} = \sum_{j=2i+2}^{2i-2} \frac{\partial}{\partial X_{2i,j}}.
\end{equation*}

It turns out that applying this operator to the $n$-point amplitude produces a much simpler object, closely related to an amplitude where gluons $i-1,i,$ and $i+1$ are converted into Tr($\phi^3$) scalars. In \cref{sec:PolConfig1}, we explain that the action of this operator is equivalent to a special choice for the polarization of gluon $i$. Additionally, if we act with this operator on $(n-2)$ of the gluons, we find that all $n$ gluons have been converted into scalars --- transforming the original, pure YM amplitude into a Tr($\phi^3$) amplitude! This is a very surprising fact, but, as we explain in \cref{sec:Splits}, one is naturally led to discover it by consecutively applying particular types of ``split factorizations'' \cite{Splits} that implement the action of this operator $\W_{2i}$.

To understand the action of these operators from splits, we need to define the scalar-scaffolded gluon amplitude via its surface integral formulation \cite{Gluons}. Doing so makes it straightforward to find other surface integrals that compactly represent the action of any number of $\W_e$'s on the amplitude. Interestingly, after applying $(n-2)$ $\W_e$'s, we do \textit{not} land on the surface integral for Tr($\phi^3$) amplitudes; acting with the $\W_e$'s on all but gluons $i,j$, we instead find 
\begin{equation*}
\begin{gathered}
    \begin{tikzpicture}[line width=1,scale=10]

    \node[regular polygon, regular polygon sides=22,  minimum size=2cm] (p) at (0,0) {};
     \draw (0,0) circle [radius=0.101cm];

    \node[scale=0.8,xshift=-15,yshift=-5] at (p.corner 8) {$i+1$};
    \node[scale=0.8,xshift=-5,yshift=-5] at (p.corner 10) {$i$};    
    \node[scale=0.8,xshift=7,yshift=7] at (p.corner 22) {$j$};
    \node[scale=0.8,xshift=17,yshift=7] at (p.corner 20) {$j+1$};
    
    \foreach \i in {20}
    {\draw[fill] (p.corner \i) circle [radius=0.1pt];}
    \foreach \i in {8}
    {\draw[fill] (p.corner \i) circle [radius=0.1pt];}
    \draw[fill] (p.corner 10) circle [radius=0.1pt];
     \draw[fill] (p.corner 22) circle [radius=0.1pt];

     \fill[Blue,opacity=0.2]  (p.corner 8) -- (p.corner 7) -- (p.corner 6) -- (p.corner 5) -- (p.corner 4) -- (p.corner 3) -- (p.corner 2) -- (p.corner 1) -- (p.corner 22)-- cycle;

     \fill[Blue,opacity=0.2]  (p.corner 10) -- (p.corner 11) -- (p.corner 12) -- (p.corner 13) -- (p.corner 14) -- (p.corner 15) -- (p.corner 16) -- (p.corner 17) -- (p.corner 18)-- (p.corner 19)-- (p.corner 20)-- cycle;

     \node[scale=0.8,xshift=18,yshift=30] at (p.corner 8) {$\mathcal{S}_L$};
     \node[scale=0.8,xshift=37,yshift=13] at (p.corner 10) {$\mathcal{S}_R$};

    \end{tikzpicture}     
\end{gathered}
\propto \quad \int_{\mathcal{S}_n} \prod \frac{d y}{y}  \prod_{(i,j) \in \mathcal{S}_n} u_{i,j}^{\alpha^\prime X_{i,j}} \frac{1}{\prod_{(k,m)\in \mathcal{S}_L} u_{k,m} \times \prod_{(k,m)\in \mathcal{S}_R} u_{k,m} },
\end{equation*}
which is the surface integral for Tr$(\phi^3)$ under the kinematic shift $X_{i,j} \to X_{i,j} - 1/\alpha^\prime$, if $(i,j)$ is a curve in $\mathcal{S}_L$ or in $\mathcal{S}_R$ (including boundary curves $(i+1,j)$ and $(i,j+1)$). Nonetheless, in \cref{sec:shiftsTrPhi3}, we show that this class of shifted integrals indeed yields Tr$(\phi^3)$ amplitudes at low energies! It is amusing that, while some simple kinematic shifts take us from Tr$(\phi^3)$ to pion and gluon amplitudes, others return us to Tr$(\phi^3)$ at low energies, while of course differing in the UV and at higher orders in the $\alpha^\prime$ expansion.  

In summary, to understand the features of these operators, we are forced to go back to the definition of gluon amplitudes via surface integrals. It was in this context that scalar-scaffolded gluon amplitudes were discovered as a simple shift of the Tr$(\phi^3)$ amplitudes. Now, we are finding an ``opposite'' connection --- by applying $(n-2)$ $\W_e$'s, we can start from the surface integral for gluons and land on one that gives Tr($\phi^3$) at low energies!

To conclude \cref{part:2}, we explain how these operators have a natural extension to the scalar-scaffolding formulation of loop integrands. We find that, by defining the one-loop integrand using \textit{surface kinematics} in the obvious way, the tree-level story beautifully extends to the YM surface integrand: acting with $n$ of these loop operators turns the gluon integrand to the Tr($\phi^3$) one! It is natural to conjecture that this statement can be similarly proven directly at the level of surface integrals using splits~\cite{Splits}, but we leave this discussion for future work. 

%% file: Sections_v2/ScalarScaff_v2.tex
\section{A Review of Scalar-Scaffolded Gluons}
\label{sec:ScalarScaffRev}

In order to make our presentation in this paper self-contained, in this section we give a review of the main aspects of the scalar-scaffolding formalism, as well as some examples of what gluon tree-level amplitudes and one-loop integrands look like when recast in this language. 

For the purposes of this paper, the most important aspect of this formalism is that it allows us to define the kinematics for gluon scattering amplitudes in a canonical way, encoding both on-shell gluon momenta $q_i^\mu$ with $q_i^2 = 0$, and transverse polarizations $\epsilon_i^\mu$ with $\epsilon_i \cdot q_i = 0$. As a result, in this  representation the amplitude has a \textit{unique} form. This is to be contrasted with standard expressions for gluon amplitudes using momenta and polarization vectors. Here, the expressions are always ambiguous. For instance, we can use momentum conservation to express one of the momenta $q_i^\mu$ in terms of the sum of all the rest, so that any occurrence of $q_i \cdot q_j$ can be replaced with $-\sum_{k \neq i, j} q_k \cdot q_j$. Similarly, $q_i \cdot \epsilon_j$ can be replaced with $-\sum_{k \neq i, j} q_k \cdot \epsilon_j$. Of course these ambiguities can be resolved by deciding to solve for $e.g.$ $q_n^\mu$ in terms of all the rest, but this is unnatural, breaking the cyclic symmetry of the problem. Scalar-scaffolding gives an invariant way to express the amplitudes without breaking any of the symmetries, by giving a physical picture for the production of the momenta and polarizations of the external gluons from pairs of scalars. 

The linearity of gluon amplitudes in polarization vectors, as well as the on-shell gauge invariance under $\epsilon_\mu \to \epsilon_\mu + \alpha q_\mu$, are reflected in a natural way in the scaffolding picture. These conditions force the amplitude to be written in a particular form --- the launching point to the discovery of important features about the amplitude in \cref{part:1} and \cref{part:2}. 

In \cref{sec:2.1}, we review how the kinematic invariants of gluon amplitudes can be recast as momentum dot products of the scalars that scaffold the gluons. At tree-level, this allows us to write the gluon amplitude in terms of planar Mandelstam variables $X_{i,j}$, each of which is naturally associated to a curve on a disk with marked points on the boundary. We review how the statements of gauge invariance and multi-linearity in the polarizations $\epsilon_i^\mu$ translate into these $X$'s, as well as give the expression for the factorization on a given cut at $X_{i,j}=0$. In \cref{sec:2.2}, we flash some explicit examples of the three- and four-point gluon amplitudes written in this language.

In \cref{sec:2.3}, we give a review of the scalar-scaffolded gluon one-loop integrand formulated in terms of \textit{surface kinematics}. As explained in Ref.~\cite{YMIntegrand}, this kinematic extension --- which comes from identifying kinematic variables with labelings of curves on the punctured disk --- will allow us to write down an object endowed with a generalized notion of gauge invariance and well-defined cuts directly at the level of the integrand. These features will play a crucial role in \cref{part:1} and \cref{part:2} when we generalize the results found at tree-level to one-loop. 

\subsection{How to scalar-scaffold gluons}
\label{sec:2.1}

As prefaced in \cref{sec:Intro}, we want to think of the gluons in a YM amplitude (or integrand) as being \textit{scalar-scaffolded}; that is, we think of each external gluon as being produced by two massless colored scalars \cite{Gluons}. Concretely, let's say we have an external gluon with momentum $q_1^\mu$ in an amplitude. As shown in \cref{eq:GluonScalarMap}, we quite literally ``scaffold'' the gluon with two scalars with momenta $p_1^\mu$ and $p_2^\mu$. Since the scalars couple to the gluon via minimal coupling, we are then able to encode the polarization vector $\epsilon_1^\mu$ and on-shell gluon momentum $q_1^\mu$ in terms of the scalar momenta $p_1^\mu$ and $p_2^\mu$ in the following way:
\begin{equation}
    \begin{gathered}
    \begin{tikzpicture}[line width=0.6,scale=1.2,baseline={([yshift=0.0ex]current bounding box.center)}]
        \coordinate (a) at (0,0);
        \coordinate (p1) at (-0.35,0.65);
        \coordinate (p2) at (-0.35,-0.65);
        \coordinate (q) at (1.2,0);
        \coordinate (b) at (1,0);
        \draw[photon] (a) -- (b);
        \draw[] (a) --++ (-35:-1);
        \draw[] (a) --++ (35:-1);
        \node[scale=0.8] at (p1) {$p_2^\mu$};
        \node[scale=0.8] at (p2) {$p_1^\mu$};
        \node[scale=0.8] at (q) {$q_1^\mu$};
    \end{tikzpicture}
\end{gathered} \propto \,g_{\text{YM}}\,p_2^\mu \quad \Rightarrow \quad \begin{cases}
q_1^\mu = p_1^\mu + p_2^\mu, \\ \epsilon_1^\mu = p_2^\mu -\alpha_1 (p_1+p_2)^\mu, \end{cases}
\label{eq:GluonScalarMap}
\end{equation}
where $\alpha_1$ is a gauge-dependent parameter that should drop out of the final amplitude. Of course, in addition to the fact that the scalar momenta are on-shell, we also need the gluon momentum to be on-shell. That is, we must require $(p_1+p_2)^2=0$, from which we automatically find that $\epsilon_1 \cdot q_1=0$. This means that, via scalar-scaffolding, we always get gluon amplitudes with null polarizations; however, this is enough since we can write any polarization as a linear combination of null ones. In this way, we can represent an $n$-point gluon amplitude in terms of the momenta of $2n$ scalars satisfying $(p_{2i-1}+p_{2i})^2=0$ for $i = 1, 2, \ldots, n$.

Thus, the kinematic variables on which the $n$-point gluon amplitude depends are the \textit{same} as those of the $2n$-point scalar amplitude, minus those lost due to the gluon on-shell conditions.
\begin{figure}[t]
    \centering
    \includegraphics[width=0.9\linewidth]{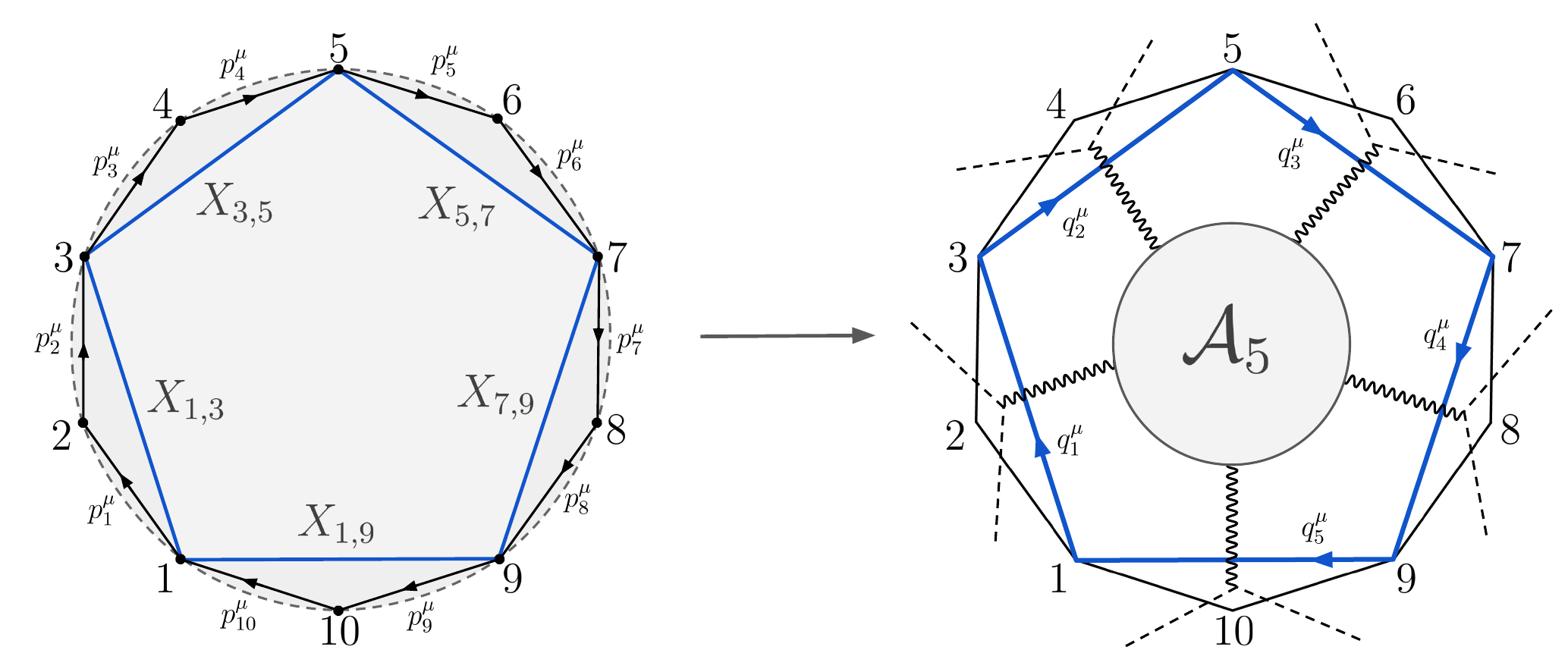}
    \caption{(Left) The momentum polygon for ten scalars (in black), drawn inside the disk with marked points on the boundary (in dashed). The scaffolding chords/ curves on the disk (in blue) are labeled by $X_{i,i+2}$ for odd $i$. (Right) In blue, we highlight the five-point gluon momentum polygon with edges $q_i^\mu$, inscribed inside the scalar polygon. After putting the gluons on-shell by taking the scaffolding curves to zero, we are left with the five-point gluon amplitude (in center).}
    \label{fig:ScalarScaff}
\end{figure}

To keep track of these scalar invariants, it is useful to introduce the so-called \textit{momentum polygon}. An example for the case of five gluons (so ten scalars) is given in \cref{fig:ScalarScaff}. To construct the momentum polygon, we draw the momentum of each scaffolded scalar tip-to-toe according to the standard ordering $1,2,\ldots,2n-1,2n$, obtaining a closed $2n$-gon that manifests momentum conservation. By construction, the vector associated with the chord from vertex $i$ to vertex $j$ in the $2n$-gon is equal to the sum of the momenta of the enclosed scalars $p_i^\mu + p_{i+1}^\mu + \cdots + p_{j-1}^\mu$. In this way, the \textit{length squared} of these chords may be identified with the set of planar Mandelstam invariants, which are exactly the variables that appear in the poles of planar diagrams:
\begin{equation}
    X_{i,j} = (p_i + p_{i+1}+ \ldots + p_{j-1})^2.
\end{equation}

Let's now do a counting exercise. For the $2n$-gon, there are $2n(2n-3)/2$ chords, matching the number of independent Mandelstam invariants appearing in a $2n$-point scalar amplitude.\footnote{This is under the assumption that $d$, the dimensionality of spacetime, is sufficiently large, so that we can ignore Gram determinant constraints. We will assume this throughout the paper unless said otherwise.} But, in order to describe gluon kinematics we must further set the so-called \textit{scaffolding chords} $X_{2i-1,2i+1}=(p_{2i-1}+p_{2i})^2=0$. This brings us down to $2n(2n-3)/2-n$ independent $X$'s, which is precisely the number of degrees of freedom of the $n$-point gluon amplitude. Therefore, the scalar variables (with the scaffolding conditions) furnish basis of kinematic space for tree-level gluon amplitudes, where momentum conservation is \textit{automatically} enforced from the fact that the $X$'s are chords of a \textit{closed} momentum polygon. 

We can equivalently think of the momentum polygon as inscribed in a disk with marked points on the boundary (depicted in dashed gray on the $l.h.s.$ of \cref{fig:ScalarScaff}), where each boundary component is assigned the momentum of the respective edge of the momentum polygon. In this picture, the kinematic invariants $X_{i,j}$ are now identified with \textit{curves} on the surface (the disk), and we can read off the momentum in a given curve by \textit{homology}, $i.e.$, by deforming the curve to a collection of boundary curves and adding their momenta. So, from this point of view, this space of curves also defines a basis for the kinematic invariants of the gluon amplitude, once we again go on the scaffolding locus $X_{2i-1,2i+1}=0$. As we will see momentarily, defining the basis of kinematic invariants in this way will turn out to be especially crucial for the one-loop integrand.

The scaffolding curves --- marked in blue in \cref{fig:ScalarScaff} --- form an inscribed $n$-gon, corresponding to the actual \textit{gluon} momentum $n$-gon, where each edge is associated with a gluon momentum $q_i^\mu$. All of the chords/ curves living purely inside this smaller $n$-gon are identified with invariants containing sums of gluon momenta. These are precisely the $X_{i,j}$'s where both $i$ and $j$ are \textit{odd}, and they are the variables allowed to appear as poles in the gluon amplitude. The remaining chords (where one of $i,j$ is even) encode information related to the polarizations, in the form of the dot products $q_k\cdot \epsilon_m$ and $\epsilon_k \cdot \epsilon_m$. The explicit correspondence depends on the gauge parameter $\alpha_i$, but one can easily read it off using the map~\eqref{eq:GluonScalarMap}.

To summarize, we can use the scalar-scaffolding representation to write the gluon $n$-point amplitude $\mathcal{A}_n(q_i\cdot q_j, \epsilon_i\cdot q_j, \epsilon_i\cdot \epsilon_j)$ purely in terms of the kinematics of $2n$ scalars $\mathcal{A}_{n}(X_{i,j})$.\footnote{For the remainder of this paper, we will denote the normal-ordered $2n$-point scalar-scaffolded gluon amplitude as $\mathcal{A}_{n}(1, 2, \ldots, 2n)$, where we always implicitly assume that we are on the support of $X_{2i-1,2i+1}=0$.} It is then natural to ask how well-known aspects of amplitudes translate into this language. For our present purposes, we will need to understand (1) what the statement of gauge invariance is in terms of the $X$'s, and (2) what the spin-sum gluing rule looks like when we localize on a particular pole. 

As explained in Ref.~\cite{Gluons}, in scalar variables the statement of gauge invariance comes together with linearity in the polarization of the $i^{\rm{th}}$ gluon. Due to these two requirements, we find that the amplitude $\mathcal{A}_n$ must be linear in each $X_{2i,j}$, and that this linearity must be such that the full amplitude can be written in the forms \cite{Gluons}
\begin{equation}
\begin{cases}
\label{eq:GgInvariance}
    \mathcal{A}_{n}(1, 2, \ldots, 2n) = \sum_{j \neq\{2i,2i\pm 1\}} (X_{2i,j}-X_{2i-1,j}) Q_{2i,j}, \\
    \mathcal{A}_{n}(1, 2, \ldots, 2n) = \sum_{j \neq\{2i,2i\pm 1\}}(X_{2i,j}-X_{2i+1,j}) Q_{2i,j}, 
\end{cases} \, \, \,  \text{with } \,\, Q_{2i,j} = \frac{\partial \mathcal{A}_{n}}{\partial X_{2i,j}}.
\end{equation}
These forms make manifest both invariance under a gauge transformation in the $i^{\rm{th}}$ gluon, $\mathcal{A}_n(\epsilon_i^\mu \to \epsilon_i^\mu + \alpha q_i^\mu) = \mathcal{A}_n(\epsilon_i^\mu) $, as well as linearity in the respective polarization, $\mathcal{A}_n(\beta \epsilon_i^\mu) = \beta\mathcal{A}_n( \epsilon_i^\mu)$. We therefore see that these two statements ---  distinct in the ordinary gluon language --- are naturally unified in the scalar-scaffolding language. In particular, if we subtract the two representations in Eq.~\eqref{eq:GgInvariance}, we find the following identity that the amplitude must satisfy
\begin{equation}
     \sum_{j}(X_{2i+1,j}-X_{2i-1,j}) \frac{\partial \mathcal{A}_{n}}{\partial X_{2i,j}} =0,
     \label{eq:gauge}
\end{equation}
which is the scaffolding representation of gauge invariance. In \cref{sec:2.3}, we explain how these statements extend to the surface one-loop \cite{YMIntegrand}. 

It is likewise straightforward to translate the polarization sum appearing in gluon factorization channels into scalar language. As demonstrated in Ref.~\cite{Gluons}, the residue on a gluon propagator $X_{i,j}=0$ (with $i$ and $j$ both odd) can be written in terms of the $X$'s as follows:
\begin{figure}[t]
    \centering
    \includegraphics[width=0.85\linewidth]{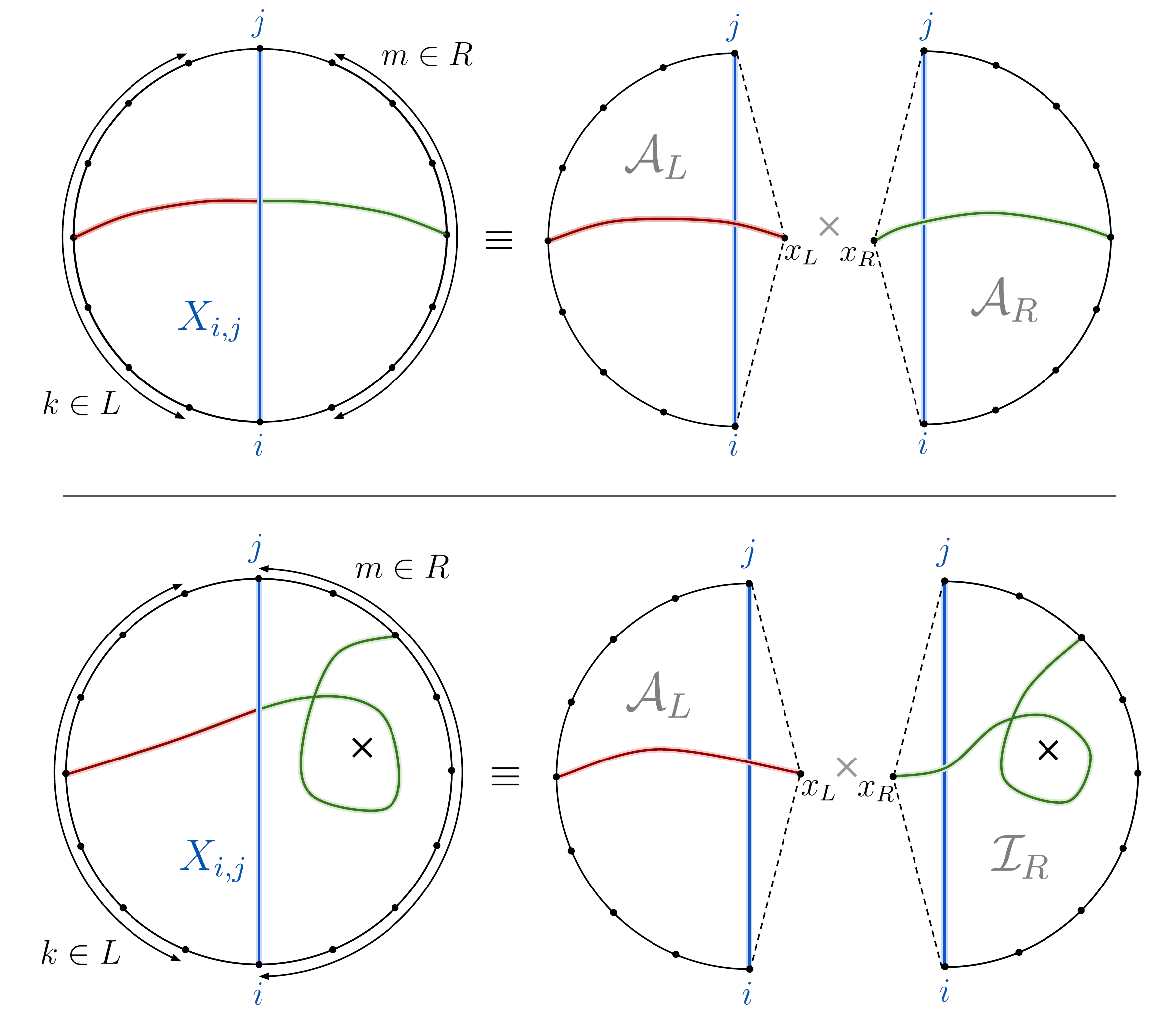}
    \caption{(Top) Tree-level factorization at $X_{i,j}=0$ into two lower-point tree amplitudes, $\mathcal{A}_L$ and $\mathcal{A}_R$ . (Bottom) Tree-loop factorization of the integrand at $X_{i,j}=0$ into a lower-point tree amplitude $\mathcal{A}_L$ and a lower-point one-loop integrand, $\mathcal{I}_R$.}
    \label{fig:factorization}
\end{figure}
\begin{equation}
\begin{aligned}
  \mathop{\mathrm{Res}}_{X_{i,j}=0} \mathcal{A}_{n}&= \sum_{k \in L, \ m\in R} (X_{k,m} -X_{k,j}-X_{m,i}) \frac{\partial \mathcal{A}_{L}}{\partial X_{k,x_L}}  \times \frac{\partial \mathcal{A}_{R}}{\partial X_{m,x_R}}, \\
  &= \sum_{k \in L, \ m\in R} (X_{k,m} -X_{k,i}-X_{m,j}) \frac{\partial \mathcal{A}_{L}}{\partial X_{k,x_L}}  \times \frac{\partial \mathcal{A}_{R}}{\partial X_{m,x_R}}
\end{aligned}
  \label{eq:factorization}
\end{equation}
where we can move from the first line to the second line by using the constraint~\eqref{eq:gauge} in both $x_L$ and $x_R$. In the above formula, we have $L = \{i+1,i+2,\cdots, j-1\}$ and $R= \{j+1,j+2,\cdots, i-1\}$, and $\mathcal{A}_L$ and $\mathcal{A}_R$ are the two lower-point gluon amplitudes we obtain when the propagator $X_{i,j}$ goes on-shell. This factorization is depicted geometrically on the $l.h.s.$ of \cref{fig:factorization}, where we see that these lower-point amplitudes depend on the indices to the left (right) of chord $X_{i,j}$ as well as on a new index associated with the summed-over polarization of the intermediate gluon. So, the kinematic dependence of these amplitudes are given by $\mathcal{A}_L \equiv \mathcal{A}_L(i,i+1,\cdots, j-1,j,x_L)$ and $\mathcal{A}_R \equiv \mathcal{A}_R(i,x_R,j,j+1,\cdots, i-1)$.

\subsection{Examples at three- and four-points}
\label{sec:2.2}

We now present some simple examples of what YM amplitudes look like when cast in terms of the scalar-scaffolded variables $X_{i,j}$. Let's start with the simplest case of the three-point gluon interaction. Applying the mapping~\eqref{eq:GluonScalarMap} to each one of the three gluons, we obtain
\begin{equation}
     \mathcal{A}_3(1,2,3,4,5,6) = X_{1,4}X_{2,6}+X_{3,6}X_{2,4}+X_{2,5}X_{4,6}-X_{2,5}X_{3,6}-X_{1,4}X_{3,6}-X_{1,4}X_{2,5}.
     \label{eq:3ptGluons}
\end{equation}
Using this expression, we can demonstrate gauge invariance and linearity in gluon $1$ by checking that dependence in the $X_{2,j}$ goes as in \cref{eq:GgInvariance}, with $i=1$. Indeed, using \cref{eq:GgInvariance}, we can make gauge invariance and linearity in gluon $1$ manifest by writing $\AA_3$ as follows:
\begin{equation}
\begin{aligned}
    \mathcal{A}_3 &=  (X_{2,4}-X_{1,4}) X_{3,6}  + X_{2,5}(X_{4,6}-X_{3,6}-X_{1,4}) + X_{2,6}X_{1,4}\\
    &=  X_{2,4} X_{3,6}  + X_{2,5}(X_{4,6}-X_{3,6}-X_{1,4}) + (X_{2,6}-X_{3,6})X_{1,4},
\end{aligned}
\end{equation}
since we set the chords $X_{1,5}=X_{1,6}=0$ and $X_{3,4}=X_{3,5}=0$. Of course, we can also write $\AA_3$ in a form that makes gauge invariance in gluon $2$ or gluon $3$ manifest; in these cases, we would examine linearity in $X_{4,j}$ or $X_{6,j}$, respectively.

As for the four-point YM amplitude, using the fact that it has a unique form in scalar-scaffolded variables, it is meaningful to decompose it in a Laurent expansion as 
\begin{equation}
    \mathcal{A}_4(1,2,3,4,5,6,7,8) = \frac{R_{1,5}}{X_{1,5}} + \frac{R_{3,7}}{X_{3,7}} + C^{(4)},
\end{equation}
where $R_{1,5}$ and $R_{3,7}$ are the residues of $\AA_4$ at $X_{1,5}=0$ and $X_{3,7}=0$, respectively, and $ C^{(4)}$ the pure contact part. To be explicit, here $R_{1,5}$ is independent of $X_{1,5}$ and $R_{3,7}$ is independent of $X_{3,7}$. It is worth stressing this basic point again: the amplitudes have a completely canonical Laurent expansion in the $X$ variables, but individual terms in the expansion (like the ones shown in the above equation) are not gauge invariant. Gauge invariance is then an interesting statement that relates different terms in the Laurent expansion to each other. 

We can write the residues in terms of $X_{i,j}$'s as follows:
\begin{equation}
\begin{aligned}
    R_{1,5}=&-X_{1,4} X_{1,6} X_{2,7}-X_{1,4} X_{5,8} X_{2,7}+X_{1,4} X_{6,8} X_{2,7}+X_{1,4} X_{1,6} X_{2,8}+X_{1,4}
   X_{1,6} X_{3,7}\\
   &-X_{1,6} X_{2,4} X_{3,7}+X_{1,6} X_{2,5} X_{3,7}-X_{1,4} X_{1,6} X_{3,8}+X_{1,6}
   X_{2,4} X_{3,8}-X_{1,6} X_{2,5} X_{3,8}\\
   &-X_{1,6} X_{2,5} X_{4,7}+X_{1,6} X_{2,5} X_{4,8}-X_{1,6}
   X_{2,4} X_{5,8}+X_{1,4} X_{2,6} X_{5,8}-X_{1,4} X_{3,6} X_{5,8}\\
   &+X_{2,4} X_{3,6} X_{5,8}-X_{2,5}
   X_{3,6} X_{5,8}+X_{1,4} X_{3,7} X_{5,8}-X_{2,4} X_{3,7} X_{5,8}+X_{2,5} X_{3,7} X_{5,8}\\
   &+X_{2,5}
   X_{4,6} X_{5,8}-X_{2,5} X_{4,7} X_{5,8}-X_{1,4} X_{2,5} X_{6,8}-X_{1,4} X_{3,7} X_{6,8}+X_{2,4}
   X_{3,7} X_{6,8}\\
   &-X_{2,5} X_{3,7} X_{6,8}+X_{2,5} X_{4,7} X_{6,8}, \\
   \\
   R_{3,7}=&-X_{1,4} X_{2,7} X_{3,6}+X_{1,5} X_{2,7} X_{3,6}+X_{1,4} X_{2,8} X_{3,6}-X_{1,5} X_{2,8} X_{3,6}-X_{1,4}
   X_{3,8} X_{3,6}\\
   &+X_{1,5} X_{3,8} X_{3,6}+X_{2,4} X_{3,8} X_{3,6}-X_{2,5} X_{3,8} X_{3,6}-X_{2,8}
   X_{4,7} X_{3,6}+X_{2,7} X_{4,8} X_{3,6}\\
   &-X_{2,7} X_{5,8} X_{3,6}-X_{1,5} X_{2,7} X_{4,6}+X_{1,5}
   X_{2,8} X_{4,6}-X_{1,5} X_{3,8} X_{4,6}+X_{2,5} X_{3,8} X_{4,6}\\
   &-X_{2,7} X_{3,8} X_{4,6}+X_{1,5}
   X_{2,7} X_{4,7}-X_{1,6} X_{2,7} X_{4,7}-X_{1,5} X_{2,8} X_{4,7}+X_{1,6} X_{2,8} X_{4,7}\\
   &+X_{1,5}
   X_{3,8} X_{4,7}-X_{1,6} X_{3,8} X_{4,7}-X_{2,5} X_{3,8} X_{4,7}+X_{2,6} X_{3,8} X_{4,7}+X_{2,7}
   X_{4,6} X_{5,8}\\
   &-X_{2,7} X_{4,7} X_{5,8}+X_{2,7} X_{4,7} X_{6,8}.
\end{aligned}
\end{equation}
The contact part is given by
\begin{equation}
\begin{aligned}
    C^{(4)} =& -X_{1,5} X_{2,7}+X_{1,4} X_{2,7}+X_{1,6} X_{2,7}+X_{5,8} X_{2,7}-X_{6,8} X_{2,7}-X_{1,4} X_{2,8}+X_{1,5}
   X_{2,8}\\
   &-X_{1,6} X_{2,8}+X_{1,4} X_{3,6}-X_{1,5} X_{3,6}-X_{2,4} X_{3,6}+X_{2,5} X_{3,6}-X_{1,4}
   X_{3,7}+X_{1,5} X_{3,7}\\
   &-X_{1,6} X_{3,7}+X_{2,4} X_{3,7}-X_{2,5} X_{3,7}+X_{1,4} X_{3,8}-X_{1,5}
   X_{3,8}+X_{1,6} X_{3,8}-X_{2,4} X_{3,8}\\
   &+X_{2,5} X_{3,8}+X_{1,5} X_{4,6}-X_{2,5} X_{4,6}+X_{2,8}
   X_{4,6}-X_{1,5} X_{4,7}+X_{1,6} X_{4,7}+X_{2,5} X_{4,7}\\
   &-X_{2,6} X_{4,8}+X_{3,6} X_{5,8}-X_{3,7}
   X_{5,8}-X_{4,6} X_{5,8}+X_{4,7} X_{5,8}+X_{2,4} X_{6,8}+X_{3,7} X_{6,8}\\
   &-X_{4,7} X_{6,8}.
\end{aligned}
\end{equation}

\subsection{Surface kinematics and the one-loop gluon integrand}
\label{sec:2.3}

Just as at tree-level, we can think of the gluons entering into a one-loop process as scalar-scaffolded, such that their momenta and polarizations are given in terms of the momenta of the scalars by~\eqref{eq:GluonScalarMap}. Similarly, we can draw the scalar momentum polygon and inscribe it in a disk with marked points on the boundary, assigning boundary components of the disk to the momenta of the edges of the polygon just as in \cref{fig:ScalarScaff}. But now, at one-loop, the disk is a punctured disk. 

At tree-level, we saw that the space of curves on the disk was in one-to-one correspondence with the basis of kinematic invariants for the $2n$-point scalar problem; $X_{i,j}$ was naturally associated to the curve going from $i$ to $j$, where we could read off its momentum by \textit{homology}. However, at one-loop, due to the presence of the puncture, the space of curves now includes not only curves going from $i$ to $j$, but also those ending on the puncture $X_{i,p}$, where $p$ stands for the puncture index. These chords are precisely those kinematic invariants that should include the loop-momentum $l^\mu$.

\begin{figure}[t]
    \centering
    \includegraphics[width=\linewidth]{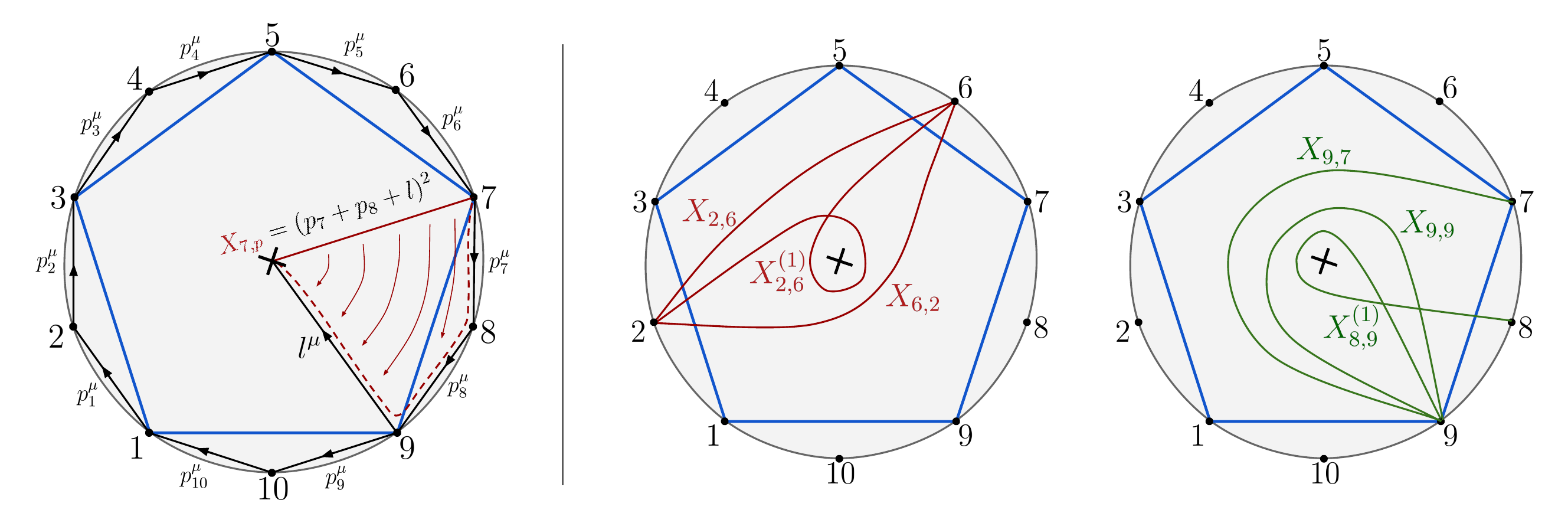}
    \caption{(Left) To define a basis of homology at one-loop, one has to further assign a momentum $l^\mu$ to one of the curves ending on the puncture. Doing this, we can then read off the momenta of the remaining curves; for example, with $X_{9,p} = l^2$, we find by homology that $X_{7,p}$ has momentum $(p_7+p_8+l)^2$. (Right) In red, we show examples of curves that are homologous to each other and therefore are assigned the same momentum in standard kinematics. In surface kinematics, we have instead $X_{2,6}^{(1)} = X_{2,6}$ and $X_{2,6} \neq X_{6,2}$. In green, we have drawn curves that are assigned zero momentum in standard kinematics, but which are given a non-zero labeling in surface kinematics. }
    \label{fig:Scaff-Oneloop}
\end{figure}

This means that, in order to be able to assign momentum to all the curves of the punctured disk, we have to associate momentum $l^\mu$ to one of the curves ending on the puncture; see the $l.h.s.$ of \cref{fig:Scaff-Oneloop}.\footnote{This is in addition to giving to each boundary curve an external gluon momentum, just as we did at tree-level.} Having done this, the situation is just like that at tree-level, where the space of curves of the punctured disk (up to homology) is in one-to-one correspondence with the space of scalar integrand invariants. By further going to the locus where the scaffolding curves $X_{2i-1,2i+1} =0$, we define a basis for the invariants of the gluon integrand for general spacetime dimension $d$ (large enough such that there are no Gram determinant constraints). 

The presence of the puncture introduces a novelty at loop-level not seen for trees: there are an infinite number of curves one can draw on the punctured disk up to homotopy, corresponding to the curves that self-intersect while going around the puncture (see the $r.h.s.$ of \cref{fig:Scaff-Oneloop}). However, since the puncture carries no momentum, these curves are all \textit{homologous} to each other, so that up to homology we still have a finite number of curves. For example, in \cref{fig:Scaff-Oneloop}, we draw three homotopically distinct curves which are all assigned the same momentum: the once-self-intersecting curve $X_{2,6}^{(1)}$, as well as $X_{2,6}$ and $X_{6,2}$, which do not self-intersect by go from $2$ to $6$ in different ways around the puncture. In particular, this means that curves which are homologous to boundary curves are assigned zero momentum by the on-shell condition. On the $r.h.s.$ of \cref{fig:Scaff-Oneloop}, we present examples of such curves in green. First, we have curve $X_{9,7}$, which is homologous to the scaffolding curve $X_{7,9} = 0$. We also show curve $X_{9,9}$ (homologous to nothing) and $X_{8,9}^{(1)}$ (homologous to a scalar boundary curve). The curves $X_{i,i}$ are dual to the propagators that appear in diagrams with tadpoles at one-loop~\cite{Gluons,YMIntegrand}. Of course, when we define the physical integrand, we usually remove these by-hand, since they give rise to scaleless integrals that vanish upon loop integration. Similarly, curves like $X_{9,7}$ are dual to the propagators of diagrams with external gluon bubbles, which we also remove from the integrand for the same reason; see Ref.~\cite{YMIntegrand} for more details.

Therefore, just like at tree-level, we can recast the gluon one-loop integrand using scalar variables $X_{i,j}$ that are naturally associated to the curves (up to homology) of the punctured disk. However, by doing this we will still run into the standard problems regarding gluon integrands: as just discussed, we will have to remove the contributions coming from tadpoles and external bubble diagrams (since these give $1/0$ divergences). This leads to a non-gauge-invariant object, whose cuts don't match gluing of lower-order amplitudes. 

It was proposed in Ref.~\cite{YMIntegrand} that these problems can be ameliorated by considering an integrand defined under a generalization of kinematics --- \textit{surface kinematics}. In this formalism, one really assigns a kinematic variable to \textit{each} curve on the surface: instead of identifying curves up to homology, we let curves $X_{i,j}$ and $X_{j,i}$ be different, but we still identify $X_{i,j}^{(q)}=X_{i,j}$ (for $q$ the self-intersection number) for all curves other than the boundary curves. For these curves, we crucially keep $X_{i,i+1}^{(1)}$ not equal to zero while still identifying with it all higher-self-intersecting curves. Thus, all the curves in green in \cref{fig:Scaff-Oneloop} have their own kinematic variable, other than $X_{2,6}^{(1)}$ which is set to be equal to $X_{2,6}$. 

The $n$-point YM integrand defined under surface kinematics --- which we will call the \textit{surface integrand} $\mathcal{I}_n \equiv \mathcal{I}_n(1, 2, \ldots, 2n)$ --- is naturally produced as the low-energy limit of a surface integral \cite{Gluons,YMIntegrand}. For most of this paper, the details of the surface integrals behind tree and loop gluon amplitudes are irrelevant, and we will review the important aspects as they occur in the rest of paper. For now, what is important is to discuss the remarkable features of this object --- namely how it manifests gauge invariance by generalizing Eq.~\eqref{eq:GgInvariance} and has a well-defined spin-sum gluing rule for the cuts as in Eq.~\eqref{eq:factorization}.\footnote{Maybe the most remarkable feature of the surface integrand is that it has a well-defined \textit{loop-cut} which matches the gluing of a tree-level object~\cite{YMIntegrand}. However, in the context of this paper, we will only need the information about tree-loop type cuts.}

According to Ref.~\cite{YMIntegrand}, gauge invariance and multi-linearity in gluon $i$ means that we can write the surface integrand as
\begin{equation}
\begin{aligned}
    \mathcal{I}_n &= \sum_{j\neq 2i} \left[ \left(X_{2i,j} - X_{2i-1,j}\right) \frac{\partial \mathcal{I}_n}{\partial X_{2i,j}} + \left(X_{j,2i} - X_{j,2i-1}\right) \frac{\partial \mathcal{I}_n}{\partial X_{j,2i}} \right] \\
    & \quad \quad \quad \quad\quad \quad +  X_{2i-2,2i-1} \times \left[ \frac{\partial \mathcal{I}_n}{\partial X_{2i-2,2i}} + \frac{\partial \mathcal{I}_n}{\partial X_{2i-2,2i-1}} \bigg \vert_{2i\rightarrow 2i-1} \right] \\
    &= \sum_{j\neq 2i} \left[ \left(X_{2i,j} - X_{2i+1,j}\right) \frac{\partial \mathcal{I}_n}{\partial X_{2i,j}} + \left(X_{j,2i} - X_{j,2i+1}\right) \frac{\partial \mathcal{I}_n}{\partial X_{j,2i}} \right] \\
    &\quad\quad\quad\quad\quad \quad  + X_{2i+1,2i+2} \times \left[ \frac{\partial \mathcal{I}_n}{\partial X_{2i,2i+2}} + \frac{\partial \mathcal{I}_n}{\partial X_{2i+1,2i+2} } \bigg \vert_{2i\rightarrow 2i+1} \right],
\end{aligned}
\label{eq:SurfaceGauge}
\end{equation}
where $j$ runs over all indices including the puncture. (Of course, when $j=p$ we only get a single factor inside the brackets since $X_{2i,p}=X_{p,2i}$.) This statement is very similar to the one found at tree-level \eqref{eq:GgInvariance}, differing only by a correction term proportional to boundary curves $X_{i,i+1} \equiv X_{i,i+1}^{(1)}$. Likewise, at one-loop the identity~\eqref{eq:gauge} generalizes nicely to the statement
\begin{equation}
\begin{aligned}
    &\sum_{j\neq 2i} \left[ \left(X_{2i+1,j} - X_{2i-1,j}\right) \frac{\partial \mathcal{I}_n}{\partial X_{2i,j}} + \left(X_{j,2i+1} - X_{j,2i-1}\right) \frac{\partial \mathcal{I}_n}{\partial X_{j,2i}} \right] +  X_{2i-2,2i-1} \left[ \frac{\partial \mathcal{I}_n}{\partial X_{2i-2,2i}} +\right. \\
    &\quad \quad \left.+ \frac{\partial \mathcal{I}_n}{\partial X_{2i-2,2i-1}} \bigg \vert_{2i\rightarrow 2i-1} \right] 
     - X_{2i+1,2i+2}  \left[ \frac{\partial \mathcal{I}_n}{\partial X_{2i,2i+2}} + \frac{\partial \mathcal{I}_n}{\partial X_{2i+1,2i+2} } \bigg \vert_{2i\rightarrow 2i+1} \right] = 0,
\end{aligned}
\label{eq:gaugeLoop}
\end{equation}
which once more is the same as \cref{eq:gauge} up to some correction terms proportional to curves $X_{i,i+1}$. 

Finally, on a cut of the integrand through a tree-level propagator (a ``tree-loop cut''), the residue $R_{i,j} \equiv \mathop{\mathrm{Res}}_{X_{i,j}=0}\mathcal{I}_n$ is given by~\cite{YMIntegrand}
\begin{equation}
\begin{aligned}    
R_{i,j}=& \sum_{k \in L, m\in R} \left[\left(X_{k,m}-X_{k,j}\hat{\theta}_{k,j-1} - X_{i,m}\right) \frac{\partial \mathcal{A}_L}{\partial X_{k,x_L}}\frac{\partial \mathcal{I}_R}{\partial X_{x_R,m}} \right. \\
&\left. \quad \quad \quad  \quad \quad+ \left(X_{m,k}-X_{k,j}\hat{\theta}_{k,j-1}  - X_{m,i}\hat{\theta}_{m,i-1} \right) \frac{\partial \mathcal{A}_L}{\partial X_{k,x_L}}\frac{\partial \mathcal{I}_R}{\partial X_{m,x_R}}\right]\\
&\quad + \sum_{k\in L} \left[\left(X_{k,p} - X_{i,p} -X_{k,j}\hat{\theta}_{k,j-1}\right)  \frac{\partial \mathcal{A}_L}{\partial X_{k,x_L}}\frac{\partial \mathcal{I}_R}{\partial X_{x_R,p}}\right] \\
& \quad + \sum_{k\in L} X_{i-1,i} \frac{\partial \mathcal{A}_L}{\partial X_{k,x_L}}\frac{\partial \mathcal{I}_R}{\partial X_{i-1,i}} \bigg\vert_{x_R \rightarrow i},
\end{aligned}  
\label{eq:tree-loop}
\end{equation}
where we are assuming the puncture is on the right of curve $X_{i,j}$ (see \cref{fig:factorization}), and have $\hat{\theta}_{i,k} = 1-\delta_{i,k}$, $L=\{i+1,i+2,\ldots,j-1\}$, and $R=\{j,j+1,\ldots,i-1,i\}$.

\subsection{One-loop one- and two- point surface integrands}

To finish this section, we provide some explicit examples of one-loop surface integrands for $n=1$ and $n=2$. 

The one-point gluon integrand is simply given by the tadpole diagram. In surface kinematics, this object has the following form: 
\begin{equation}
    \mathcal{I}_1 = \frac{2 X_{1,2}}{X_{1,p}} -\frac{(1+\Delta) 
   (X_{2,p}-X_{1,p})}{X_{1,p}},
\label{eq:one-pt-exam}
\end{equation}
where $\Delta = 1-d$ encodes the integrand's dependence on the spacetime dimensionality $d$. Note that there is manifest dependence on the boundary curve $X_{1,2}$; once we set it to zero, we are left with the standard scaleless tadpole integrand that integrates to zero. In addition, one can easily check that it satisfies the surface gauge-invariance~\eqref{eq:SurfaceGauge}.

For a slightly more interesting example, at $n=2$ we have the following surface integrand
\begin{equation}
\begin{aligned}
    \mathcal{I}_2 =& \frac{X_{2,p} \Delta }{X_{1,p}}-\frac{X_{3,p} \Delta }{X_{1,p}}+\frac{X_{4,p} \Delta }{X_{1,p}}-\frac{X_{2,p} X_{4,p}
   \Delta }{X_{1,p} X_{3,p}}+\frac{X_{4,p} \Delta }{X_{3,p}}+\frac{X_{3,p} X_{1,2} \Delta }{X_{1,p}
   X_{1,1}}-\frac{X_{4,p} X_{1,2} \Delta }{X_{1,p} X_{1,1}}+\frac{X_{1,2} X_{1,4} \Delta }{X_{1,p}
   X_{1,1}}\\
   &-\frac{X_{2,p} X_{1,4} \Delta }{X_{1,p} X_{1,1}}+\frac{X_{3,p} X_{1,4} \Delta }{X_{1,p}
   X_{1,1}}-\frac{X_{2,4} \Delta }{X_{1,p}}-\frac{X_{2,4} \Delta }{X_{3,p}}-\frac{X_{3,p} X_{2,4}
   \Delta }{X_{1,p} X_{1,1}}+\frac{X_{2,4} \Delta }{X_{1,1}}+\frac{X_{2,3} X_{3,4} \Delta
   }{X_{3,p} X_{3,3}}\\
   &+\frac{X_{1,p} X_{3,4} \Delta }{X_{3,p} X_{3,3}}-\frac{X_{2,p} X_{3,4} \Delta }{X_{3,p}
   X_{3,3}}-\frac{X_{1,p} \Delta }{X_{3,p}}+\frac{X_{2,p} \Delta }{X_{3,p}}-\frac{X_{4,p} X_{2,3} \Delta
   }{X_{3,p} X_{3,3}}+\frac{X_{1,p} X_{2,3} \Delta }{X_{3,p} X_{3,3}}-\frac{X_{1,p} X_{2,4} \Delta }{X_{3,p}
   X_{3,3}}\\
   &+\frac{X_{2,4} \Delta }{X_{3,3}}
   +\frac{X_{2,p}}{X_{1,p}}-\frac{X_{3,p}}{X_{1,p}}+\frac{X_{4,p}}{X_{1,p}}-\frac{X_{2,p} X_{4,p}}{X_{1,p}
   X_{3,p}}+\frac{X_{4,p}}{X_{3,p}}+\frac{X_{3,p} X_{1,2}}{X_{1,p} X_{1,1}}-\frac{X_{4,p} X_{1,2}}{X_{1,p}
   X_{1,1}}+\frac{3 X_{1,2} X_{1,4}}{X_{1,p} X_{1,1}}\\
   &-\frac{X_{2,p} X_{1,4}}{X_{1,p}
   X_{1,1}}+\frac{X_{3,p} X_{1,4}}{X_{1,p} X_{1,1}}-\frac{X_{1,1} X_{2,3}}{X_{1,p} X_{3,p}}+\frac{X_{1,4}
   X_{2,3}}{X_{1,p} X_{3,p}}+\frac{X_{1,1} X_{2,4}}{X_{1,p} X_{3,p}}-\frac{2 X_{2,4}}{X_{1,p}}-\frac{2
   X_{2,4}}{X_{3,p}}\\
   &-\frac{X_{3,p} X_{2,4}}{X_{1,p} X_{1,1}}+\frac{X_{2,4}}{X_{1,1}}+\frac{X_{1,1}
   X_{3,3}}{X_{1,p} X_{3,p}}-\frac{X_{1,2} X_{3,3}}{X_{1,p} X_{3,p}}-\frac{X_{1,4} X_{3,3}}{X_{1,p}
   X_{3,p}}+\frac{X_{2,4} X_{3,3}}{X_{1,p} X_{3,p}}-\frac{X_{1,1} X_{3,4}}{X_{1,p} X_{3,p}}\\
   &+\frac{X_{1,2}
   X_{3,4}}{X_{1,p} X_{3,p}}+\frac{3 X_{2,3} X_{3,4}}{X_{3,p} X_{3,3}}+\frac{X_{1,p} X_{3,4}}{X_{3,p}
   X_{3,3}}-\frac{X_{2,p} X_{3,4}}{X_{3,p} X_{3,3}}-\frac{X_{1,p}}{X_{3,p}}+\frac{X_{2,p}}{X_{3,p}}-\frac{X_{4,p}
   X_{2,3}}{X_{3,p} X_{3,3}}+\frac{X_{1,p} X_{2,3}}{X_{3,p} X_{3,3}}\\
   &-\frac{X_{1,p} X_{2,4}}{X_{3,p}
   X_{3,3}}+\frac{X_{2,4}}{X_{3,3}}-1-\Delta,
\label{eq:two-pt-exam}
\end{aligned}
\end{equation}
where, in addition to the boundary curves $X_{i,i+1} = X_{i,i+1}^{(1)}$, we now see the tadpole singularities $X_{1,1}=0$ and $X_{3,3}=0$.

%% file: Sections_v2/Soft_v2.tex
\newpage
\titlespacing*{\part}{10pt}{*4}{*2}
\part{ Soft Limits at Tree-Level and One-Loop}

\renewcommand\thesection{\arabic{section}} 
\titlespacing*{\section}{5pt}{*4}{*2}
\label{part:1}

We now proceed to the study of the soft expansion of tree-level YM amplitudes and of the one-loop integrand, both defined via scalar-scaffolding. As we will see, working in the scalar-scaffolded representation makes it especially easy both to rigorously define the soft limit and transparently understand how the soft expansion follows from factorization and gauge invariance.   

Before we actually take the soft limit, we must carefully define it. The most naive way of defining a soft limit in (say) the $n^{\rm{th}}$ particle is simply to take the momentum $q_n^\mu \to 0$. But this is not precise enough: for instance, if we do this by rescaling $q^\mu_n \to \varepsilon q^\mu_n$ and sending $\varepsilon \to 0$, momentum is not conserved for any $\varepsilon \neq 1$. So, instead, we must give a parametrization of $n$-point kinematics $q^\mu_i(\varepsilon),
\epsilon_i^\mu(\varepsilon)$ which starts with the $n$-point kinematics at $\varepsilon=1$ and ends with the $(n-1)$-point kinematics at $\varepsilon=0$, ensuring that both momentum conservation $\sum_i q_i^\mu =0$ and the on-shell conditions $q_i^2=0$, $\epsilon_i \cdot q_i =0$ are true for all $\varepsilon$. We will see in this \cref{part:1} that the scalar-scaffolding picture suggests a simple and natural way of doing this, by directly deforming the momentum polygon. As we emphasized in our discussion of kinematics, working with $X$ variables trivializes momentum conservation, and maintaining the on-shell conditions only forces that the $X$'s for boundary and scaffolding are kept at zero. 

As also emphasized in the previous section, the scalar-scaffolded representation of the gluon amplitude has a completely \textit{locked} form, free of gauge redundancies while retaining cyclic invariance; instead, gauge invariance is reflected in the form of the amplitude given in \cref{eq:GgInvariance} at tree-level and in \cref{eq:SurfaceGauge} at one-loop. This allows us to canonically define our soft limit as a Laurent expansion in a set of ``soft factors.'' 

We will start at tree-level, reproducing the leading Weinberg soft factor as a consequence of factorization (the gluing rule of  Eq.~\eqref{eq:factorization}) together with gauge invariance. We then use the fact that the amplitude satisfies the gauge-invariance identity~\eqref{eq:gauge}
to determine the subleading term in the expansion as well as put certain constraints (in the form of sum rules) on all higher order terms. (We describe these derivations in more detail in \cref{app:HigherOrdSoft}.)

We then turn to looking at the soft expansion of the one-loop integrand. As we have mentioned, ordinary loop integrands for gluons are not in general well-defined, due to the $1/0$ divergences associated with tadpoles and external bubbles. Even if these are manually removed in some way, the integrands are no longer gauge invariant and do not factorize correctly on cuts. Thus, we do not expect any nice behavior in the soft limit for integrands, only for amplitudes post-loop integration. But, as we reviewed in the previous section, the one-loop surface integrand offers us hope, as it {\it does} satisfy the two properties crucial to obtaining our results at tree-level: (surface) gauge invariance and factorization. By generalizing the soft expansion at tree-level to surface kinematics at one-loop, we find that, at leading order, the surface integrand yields precisely the Weinberg soft term plus a correction which manifestly vanishes upon integration. This is a remarkable fact: while the Weinberg soft theorem is known to be all-loop-orders exact, it is surface kinematics that gives us the kinematical structure necessary to promote this statement to loop integrand level! We also discuss higher-order terms in the one-loop soft expansion in App.~\ref{app:highorderLoop}.

\section{Defining the Soft Limit at Tree-Level}
\label{sec:3-1}

As referenced in the preamble, here we will explicitly define a (minimal) soft limit in scalar variables which we will use to expand the YM amplitude and one-loop integrand. However, before doing this, let's first analyze this question in the standard language of polarization vectors and gluon momenta, where the on-shell conditions ($q_i^2=0$ and $\epsilon_i \cdot q_i =0$) are not made manifest. 

To do this, consider the $n$-point gluon momentum polygon \textit{before} scaffolding, where each edge corresponds to the momentum of an external gluon. Label the vertices by dual coordinates $x_i^\mu$, such that each gluon momentum is given by $q_i^\mu = (x_{i+1}-x_{i})^\mu$. In this geometric picture, we can naively take the $n^{\rm{th}}$ gluon soft in infinitely many ways: all we must do is map this $n$-point polygon to \textit{any} $(n-1)$-point polygon, where our only requirement is that $x_1^\mu$ and $x_n^\mu$ must move together to a common vertex $x_s^\mu$ of the smaller polygon so that $x_1^\mu - x_n^\mu = q_n^\mu \to 0$. Minimally, however, we can achieve this limit by leaving all $x_i^\mu$ for $i \neq 1, n$ fixed and only moving $x_1^\mu$ and $x_n^\mu$ together to $x_s^\mu$. Defined as a deformation on the closed momentum polygon, this limit guarantees momentum conservation holds at all times, but we will still need to enforce the on-shell conditions.

In the simplest incarnation, we have that $x_1^\mu$ and $x_n^\mu$ follow a straight line until $x_s^\mu$, giving us:
\begin{equation}
\begin{gathered}
    \begin{tikzpicture}[line width=1,scale=10]

    \node[regular polygon, regular polygon sides=22,  minimum size=3cm] (p) at (0,0) {};

   \node[scale=0.8,xshift=-12,yshift=-5] at (p.corner 7) {$x_2^\mu$};
   \node[scale=0.8,xshift=-10,yshift=-10] at (p.corner 11) {$x_1^\mu$};
   \node[scale=0.8,xshift=10,yshift=-10] at (p.corner 15) {$x_n^\mu$};
   \node[scale=0.8,xshift=17,yshift=0] at (p.corner 19) {$x_{n-1}^\mu$};
    \node[scale=0.8,xshift=0,yshift=10,Blue] at (p.center) {$x_s^\mu$};
    \node[scale=0.8,xshift=-20,yshift=-20,Blue] at (p.center) {$v_1^\mu$};
     \node[scale=0.8,xshift=25,yshift=-13,Blue] at (p.center) {$v_n^\mu$};

    \draw[fill] (p.corner 7) circle [radius=0.1pt];
    \draw[fill] (p.corner 11) circle [radius=0.1pt];
    \draw[fill] (p.corner 15) circle [radius=0.1pt];
    \draw[fill] (p.corner 19) circle [radius=0.1pt];
    \draw[fill,Blue] (p.center) circle [radius=0.1pt];

    \path[Black] (p.corner 6) edge (p.corner 7);
    \path[Black] (p.corner 7) edge (p.corner 11);
    \path[Black] (p.corner 15) edge (p.corner 11);
    \path[Black] (p.corner 15) edge (p.corner 19);
    \path[Black] (p.corner 19) edge (p.corner 20);
    \draw[Gray,dashed,opacity=0.5] (p.corner 19) -- (p.corner 20)-- (p.corner 21) --(p.corner 22)--(p.corner 1)--(p.corner 2)--(p.corner 3)--(p.corner 4)--(p.corner 5)--(p.corner 6)--(p.corner 7);

    \begin{scope}[every node/.style={allow upside down, sloped}]
  \draw (p.corner 11)-- node {\midarrow} (p.corner 7);
  \draw (p.corner 15)-- node {\midarrow} (p.corner 11);
  \draw (p.corner 19)-- node {\midarrow} (p.corner 15);
  \draw[Blue] (p.corner 15)-- node {\midarrow} (p.center);
  \draw[Blue] (p.corner 11)-- node {\midarrow} (p.center);
    \end{scope}
    \end{tikzpicture}     
\end{gathered} \quad \Rightarrow \quad \begin{cases}
    x_1^{\mu}(\varepsilon) = x_s^\mu - \varepsilon \underbracket[0.4pt]{(x_s^\mu - x_1^\mu)}_{v_1^\mu} \\
    x_n^{\mu}(\varepsilon) =  x_s^\mu - \varepsilon \underbracket[0.4pt]{(x_s^\mu - x_n^\mu)}_{v_n^\mu}
\end{cases},
\label{eq:soft-limit}
\end{equation}
where $v_1^\mu$ and $v_n^\mu$ are the vectors connecting $x_1^\mu$ and $x_n^\mu$ to $x_s^\mu$, respectively, which satisfy $v_n^\mu = v_1^\mu + q_n^\mu$. The soft limit itself corresponds to changing the parameter $\varepsilon$ smoothly from $1$ to $0$. Note that, since we are only deforming $x_1^\mu$ and $x_n^\mu$, we only affect the momenta of the two gluons adjacent to the soft gluon, $q_1^\mu$ and $q_{n-1}^\mu$. This means that our minimal mapping trivially keeps all gluons $2$ through $n - 2$ on-shell at all times. We therefore only need to worry about what happens with gluons $1$, $n$, and $n - 1$.

Starting with gluon $n$, under this path the momentum $q_n^\mu(\varepsilon)$ is simply given by
\begin{equation}
  q_n^\mu(\varepsilon) = (x_1-x_n)^\mu(\varepsilon) = \varepsilon  (x_1-x_n)^\mu =  \varepsilon q_n^\mu,
\end{equation}
so gluon $n$ remains on-shell during the limit. For gluon $1$, we have that 
\begin{equation}
    q_1^\mu(\varepsilon) = (x_2 - x_s)^\mu + \varepsilon(x_s^\mu - x_1^\mu) = q_1^\mu + (\varepsilon-1) v_1^\mu. 
\end{equation}
As such, keeping $q_1^2(\varepsilon)=0$ for any $\varepsilon$ requires us to constrain $v_1^\mu$ with $v_1^2=0$ and $v_1\cdot q_1=0$. Further ensuring that $\epsilon_1 \cdot q_1(\varepsilon) =0$, we also need $v_1 \cdot \epsilon_1 =0$. 

For gluon $n-1$, we analogously find
\begin{equation}
    q_{n-1}^\mu (\varepsilon) = (x_s - x_{n-1})^\mu + \varepsilon(x_n - x_s)^\mu = q_{n-1}^\mu + (1-\varepsilon) v_n^\mu,  
\end{equation}
so asking for $q_{n-1}^2(\varepsilon)=0$ similarly implies that $v_n^2=0$ and $v_n\cdot q_{n-1}=0$. Finally, imposing $\epsilon_{n-1}\cdot q_{n-1}(\varepsilon)=0$ gives the condition that $\epsilon_{n-1} \cdot v_n =0$. Therefore, in order to keep everyone on-shell while changing only the momenta of the adjacent gluons, we must impose the following constraints on $v_1^\mu$ and $v_n^\mu$:
\begin{equation}
\begin{aligned}
   \underline{\text{Gluon 1}}:& \quad  v_1^2 =0, \quad q_1 \cdot v_1 =0, \quad \epsilon_1 \cdot v_1 =0,\\
    \underline{\text{Gluon }n-1}:& \quad  v_n^2 =0, \quad q_{n-1} \cdot v_n =0, \quad \epsilon_{n-1} \cdot v_n =0.
\end{aligned}
\label{eq:soft-conds}
\end{equation}
Note that these constraints are gauge invariant, as required. 

Since, $e.g.$, $v_n^\mu$ can be written as a function of $v_1^\mu$ through the equality $v_n^\mu = v_1^\mu + q_n^\mu$, Eq.~\eqref{eq:soft-conds} constitutes a total of six constraints on the vector $v_1^\mu$. As such, there certainly exists a solution for $x_s^\mu$ in six dimensions or higher. But, given our universe, it is natural to ask if there is a way to realize this limit in $d = 4$. To explore this question, it is best to work in spinor-helicity variables, where we will represent the momenta of gluons $1$, $n$ and $n-1$ as
\begin{equation}
    (q_1)_{\alpha,\dot{\alpha}} = \lambda_{1,\alpha}\tilde{\lambda}_{1,\dot{\alpha}},\quad (q_n)_{\alpha,\dot{\alpha}} = \lambda_{n,\alpha}\tilde{\lambda}_{n,\dot{\alpha}},  \quad (q_{n-1})_{\alpha,\dot{\alpha}} = \lambda_{n-1,\alpha}\tilde{\lambda}_{n-1,\dot{\alpha}},
\end{equation}
and their polarizations as
\begin{equation}
    (\epsilon_i^{-})_{\alpha,\dot{\alpha}}= \frac{\lambda_{i,\alpha}\tilde{\mu}_{i,\dot{\alpha}}}{[\tilde{\lambda}_i \tilde{\mu}_i]}, \quad (\epsilon_i^{+})_{\alpha,\dot{\alpha}}= \frac{\mu_{i,\alpha}\tilde{\lambda}_{i,\dot{\alpha}}}{\langle \lambda_i \mu_i \rangle},
\end{equation}
with $\mu_i, \tilde{\mu}_i$ reference spinors. Since $v_1^\mu$ and $v_n^\mu$ are also null vectors by the conditions in \cref{eq:soft-conds}, we can similarly write them as spinors:
\begin{equation}
    (v_1)_{\alpha,\dot{\alpha}} = \lambda^{v_1}_{\alpha}\tilde{\lambda}^{v_1}_{\dot{\alpha}}, \quad (v_n)_{\alpha,\dot{\alpha}} = \lambda^{v_n}_{\alpha}\tilde{\lambda}^{v_n}_{\dot{\alpha}}.
\end{equation}
Then, the condition $v_n^\mu = v_1^\mu + q_n^\mu$ nicely turns into the standard three-point kinematics constraint, which means that we have either
\begin{equation}
    \lambda^{v_1} \propto \lambda^{v_n}\propto \lambda_{n}, \quad \text{ or }\quad \tilde{\lambda}^{v_1} \propto \tilde{\lambda}^{v_n}\propto \tilde{\lambda}_{n}.
\label{eq:3partConstraint}
\end{equation}
Let's now consider the situation where both gluons $1$ and $n-1$ have positive helicity, and choose the $\lambda$'s to be parallel. Then, we can satisfy the conditions in \cref{eq:soft-conds} by picking
\begin{equation}
    v_1 \propto \lambda_n  \tilde{\lambda}_1, \quad  v_n \propto \lambda_n  \tilde{\lambda}_{n-1}. 
\end{equation}
However, if we had instead chosen the $\tilde{\lambda}$'s to be parallel, it is easy to conclude that there is no solution: we would need to pick the forbidden values of the reference spinors to satisfy all the conditions.

In contrast, if we take gluons $1$ and $n-1$ to both have negative helicity, then we find there is \textit{no} solution if we choose the $\lambda$'s to be parallel, but there is a solution for $\tilde{\lambda}$'s parallel given by 
\begin{equation}
     v_1 \propto \lambda_1  \tilde{\lambda}_n, \quad  v_n \propto \lambda_{n-1}  \tilde{\lambda}_{n}. 
\end{equation}
Finally, there is unsurprisingly no solution when the helicities of gluons $1$ and $n-1$ are opposite. Therefore, we conclude that the soft limit as defined above is realizable in $d = 4$ as long as gluons $1$ and $n-1$ have the \textit{same} helicity.

So, having carefully defined our limit in the standard gluon-amplitude language, let us understand how this particular limit translates into the scalar variables $X_{i,j}$ introduced in the previous section. For now, we'll do this for the case at tree-level, and we will explain how it generalizes to the one-loop surface integrand in \cref{sec:3.1}. We start by placing the gluon momentum polygon inside the $2n$-point scalar polygon, where we map $x_i^\mu \to x_{2i - 1}^\mu$.

Since, by definition, we are only moving $x_1^\mu$ and $x_{2n-1}^\mu$ during the soft limit, the only chords that change are $X_{1,j}$ and $X_{j,2n-1}$. They evolve as
\begin{equation}
\begin{aligned}
    &X_{1,j}(\varepsilon) = [x_j-x_1(\varepsilon)]^2 = [(x_j-x_s) + \varepsilon (x_s - x_1)]^2 = X_{s,j} + \varepsilon (X_{1,j} - X_{j,s}), \\
    &X_{j,2n-1}(\varepsilon) = [x_j-x_{2n-1}(\varepsilon)]^2 = [(x_j-x_s) + \varepsilon (x_s - x_{2n-1})]^2 = X_{s,j} + \varepsilon (X_{j,2n-1} - X_{j,s}), 
\end{aligned}
\label{eq:softlim}
\end{equation}
where $X_{s,j}$ are the chords on the final polygon we obtain after colliding points $x_1^\mu$ and $x_{2n-1}^\mu$ together to $x_s^\mu$. To ensure the gluons remain on-shell throughout the soft limit we must have: $X_{1,3}(\varepsilon) = X_{2n-1,2n-3}(\varepsilon) = 0$ and $X_{1,2}(\varepsilon) = X_{2n-1,2n-2}(\varepsilon) = 0$. This constrains $x_s$ to be null separated from $(1,2,3)$ as well as $(2n-3,2n-2,2n-1)$, which is exactly what we found in \eqref{eq:soft-conds}. In scalar variables, we see from the above that the objects controlling the perturbative soft expansion are
\begin{equation}
    \delta_j^{(1)} = \varepsilon (X_{1,j} - X_{s,j}), \quad \delta_j^{(2n-1)} = \varepsilon (X_{j,2n-1} - X_{s,j}),
\end{equation}
which we will refer to as our ``soft factors.'' Note that the $\delta_j^{(1)}$ are non-vanishing only for $j = 4, 5, \ldots, 2n - 2$ and the $\delta_j^{(2n-1)}$ for $j = 2, 3, \ldots, 2n-4$. Therefore, we  \textit{define} the soft expansion of the scalar-scaffolded YM amplitude $\mathcal{A}_n$ as a Laurent expansion in $\delta^{(1)}_j,\delta^{(2n-1)}_j$. 

We stress that, given that the $X$'s provide a basis for the full gluon kinematic space, it is not necessary to refer directly to the momentum polygon picture to define soft limits. {\it Any} motion in $X$ space that leaves the scaffolding $X$'s untouched (at zero) and where $X_{1,j}, X_{2n-1,j}$ are identified (at the end) {\it defines} a consistent soft limit. Of course, as discussed above, in any fixed spacetime dimension the $X$'s are {\it not} in general all independent, and therefore soft maps in $X$ space might not be reachable given the dimension constraints. We can also understand that any soft map in $X$ space which does not touch $X_{2,k}$ and $X_{2n-2,k}$ also leaves the polarizations of the adjacent gluons unchanged, since we are free to pick $\epsilon_1^\mu = p_2^\mu =(x_3 - x_2)^\mu $ and $\epsilon_{n-1}^\mu  = p_{2n-3}^\mu  = (x_{2n-2}-x_{2n-3})^\mu $.

In some important cases, these soft factors are not actually differences in $X$ variables but rather equivalent to a single $X$. In particular, since $X_{s,2n-3}= X_{s,3}=0$ and $X_{s,2}= X_{s,2n-2}=0$, we have during the soft limit that
\begin{equation}
\begin{aligned}
    X_{2,2n-1}(\varepsilon) = \delta_2^{(2n-1)}, \\
    X_{3,2n-1}(\varepsilon) = \delta_3^{(2n-1)},
\end{aligned} \quad \quad 
\begin{aligned}
    X_{1,2n-2}(\varepsilon)  = \delta_{2n-2}^{(1)}, \\
    X_{1,2n-3}(\varepsilon)  = \delta_{2n-3}^{(1)}.
\end{aligned}
\label{eq:softlimit2}
\end{equation}
The tree-level YM amplitude has poles when $X_{1,2n-3}, X_{3,2n-1} \to 0$, which correspond to gluon $n$ becoming collinear with the neighboring gluons $1$ and $n - 1$, respectively. This means --- just as with the famous leading Weinberg pole --- the leading order term in the soft expansion will localize on these two factorization channels.

\subsection{Setting up the soft expansion}
\label{sec:1.1}

Having defined the soft limit in scalar-scaffolded variables, let's now look a bit closer at the structure of the amplitude $\mathcal{A}_n$. As we just explained, the only two soft factors which appear as poles in $\mathcal{A}_n$ are $\delta_3^{(2n-1)}=X_{3,2n-1}(\varepsilon)$ and $\delta_{2n-3}^{(1)}=X_{1,2n-3}(\varepsilon)$. Therefore, to perform the soft expansion, we will start by explicitly writing out the dependence on these two poles as follows:
\begin{equation}
    \mathcal{A}_{n} = \frac{R_{1,2n-3}}{X_{1,2n-3}} +  \frac{R_{3,2n-1}}{X_{3,2n-1}} + \mathcal{R},
    \label{eq:LaurentSeries}
\end{equation}
where $R_{1,2n-3}$ and $R_{3,2n-1}$ are the residues of $\mathcal{A}_{n}$ at $X_{1,2n-3}=0$ and $X_{3,2n-1}=0$, respectively, and $\mathcal{R}$ is everything else, which does \textit{not} have poles in either $X_{1,2n-3}$ or $X_{3,2n-1}$. Note that, even though $R_{1,2n-3}$ and $R_{3,2n-1}$ are only gauge invariant on the support of $X_{1,2n-3}=0$ and $X_{3,2n-1}=0$, respectively, since the amplitude has a locked form, decomposing it in this way is completely canonical. 

In the next subsection, we use the explicit form of the residues, $R_{1,2n-3}$ and $R_{3,2n-1}$, as given in \cref{eq:factorization}, to obtain the leading Weinberg soft theorem, which is of order  $\mathcal{O}(\delta^{-1})$. But let us start with some remarks on how higher order terms in the soft expansion can also be derived (in the case of the subleading) and constrained (in the case of sub-subleading and higher) by using gauge invariance in gluon $n$. We will leave most of the technical aspects of this discussion for App.~\ref{app:HigherOrdSoft}, but we show here the main result which constrains higher order terms in the soft expansion. 

Starting at $\mathcal{O}(\delta^0)$, $\mathcal{R}$ plays a role in the soft expansion of $\mathcal{A}_n$. And, unlike for the residues $R_{1,2n-3}$ and $R_{3,2n-1}$, which are known to all soft orders (and written out in App.~\ref{app:HigherOrdSoft}), we seem to have no formula to control its soft expansion. However, knowing that $\mathcal{A}_n$ must be gauge invariant in the $n^{\rm{th}}$ gluon means we must be able to write $\mathcal{A}_{n}$ as
\begin{equation}
\label{eq:An-form-gaug-1}
    \frac{\sum_j (X_{j,2n}-X_{1,j}) \partial_{X_{j,2n}} R_{1,2n-3}}{X_{1,2n-3}} + \frac{\sum_j (X_{j,2n}-X_{1,j}) \partial_{X_{j,2n}} R_{3,2n-1}}{X_{3,2n-1}} + \sum_j (X_{j,2n}-X_{1,j}) C_j,
\end{equation}
where $j = 2, 3, \ldots, 2n-2$ (here and everywhere else in this subsection) and the $C_j$ are defined by
\begin{equation}
    C_j = \frac{\partial}{\partial X_{j,2n}} \left(\mathcal{A}_{n} - \frac{R_{1,2n-3}}{X_{1,2n-3}} -  \frac{R_{3,2n-1}}{X_{3,2n-1}} \right),
\end{equation}
where we use the fact that gauge invariance in $n$ implies linearity in all $X_{j,2n}$.

Now, since $R_{3,2n-1}$ is gauge invariant in the $n^{\rm{th}}$ gluon (on the locus of $X_{3,2n-1}=0$), we can write $R_{3,2n-1}$ as $\sum_j (X_{j,2n}-X_{1,j}) \partial_{X_{j,2n}} R_{3,2n-1}$. So, the term in the numerator of $X_{3,2n-1}$ in \cref{eq:An-form-gaug-1} is precisely $R_{3,2n-1}$. Similarly, we can write $R_{1,2n-3}$, on the locus of $X_{1,2n-3}=0$,  as  $\sum_j (X_{j,2n}-X_{1,j}) \partial_{X_{j,2n}} R_{1,2n-3}$. However, in this case we have to correct for the fact that the term in the numerator of $X_{1,2n-3}$ in \cref{eq:An-form-gaug-1} does \textit{not} have $X_{1,2n-3} = 0$. Therefore, the numerator of $X_{1,2n-3}$ comes out to $R_{1,2n-3} - X_{1,2n-3} \partial_{X_{2n-3,2n}} R_{1,2n-3}$. Our amplitude then takes the form
\begin{equation}
    \mathcal{A}_{n} =  \frac{R_{1,2n-3}}{X_{1,2n-3}} +  \frac{R_{3,2n-1}}{X_{3,2n-1}} + \sum_j (X_{j,2n}-X_{1,j}) C_j - \frac{\partial R_{1,2n-3}}{\partial X_{2n-3,2n}}.
    \label{eq:rep1}
\end{equation}
But, recall that gauge invariance and linearity in the polarization of the $n^{\rm{th}}$ gluon means that we can also write the amplitude in the form of the first line in \cref{eq:GgInvariance}. Using the exact same line of reasoning as above, this gives that
\begin{equation}
    \mathcal{A}_{n} =  \frac{R_{1,2n-3}}{X_{1,2n-3}} +  \frac{R_{3,2n-1}}{X_{3,2n-1}} + \sum_j (X_{j,2n}-X_{j,2n-1}) C_j - \frac{\partial R_{3,2n-1}}{\partial X_{3,2n}}.
    \label{eq:rep2}
\end{equation}
Therefore, if we \textit{subtract} \cref{eq:rep1} and \cref{eq:rep2}, we obtain 
\begin{equation}
 \sum_j (X_{j,2n-1}-X_{1,j}) C_j =   -\frac{\partial R_{3,2n-1}}{\partial X_{3,2n}} + \frac{\partial R_{1,2n-3}}{\partial X_{2n-3,2n}}.
    \label{eq:constContact}
\end{equation}
Note that, so far, this equation is exact; however, if we apply the soft limit to Eq.~\eqref{eq:constContact}, we can replace $(X_{j,2n-1} - X_{1,j}) = (\delta_j^{(2n-1)} - \delta_j^{(1)})$, and then we find
\begin{equation}
     \sum_j \delta_j^{(2n-1)} C_j - \sum_j \delta_j^{(1)} C_j=   -\frac{\partial R_{3,2n-1}}{\partial X_{3,2n}} + \frac{\partial R_{1,2n-3}}{\partial X_{2n-3,2n}}.
    \label{eq:constraintContact2}
\end{equation}
Taylor-expanding the $C_j$ in soft factors gives us
\begin{equation}
\begin{aligned}
    C_j = C_j^{(0)} + \sum_k \left(\delta^{(1)}_k  C_{j,\{(k;1)\}} + \delta^{(2n-1)}_k  C_{j,\{(k;2n-1)\}} \right)  + \cdots,     
\end{aligned}
\end{equation}
where $C_{j,\{(k_1,\ldots,k_m;1),(l_1,\ldots,l_r;2n-1)\}} = ( \partial_{X_{1,k_1}} \cdots  \partial_{X_{1,k_m}} ) ( \partial_{X_{l_1,2n-1}} \cdots  \partial_{X_{l_r,2n-1}} ) C_j$. Plugging this into Eq.~\eqref{eq:constraintContact2}, we see that the leading $C_j^{(0)}$ are fully determined by matching to the soft expansion of the $r.h.s.$. As such, it is possible --- using just factorization and gauge invariance --- to derive the subleading term in the soft expansion of $\mathcal{A}_n$.

However, clearly such a strategy will not work for higher order terms in the expansion. Indeed, already at the next order, on the $l.h.s.$ we find: 
\begin{equation}
   \sum_{j,k} \left( \delta_j^{(2n-1)}\delta_k^{(2n-1)} C_{j,\{(k;2n-1)\}} -  \delta_j^{(1)}\delta_k^{(1)} C_{j,\{(k;1)\}}
    + \delta_{j}^{(1)}\delta_{k}^{(2n-1)} ( C_{k,\{(j;1)\}}  - C_{k,\{(j;2n-1)\}}) \right).
\end{equation}
Matching with $r.h.s.$, we are thus only able to determine the totally symmetric part of the tensors $C_{j,\{(k;2n-1)\}}$ and $C_{j,\{(k;1)\}}$, as well as the full linear combination $(-C_{k,\{(j;1)\}}  - C_{j,\{(k;2n-1)\}})$. Now, clearly this doesn't entirely fix these matrices: in particular, if we shift them as $C_{k,\{(j;1)\}} \to C_{k,\{(j;1)\}} +A_{k,j}$ and $C_{j,\{(k;2n-1)\}} \to C_{j,\{(k;2n-1)\}} + A_{k,j}$ with $A_{k,j}$ skew-symmetric, then $A_{k,j}$ just drops out from Eq.~\eqref{eq:constraintContact2}. We therefore see that --- using just gauge invariance in the $n^{\rm{th}}$ gluon --- we can only determine the soft expansion up to $\mathcal{O}(\delta^0)$, corresponding to the leading and first subleading terms.\footnote{Such a result is unsurprising given recent work in Ref.~\cite{Wei:2024ynm} demonstrating the non-existence of universal soft terms past sub-leading order.}  However, despite this restriction, we should pause to appreciate the remarkable fact that we can say \textit{anything} about the $C_j$. Since gauge invariance gave us the constraint \eqref{eq:constContact}, such a statement would be \textit{impossible} to make if we were dealing with, say, scalars and not gluons. So, it is clear that gauge invariance plays a critical role in the structure and universality of the soft expansion.

\section{The Leading Soft Theorem at Tree-Level}

Let us now derive the leading order term in the soft expansion of $\mathcal{A}_n$. Starting with $R_{1,2n-3}$, we use Eq.~\eqref{eq:factorization} to write
\begin{equation}
    R_{1,2n-3}= \sum_{j \in L, \ J\in \{2n-2,2n-1,2n\}} (X_{j,J}-X_{1,j} -X_{2n-3,J}) \frac{\partial M_L}{\partial X_{j,x_L}}  \times \frac{\partial M_R}{\partial X_{J,x_R}},
    \label{eq:factX1}
\end{equation}
where $L = \{ 2, 3, \ldots, 2n -4 \}$, $M_L = \mathcal{A}_{n-1}(1,2,\ldots,2n-4,2n-3,x_L)$ and $M_R = \mathcal{A}_3(1,x_R,2n-3,2n-2,2n-1,2n)$, which is simply a three-point gluon amplitude. We can therefore use the explicit form of $M_R$ given in \cref{eq:3ptGluons} to write out the derivatives $\partial_{X_{J,x_R}} M_R$ appearing in Eq.~\eqref{eq:factX1}. We'll do exactly this in detail in App.~\ref{app:HigherOrdSoft}; for now, since we only care about the leading order contribution, we can note that the only derivative that contributes is for $J = 2n-2$, $\partial_{X_{2n-2,x_R}} M_R= X_{2n-3,2n}$. So, we find
\begin{equation}
    R_{1,2n-3} = X_{2n-3,2n} \times  \underbracket[0.4pt]{\sum_{j = 2}^{2n-4}(X_{j,2n-2} - X_{s,j})\frac{\partial M_L}{\partial X_{j,2n-2}}}_{M_L(1,2,\ldots, 2n-2)} + \mathcal{O}(\delta),
\end{equation}
where we have replaced $x_L \to 2n-2$. Now, using the gauge-invariance statement in gluon $n - 1$, we can recognize the sum above to be exactly the amplitude $M_L$. Further noting that, at leading order in $\delta_j^{(1)}$, $M_L(1,2,\ldots, 2n-2) \to \mathcal{A}_{n-1}(s, 2, 3, \ldots, 2n-2) \equiv \mathcal{A}_{n-1}$, we find
\begin{equation}
    R_{1,2n-3} = X_{2n-3,2n} \times \mathcal{A}_{n-1} + \mathcal{O}(\delta).
\label{eq:LeadR1}
\end{equation}
In other words, at leading order, the spin sum coming from a cut on an intermediate gluon precisely turns into the gauge-invariance statement, which allows us to recover just the lower-point amplitude!

Repeating this analysis for $R_{3,2n-1}$, where now the two lower-point amplitudes are $M_L=\mathcal{A}_3(1,2,3,x_L,2n-1,2n)$ and $M_R = \mathcal{A}_{n-1}(2n-1, x_R, 3,4,\ldots,2n-2)$, the residue formula tell us that
\begin{equation}
     R_{3,2n-1}= \sum_{j \in R, \ J\in \{2,1,2n\}} (X_{j,J}-X_{j,2n-1} -X_{3,J}) \frac{\partial M_L}{\partial X_{J,x_L}}  \times \frac{\partial M_R}{\partial X_{j,x_R}},
\end{equation}
with $R = \{4,5,\ldots,2n-2\}$. Using again the explicit form of the three-point amplitude $M_L$, the only term that contributes at $\mathcal{O}(\delta^0)$ in the $J$-sum (see App.~\ref{app:HigherOrdSoft} for details) comes from $J=2$. We find 
\begin{equation}
\begin{aligned}
    R_{3,2n-1}&= X_{3,2n}\times \sum_j(X_{2,j} - X_{j,2n-1})\frac{\partial M_R}{\partial X_{j,2}} + \mathcal{O}(\delta),
\end{aligned}
\end{equation}
where we have relabeled $x_R = 2$. Now, just as for the previous residue, the sum over $j$ above is precisely equal to $M_R$ due to gauge invariance in gluon $2$. Further expanding $M_R(2n-1, 2, 3,\ldots, 2n-2)$ in powers of $\delta_j^{(2n-1)} = X_{j,2n-1} - X_{j,s}$ gives us
\begin{equation}
    R_{3,2n-1} = X_{3,2n} \times \mathcal{A}_{n-1} + \mathcal{O}(\delta),
\end{equation}
where, unsurprisingly, we see the appearance of precisely the same $(n-1)$-point amplitude as in Eq.~\eqref{eq:LeadR1}.

Therefore, at leading order in soft expansion we find
\begin{equation}
    \mathcal{A}_{n} \to \left( \frac{X_{2n-3,2n}}{X_{1,2n-3}} + \frac{X_{3,2n}}{X_{3,2n-1}} \right) \mathcal{A}_{n-1}(s,2,3,4,\cdots,2n-2) + \mathcal{O}(\delta^0).
    \label{eq:LeadSoft}
\end{equation}

As we emphasized earlier, when written in terms of the scalar $X_{i,j}$ variables, the gluon amplitude has a completely unique form. Thus, the soft expansion is well-defined. However, the leading order term (as well as the subleading term discussed in App.~\ref{app:HigherOrdSoft}) given above is \textit{not} completely gauge invariant: while trivially gauge invariant in gluons $1$ through $n - 1$ (due to $\mathcal{A}_{n-1}$), we'll show in the next section that it is only gauge invariant in the $n^{\rm{th}}$ gluon up to a subleading correction. This is not a problem: we will describe a simple operation which allows us to ``gauge-invariantify'' the leading soft term. 
This reflects a general fact of our soft expansion in scalar-scaffolded variables, that follows from the more general statement about the Laurent expansion of the amplitude we have already alluded to, independent of soft limits. While the terms in the soft expansion are canonical, the terms are not individually gauge-invariant. Instead gauge invariance relates different terms in the Laurent expansion to each other, allowing us ` ``gauge-invariantify",  determining parts of the higher order terms in the expansion in terms of the lower-order ones.  

\subsection{Gauge-invariantifying the soft expansion}

Looking back to the Weinberg soft pole \cite{Weinberg}, we can easily translate it into the $X$'s and compare with what we've derived in the previous section:
\begin{equation}
    \left(\frac{\epsilon_n \cdot q_1}{q_1 \cdot q} -\frac{\epsilon_n \cdot q_n}{q_n \cdot q}\right) = 2 \left( \frac{X_{3,2n}}{X_{3, 2n-1}} + \frac{X_{2n-3, 2n}}{X_{1,2n-3}} - 1\right).
\end{equation}

Up to a multiplicative factor, this agrees with what we found at $\mathcal{O}(\delta^{-1})$, but curiously it also includes a term subleading in the soft limit. This reflects a nice fact about the Weinberg soft term: it is {\it exactly} gauge invariant in the soft gluon, even away from the soft limit. In $X$ language, while the terms with poles are gauge invariant up to ${\cal O}(\delta^{-1})$, they are not fully gauge invariant in gluon $n$ simply because the variation of $e.g.$ $X_{3,2n}$ is proportional to $X_{3,2n-1}$. But beautifully, the extra $-1$ term makes the expression fully gauge-invariant in $2n$. 

So, a simple question is whether there is any operation we can do to each order $\mathcal{O}(\delta^k)$ in our expansion to make it gauge invariant in the $n^{\rm{th}}$ polarization on its own. To pursue this, let us the define an operator
\begin{equation}
    \mathcal{G}_{(e,o)}\left[F \right] = \sum_{j} (X_{j,e} - X_{j,o}) \partial_{X_{j,e}} F - F,
\end{equation}
where $j$ ranges over all indices other than $e$ and $o$, for $e$ an even index and $o$ an odd index. The function $F$ is gauge invariant in index $e$ if and only if it satisfies
\begin{equation}
    \mathcal{G}_{(e,e+1)}\left[F \right]=0, \quad  \text{ and }\quad \mathcal{G}_{(e,e-1)}\left[F \right]=0. 
\end{equation}
Furthermore, if we have a function $F$ that, while \textit{not} gauge invariant in index $e$, satisfies $\mathcal{G}_{(e,e+1)}\left[F \right] = \mathcal{G}_{(e,e-1)}\left[F \right] \neq 0$, then we can gauge-invariantify it by considering a new function
\begin{equation}
    F^{\mathcal{G}_e} = F + \mathcal{G}_{(e,e+1)}\left[F \right].
\label{eq:gaug-inv-form}
\end{equation}
One can check that $F^{\mathcal{G}_e}$ is, by construction, gauge invariant in $e$. Of course, by performing this operation, we generically interfere with gauge invariance in some other indices $e^\prime$. Let's now turn back to the leading term we obtained in our soft expansion, and ask whether we can gauge-invariantify it. It is simple to show that
\begin{equation}
    \mathcal{G}_{2n,1} \left[ \mathcal{S}^{-1}\right]= - \mathcal{A}_{n-1}, \quad \text{and }\quad    \mathcal{G}_{2n,2n-1} \left[ \mathcal{S}^{-1}\right]= - \mathcal{A}_{n-1},
    \label{eq:condGauge}
\end{equation}
so, upon gauge-invariantifying, we find
\begin{equation}
    \mathcal{S}^{-1,\mathcal{G}_{2n}} = \mathcal{S}^{-1} - \AA_{n-1} =  \left( \frac{X_{2n-3,2n}}{X_{1,2n-3}} + \frac{X_{3,2n}}{X_{3,2n-1}} - 1\right)\times \AA_{n-1}.
\end{equation}
This leading soft factor --- now gauge invariant in \textit{all} gluons --- agrees precisely with the standard Weinberg soft factor! 

In App.~\ref{app:HigherOrdSoft}, we explain how higher terms in the soft expansion also satisfy the condition $\mathcal{G}_{(e,e+1)}\left[F \right] = \mathcal{G}_{(e,e-1)}\left[F \right] \neq 0$, and thus can be gauge-invariantified using exactly this simple procedure.

\section{The One-Loop Leading Soft Theorem for the Surface Integrand}
\label{sec:one-loop-soft}

We now proceed to discussing how to extend the soft expansion at tree-level to the one-loop surface integrand. We start in \cref{sec:3.1} by generalizing the tree-level soft limit to surface kinematics at one-loop. Then, in \cref{sec:3.2}, we use the gauge invariance statement~\eqref{eq:SurfaceGauge} as well as the factorization rule~\eqref{eq:tree-loop} to show that, at leading order in the soft expansion, the surface integrand obeys its own leading soft term, a simple generalization of that found at tree-level.

\subsection{Defining the soft expansion in surface kinematics}
\label{sec:3.1}

In the tree-level case, we started by deriving the soft limit in momentum space, and then we translated it into the scalar variables $X_{i,j}$. In the end, we obtained a map defined purely on the curves of the disk, corresponding to the choice where we make points $x_1^\mu$ and $x_{2n-1}^\mu$ collide into a final point $x_s^\mu$ as in \cref{eq:softlim}. Thought of as a map directly on the curves living on the surface, it is then trivial to extend this map to the punctured disk:
\begin{equation}
\begin{aligned}
    X_{1,j}(\varepsilon) = X_{s,j} + \varepsilon (X_{1,j} - X_{s,j}), \\
    X_{j,1}(\varepsilon) = X_{j,s} + \varepsilon (X_{j,1} - X_{j,s}), \\
\end{aligned} \quad \quad 
\begin{aligned}
    X_{2n-1,k}(\varepsilon) = X_{s,k} + \varepsilon (X_{2n-1,k} - X_{s,k}), \\
    X_{k,2n-1}(\varepsilon) = X_{k,s} + \varepsilon (X_{k,2n-1} - X_{k,s}),
\end{aligned}
\label{eq:softlim}
\end{equation}
for $j, k \in \{2,3,\ldots,2n-2,p,2n\}$, where we impose that the scaffolding chords $X_{1,3}$ and $X_{2n-3,2n-1}$ remain zero at all times by setting $X_{s,3}=X_{2n-3,s}=0$. For curves that touch only points $1$ and $2n-1$, we additionally find
\begin{equation}
\begin{aligned}
    X_{1,1}(\varepsilon) = X_{s,s} + \varepsilon (X_{1,1} &- X_{s,s}), \quad X_{2n-1,2n-1}(\varepsilon) = X_{s,s} + \varepsilon (X_{2n-1,2n-1} - X_{s,s}), \\
    &X_{1,2n-1}(\varepsilon) = X_{s,s} + \varepsilon (X_{1,2n-1} - X_{s,s}).
\end{aligned}
\end{equation}
In the above, we have made manifest the difference between curves $X_{i,j}$ and $X_{j,i}$, since, in surface kinematics, these variables are not identified. 
\begin{figure}[t]
    \centering
    \includegraphics[width=0.85\linewidth]{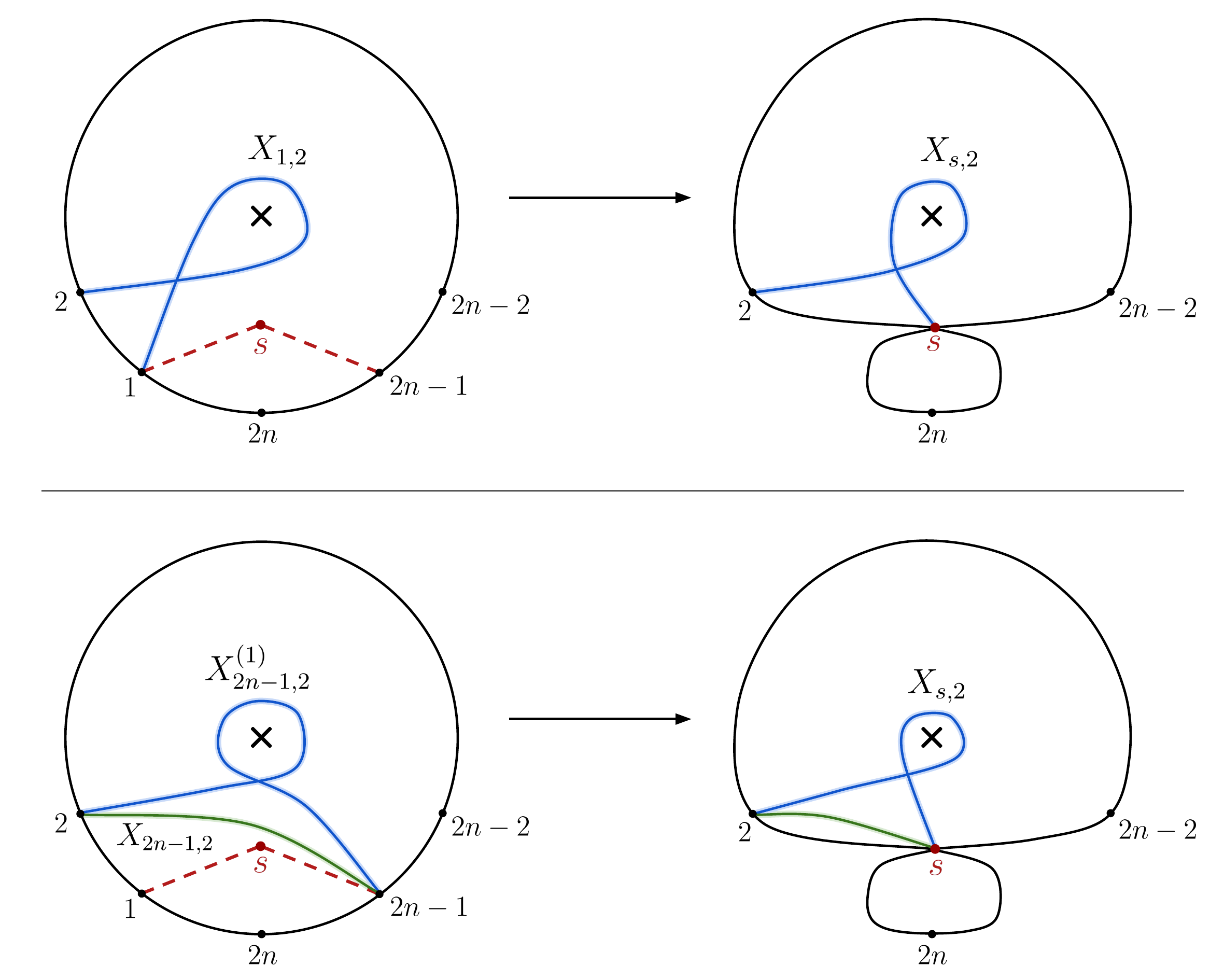}
    \caption{(Top) The self-intersecting boundary curve, $X_{1,2}$, under the soft limit is mapped into the boundary self-intersecting boundary curve $X_{s,2}$. (Bottom) Mapping of the non-self intersecting curve $X_{2n-1,2}$ and its self-intersecting version, $X_{2n-1,2}^{(1)}$, under the soft limit. As we can see, $X_{2n-1,2}$ is mapped to an honest boundary curve, which is zero, while $X_{2n-1,2}^{(1)}$ is mapped into the boundary curve $X_{s,2}$. Therefore, the only consistent way of defining the soft limit is mapping $X_{s,2} =0$. }
    \label{fig:SoftLoop}
\end{figure}

However, there is now a slight subtlety at one-loop, as it turns out that surface kinematics forces an inconsistent mapping of certain curves (boundary curves) from the $2n$-point punctured disk onto the $(2n-1)$-point one. We illustrate this in \cref{fig:SoftLoop}. In particular, consider curve $X_{1,2}$, which is associated with the self-intersecting curve shown on the left of \cref{fig:SoftLoop}. In the soft limit, this curve gets mapped into the self-intersecting boundary curve $X_{s,2}$. Of course, curve $X_{2n-1,2}$ should also map to $X_{s,2}$. However, on the surface there are two distinct curves labeled by $X_{2n-1,2}$ --- the one that self-intersects $X_{2n-1,2}^{(1)}$, represented in blue on the right of \cref{fig:SoftLoop}, and the one which does not $X_{2n-1,2}$, represented in green. In the soft limit, $X_{2n-1,2}^{(1)}$ maps into the self-intersecting boundary curve $X_{s,2}$, and the non-self-intersecting curve $X_{2n-1,2}$ maps into a genuine boundary curve that therefore should go to zero in the soft limit. However, in surface kinematics, we \textit{don't} distinguish between $X_{2n-1,2}^{(1)}$ from $X_{2n-1,2}$, and thus the only consistent thing to do is say that these three curves (including $X_{1,2}$) get mapped to zero; this means that, in the soft limit, we must enforce $X_{s,2}=0$. A similar story holds for curves $X_{2n-2,2n-1}$ and $X_{2n-2,1}$, and can be resolved in the same way by setting $X_{2n-2,s}=0$. By defining the limit like this, the lower-point integrand we land on after taking the soft limit is the one where the boundary curves $X_{s,2}$ and $X_{2n-2,s}$ have been set to zero. 

Keeping this in mind, we will organize the one-loop soft limit as a Laurent expansion in soft factors $\delta_{1,j} = \varepsilon (X_{1,j} - X_{s,j})$ and $\delta_{2n-1,j} = \varepsilon ( X_{2n-1,j} - X_{s,j})$, with the analogous for $\delta_{j,1}, \delta_{j,2n-1}$. As discussed, we need to set $X_{s,2} = X_{2n-2,s} = 0$ and $X_{s, 3} = X_{2n-3,s} = 0$. Therefore, the $X$ kinematical variables that vanish in the soft limit are given by
\begin{equation}
\begin{aligned}
    &X_{2n-3,1}(\varepsilon) = \delta_{2n-3,1},\\
    &X_{1,2}(\varepsilon) = \delta_{1,2},\\
    &X_{2n-2,1}(\varepsilon) = \delta_{2n-2,1},\\    
\end{aligned}\quad \quad 
\begin{aligned}
    &X_{2n-1,3}(\varepsilon) = \delta_{2n-1,3},\\
    &X_{2n-2,2n-1}(\varepsilon) = \delta_{2n-2,2n-1},\\ 
    &X_{2n-1,2}(\varepsilon) = \delta_{2n-1,2}.
\end{aligned}
\end{equation}

Note that, if we were dealing with standard momentum space variables, we would also have to take $X_{1,2n-3} \to 0$ and $X_{3,2n-1} \to 0$ in the soft limit by momentum conservation, since these are connected to propagators $X_{2n-3,1}$ and $X_{3,2n-1}$ respectively by a bubble diagram. However, in surface kinematics, the limit we get \textit{only} sends $X_{2n-3,1}$ and $X_{2n-1,3}$ to zero. Therefore, in the soft limit, the answer will localize solely onto these two factorization channels, just like we saw at tree-level. So, once again it is convenient to decompose the surface loop integrand $\mathcal{I}_n$ as
\begin{equation}
\label{eq:int-exp}
    \mathcal{I}_n = \frac{R_{2n-3,1}}{X_{2n-3,1}} + \frac{R_{2n-1,3}}{X_{2n-1,3}} + \mathcal{R},
\end{equation}
where $R_{2n-3,1}$ and $R_{2n-1,3}$ are the residues of $\mathcal{I}_n$ at $X_{2n-3,1} = 0$ and $X_{2n-1,3} = 0$, respectively, and $\mathcal{R}$ is everything else. Like at tree-level, $\mathcal{R}$ manifestly starts at $\mathcal{O}(\delta^0)$ in the soft expansion, since it has poles in neither $X_{2n-3,1}$ nor $X_{2n-1,3}$. 

To derive the leading contribution to the soft limit, we again use the explicit expressions for $R_{2n-3,1}$ and $R_{2n-1,3}$, given by the formula~\eqref{eq:tree-loop}. We leave this for the next section and \cref{app:highorderLoop}, and for now just discuss the generalization of the gauge-invariance constraint \eqref{eq:constContact} to one-loop-order. 

As we see from \cref{eq:SurfaceGauge} the one-loop gauge invariance statement for $\mathcal{I}_n$ is very similar to the tree-level case, up to corrections proportional to boundary curves. Therefore, proceeding exactly in the same way as in the tree-level case, we find that gauge invariance in $2n$ implies that we can write the integrand as
\begin{equation}
\begin{aligned}
     \mathcal{I}_n = &\frac{R_{2n-3,1}}{X_{2n-3,1}} + \frac{R_{2n-1,3}}{X_{2n-1,3}} + \sum_{j} C_{2n,j} (X_{2n,j} - X_{1,j}) + \sum_{j} C_{j,2n} (X_{j,2n} - X_{j,1}) \\
     &\quad +X_{1,2} \left[ C_{2n,2} + \frac{\partial \mathcal{R}}{\partial X_{1,2}}\bigg \vert_{2n \to 1}\right] - \frac{\partial R_{2n-3,1}}{\partial X_{2n-3,2n}},
\end{aligned}
\label{eq:looprep1}
\end{equation}
and, by interchanging $1 \leftrightarrow 2n-1$, as
\begin{equation}
\begin{aligned}
     \mathcal{I}_n = &\frac{R_{2n-3,1}}{X_{2n-3,1}} + \frac{R_{2n-1,3}}{X_{2n-1,3}} + \sum_{j} C_{2n,j} (X_{2n,j} - X_{2n-1,j}) + \sum_{j} C_{j,2n} (X_{j,2n} - X_{j,2n-1}) \\
     &\quad +X_{2n-2,2n-1} \left[ C_{2n-2,2n} + \frac{\partial \mathcal{R}}{\partial X_{2n-2,2n-1}}\bigg \vert_{2n \to 2n-1} \right] - \frac{\partial R_{2n-1,3}}{\partial X_{2n,3}},
\end{aligned}
\label{eq:looprep2}
\end{equation}
where, just like in the tree-level case, the $C$'s are completely well defined by
\begin{equation}
    C_{2n,j} = \frac{\partial }{\partial X_{2n,j}} \left(\mathcal{I}_{n} - \frac{R_{2n-3,1}}{X_{2n-3,1}} - \frac{R_{2n-1,3}}{X_{2n-1,3}}\right),
\end{equation}
and similarly for $C_{j,2n}$. Therefore, subtracting Eqns.~\eqref{eq:looprep1} and \eqref{eq:looprep2}, the one-loop generalization of Eq.~\eqref{eq:constContact} is
\begin{equation}
\begin{aligned}
    &\sum_{j} C_{2n,j} (X_{2n-1,j} - X_{1,j})+ \sum_{j} C_{j,2n} (X_{j,2n-1} - X_{j,1})   +X_{1,2} \left[ C_{2n,2} + \frac{\partial \mathcal{R}}{\partial X_{1,2}}\bigg \vert_{2n \to 1}\right] \\
    &- X_{2n-2,2n-1} \left[ C_{2n-2,2n} + \frac{\partial \mathcal{R}}{\partial X_{2n-2,2n-1}}\bigg \vert_{2n \to 2n-1} \right] = -\frac{\partial R_{2n-1,3}}{\partial X_{2n,3}} + \frac{\partial R_{2n-3,1}}{\partial X_{2n-3,2n}}.  
\end{aligned}
\label{eq:constrainContactLoop}
\end{equation}
Due to its similarity to \cref{eq:factorization}, one might expect that we can once again extract the leading order behavior of the $C$'s from the above constraint. Unfortunately, this naive strategy will not work here. This is because, if we expand the $l.h.s.$ of the above equation in soft factors, we now find a term $X_{s,s} ( C_{2n-1,2n} - C_{2n,1})$ starting at $\mathcal{O}(\delta^0)$, due to the fact $X_{2n-1,1} = 0$ on the scaffolding residue. Since $X_{s,s}$ is \textit{not} a soft factor, we cannot exactly match terms at $\mathcal{O}(\delta)$ as we did at tree-level. We leave to future investigations the question of whether there is some other way to extract the leading $C$'s out of \cref{eq:constrainContactLoop}, or if additional inputs are needed. 

In App.~\ref{app:highorderLoop}, we give some explicit formulae for the residues $R_{2n-3,1}$ and $R_{2n-1,3}$ that are needed to derive the one-loop soft expansion, particularly the leading term we explore next. 

\subsection{The leading soft theorem for the one-loop integrand}
\label{sec:3.2}

To derive the leading soft theorem at one-loop, we want to extract the leading order of $R_{2n-3,1}$ and $R_{2n-1,3}$. Let's start by looking at $R_{2n-1,3}$. We can write it explicitly by using the gluing rule \eqref{eq:tree-loop}, where, in this case, $\mathcal{A}_L = \mathcal{A}_3(3,x_L,2n-1,2n,1,2)$ and $\mathcal{I}_R = \mathcal{I}_{n-1}(3,4,\cdots 2n-2,2n-1,x_R)$. In App.~\ref{app:highorderLoop}, proceeding just as we did at tree-level, we present the different terms in the $j$-sum over $\partial_{X_{j,x_L}}\mathcal{A}_L$. There, we derive the following leading order term in the expansion of $R_{2n-1,3}$:
\begin{equation}
    R_{2n-1,3} = X_{2n,3} \times \mathcal{I}_{n-1}(s,2,3,\ldots,2n-2)\vert_{\mathcal{S}} + (X_{2n,3} -X_{2n,2}) X_{s,s}\frac{\partial \mathcal{I}_{n-1}}{\partial X_{s,2}}\bigg \vert_{\mathcal{S}} +\mathcal{O}(\delta),
\end{equation}
where ``$\vert_{\mathcal{S}}$'' means we are setting the boundary curves $X_{s,2},X_{2n-2,s}$ to zero. Doing the same thing for $R_{2n-3,1}$, we analogously find 
\begin{equation}
    R_{2n-3,1} = X_{2n-3,2n} \times \mathcal{I}_{n-1}(s,2,3,\ldots,2n-2)\vert_{\mathcal{S}} +(X_{2n-3,2n} -X_{2n-2,2n}) X_{s,s}\frac{\partial \mathcal{I}_{n-1}}{\partial X_{2n-2,s}}\bigg \vert_{\mathcal{S}} +\mathcal{O}(\delta).
\end{equation}
So, in both cases, we see the appearance of the lower-point gluon integrand $\mathcal{I}_{n-1}$ that we obtain by collapsing $x_1^\mu$ and $x_{2n-1}^\mu$ into the point $x_s^\mu$. In fact, as we show in App.~\ref{app:highorderLoop}, it is surface gauge invariance that let's us recognize the lower-point integrand at leading order in the spin-sum formula --- exactly mirroring the situation at tree-level! 
With both of these expressions in-hand, we can write the one-loop integrand soft factor as
\begin{equation}
\begin{aligned}
    \mathcal{I}_n =& \left( \frac{X_{2n,3}}{X_{2n-1,3}} +   \frac{X_{2n-3,2n}}{X_{2n-3,1}}\right)\mathcal{I}_{n-1}(s,2,3,\cdots,2n-2)\vert_{\mathcal{S}} \\
    &+X_{s,s}\left[\frac{X_{2n,3} - X_{2n,2}}{X_{2n-1,3}} \frac{\partial \mathcal{I}_{n-1}}{\partial X_{s,2}}\bigg \vert_{\mathcal{S}} + \frac{X_{2n-3,2n} - X_{2n-2,2n}}{X_{2n-3,1}} \frac{\partial \mathcal{I}_{n-1}}{\partial X_{2n-2,s}} \bigg \vert_{\mathcal{S}}\right]  + \mathcal{O}(\delta^0).
\end{aligned}
\end{equation}

All told, the integrand has precisely the same behavior as at tree-level, up to correction terms proportional to $X_{s,s}$. Nonetheless, when we go to physical kinematics, we should set $X_{s,s}=0$. To consider this limit, we separate the term in the bracket into two categories: those that have $X_{s,s}$ in the denominator, and those that do not. Terms in the latter case vanish, while, in the former case, the $X_{s,s}$'s cancel, and so they survive when we set $X_{s,s} \to 0$. However, these correspond precisely to scaleless terms that therefore vanish after loop-integration. So, as required, only the Weinberg term plays a physical role. It is remarkable that, within surface kinematics, the leading term is not only a property of the one-loop amplitude but of the one-loop integrand!

Just like at tree-level, we can easily gauge-invariantify the part that survives post-loop-integration to exactly match it with the leading Weinberg term:
\begin{equation}
\begin{aligned}
    \mathcal{I}_n =& \left( \frac{X_{2n,3}}{X_{2n-1,3}} +   \frac{X_{2n-3,2n}}{X_{2n-3,1}} - 1\right)\mathcal{I}_{n-1}(s,2,3,\cdots,2n-2)\vert_{\mathcal{S}} \\
    &+X_{s,s}\left[\frac{X_{2n,3} - X_{2n,2}}{X_{2n-1,3}} \frac{\partial \mathcal{I}_{n-1}}{\partial X_{s,2}}\bigg \vert_{\mathcal{S}} + \frac{X_{2n-3,2n} - X_{2n-2,2n}}{X_{2n-3,1}} \frac{\partial \mathcal{I}_{n-1}}{\partial X_{2n-2,s}} \bigg \vert_{\mathcal{S}}\right]  + \mathcal{O}(\delta^0),
\end{aligned}
\label{eq:one-loop-lead}
\end{equation}

Quite beautifully, we have found that the canonical Yang-Mills loop integrand --- crucially defined using surface kinematics including tadpoles and external bubbles --- also enjoys a canonical behavior in the soft limit.

%% file: Sections_v2/Transmutation_v2.tex
\newpage
\part{ Transmutation of Gluons into Scalars}
\label{part:2}
\renewcommand\thesection{\arabic{section}}
\titlespacing*{\section}{5pt}{*4}{*2}

We now switch to exploring a different topic--- one which naturally arises from the fact that gauge redundancies enforce a locked form for the scalar-scaffolded representation of gluon amplitudes. As we reviewed in \cref{sec:ScalarScaffRev}, gauge invariance and linearity in the polarization of the $i^{\rm{th}}$ gluon restrict the form of $\mathcal{A}_n$ to 
\begin{equation}
\begin{cases}
\label{eq:GgInvarianceII}
    \mathcal{A}_{n}(1, 2, \ldots, 2n) = \sum_{j \notin\{2i,2i\pm 1\}} (X_{2i,j}-X_{2i-1,j}) \partial_{X_{2i,j}} \mathcal{A}_{n}, \\
    \mathcal{A}_{n}(1, 2, \ldots, 2n) = \sum_{j \notin\{2i,2i\pm 1\}}(X_{2i,j}-X_{2i+1,j}) \partial_{X_{2i,j}} \mathcal{A}_{n}. 
\end{cases} 
\end{equation}
Staring at these expressions immediately suggests a simple kinematic locus, where we set 
\begin{equation}
    X_{2i,j} \to 1 + X_{2i+1,j} + \alpha_i (X_{2i-1,j} - X_{2i+1,j}), \quad \text{ for all }j\notin\{2i,2i\pm1\}. 
\end{equation}
Here, the freedom in $\alpha_i$ is associated with gauge redundancy, and the $1$ could be replaced by any constant $c$, which would simply rescale the final answer.
Note also that since the $X$'s have units of $p^2$, in making this choice we are working in units where this (dimensionful) $c$ is set to 1. 

Since this limit changes only $X$ variables dependent on the index $2i$, it corresponds to some particular choice of polarization $\epsilon_i$. We will explicitly define this configuration in the following section. However, in scalar-scaffolded variables, one can simply use the forms in Eq.~\eqref{eq:GgInvarianceII} to see that this choice reduces the amplitude to 
\begin{equation}
    \mathcal{A}_n( X_{2i,j} \to 1 + X_{2i+1,j} + \alpha_i (X_{2i-1,j} - X_{2i+1,j})) = \sum_{j} \frac{\partial \mathcal{A}_n }{\partial X_{2i,j}}, 
    \label{eq:KinLocusX}
\end{equation}
which is then equivalent to acting on $\mathcal{A}_n$ with the operator
\begin{equation}
    \W_{2i}[F] = \sum_{j\notin\{2i,2i\pm1\}} \frac{\partial}{\partial X_{2i,j}} F.
    \label{eq:defOp}
\end{equation}
Note that this operator --- a quite natural object within the scalar-scaffolded picture --- eliminates dependence on the index $2i$ (and therefore on the polarization $\epsilon_i$) in the simplest and most symmetric way possible. This fact alone suggests that, by acting with $\mathcal{W}_{2i}$, we might somehow be transmuting the $i^{\rm{th}}$ gluon into a scalar. 

This is the idea we will explore systematically both at tree-level and one-loop here in Part II. But before launching into this analysis, we can see that our results in Part I already hint that the object we land on after applying $\mathcal{W}_{2i}$ has intriguing  structure suggestive of some kind of ``scalarization.''
When we performed the soft expansion, we found it useful to decompose $\mathcal{A}_n$ in terms of the two collinear residues, $R_{1,2n-3}$ and $R_{3,2n-1}$, plus a part without these poles. The latter could further be written in terms of the $C_j$ and a derivative of one of the residues, as shown in \cref{eq:rep1} and, equivalently, in \cref{eq:rep2}. Using the explicit formulae for $R_{1,2n-3}$ and $R_{3,2n-1}$ given in App.~\ref{app:HigherOrdSoft}, it is trivial to check that
\begin{equation}
\begin{aligned}
    \W_{2n} &\left[\frac{R_{1,2n-3}}{X_{1,2n-3}} +  \frac{R_{3,2n-1}}{X_{3,2n-1}} + {\textstyle\sum}_{j=2}^{2n-2} (X_{j,2n}-X_{1,j}) C_j - \frac{\partial R_{1,2n-3}}{\partial X_{2n-3,2n}}\right] =\\
    &\quad \quad \frac{\mathcal{A}_{n-1}(1,\ldots,2n-3,2n-2)}{X_{1,2n-3}}
    + \frac{\mathcal{A}_{n-1}(2n-1,2, 3,\ldots,2n-2)}{X_{3,2n-1}} +{\textstyle\sum}_{j=2}^{2n-2} C_j.
\label{eq:W-on-amp}
\end{aligned}
\end{equation}
That is, the residues turn into precisely lower-point gluon amplitudes, defanged of the standard spin-sum we get when we glue three-point and $(n-1)$-point gluon amplitudes. In App.~\ref{app:HigherOrdSoft}, we also show that the sum of $C_j$ given above obeys simple sum rules as an expansion in soft factors. What's more, for the special cases of $\alpha_i=0$ or $\alpha_i=1$, we can recognize the kinematic locus \eqref{eq:KinLocusX} as that of one of the simplest ``split'' patterns -- \textit{skinny rectangles} -- discussed in Refs.~\cite{Zeros,Splits}. These splits are true not only at the level of the field theoretic amplitude but also at full stringy level, where they are most natural.

We now undertake a more systematic treatment, starting by rephrasing the kinematical locus \eqref{eq:KinLocusX} in terms of constraints on the scalar $p_i \cdot p_j$, and ultimately translating these into constraints on the dot products of the polarization and momenta of the gluons. We then study gluon amplitudes using  \textit{surface integrals} \cite{Gluons}, which we briefly review here to set notation and explain our basic strategy. As proposed in \cite{Zeros,Gluons}, the tree-level surface integral associated to the disk with $n$ marked points on the boundary $\mathcal{S}_n$ is given by \cite{Arkani-Hamed:2019mrd,Arkani-Hamed:2023lbd}: 
\begin{equation}    \mathcal{A}_{\mathcal{S}_n}^{\text{Tr(}\phi^3)}[X_{i,j}] = \int_0^{+\infty} \Omega_{y_P} \prod_{(i,j) \in \mathcal{S}_n} u_{i,j}^{X_{i,j}}[\{y_P\}],
\label{eq:SurfaceIntTr}
\end{equation}
where $\Omega_{y_P} = \prod_{P\in \mathcal{T}} d y_P/y_P$ for $\mathcal{T}$ some triangulation of $\mathcal{S}_n$, and we have taken the string scale $\alpha^\prime \to  1$. This integral is usually called the ``stringy'' Tr$(\phi^3)$ integral as it yields tree-level Tr$(\phi^3)$ amplitudes at low energies. The $u_{i,j}$'s are the so-called $u$-variables, where $u_{i,j}$ is associated with the curve from marked point $i$ to marked point $j$;\footnote{In terms of the standard worldsheet coordinates, $z_{i,j} = z_j - z_i$, the $u_{i,j}$ are simply the SL$(2,\mathbb{R})$-invariant cross-ratios, $u_{i,j} = z_{i,j-1}z_{i-1,j}/(z_{i,j}z_{i-1,j-1}).$} they are functions of the \textit{positive coordinates}, $y_P$, such that for any value of $y_P \in[0,+\infty)$, we have $u_{i,j} \in [0,1]$, for any $(i,j)$.

The $u$-variables satisfy a set of non-linear equations --- the $u$-equations \cite{Arkani-Hamed:2019plo,Arkani-Hamed:2019mrd} --- which at tree-level are simply
\begin{equation}
    u_{i,j} + \prod_{(k,m) \text{ crossing } (i,j)} u_{k,m} =1. 
\end{equation}
Remarkably, although there are as many equations as $u$-variables, the space of solutions to these equations is quite non-trivial. Namely, at tree-level, these equations define an $(n-3)$-dimensional space, which we can parametrize using the $y_P$'s. 

We then obtain the string amplitude for scalar-scaffolded gluons by a two-step procedure. We first perform the $\delta$-shift in \cref{eq:SurfaceIntTr}, where we shift $X_{i,j}$ with $+\delta$ if $(i,j)$ are both even, and $-\delta$ if they are both odd, to obtain
\begin{equation}    \mathcal{A}_{\mathcal{S}_{2n}}[X_{i,j}^\delta] = \int_0^{+\infty} \Omega_{y_P} \prod_{(i,j) \in \mathcal{S}_{2n}} u_{i,j}^{ X_{i,j}}[\{y_P\}] \times \left( \frac{ \prod_{(i,j) \text{ even}} u_{i,j}}{ \prod_{ (i,j) \text{ odd}} u_{i,j}} \right), 
\label{eq:2nScalarInt}
\end{equation}
where we have further set $\delta=1$. At low energies, this integral describes the scattering of the $2n$ scalars scaffolding the gluons. To finally obtain the gluon amplitude, we must further put the gluons on-shell, which is done by taking the \textit{scaffolding residues} on \cref{eq:2nScalarInt}\footnote{The result of this residue matches the standard open bosonic string gluon amplitude, where the gluon polarization and momenta are scalar-scaffolded \cite{Green:1987sp,Gluons}.}
\begin{equation}    \mathcal{A}_{\mathcal{S}_{2n}}^{\text{Gluon}}[ X_{i,j}] = \mathop{\mathrm{Res}}_{X_{1,3}=0} \left[\mathop{\mathrm{Res}}_{X_{3,5}=0} \left[ \cdots \mathop{\mathrm{Res}}_{X_{1,2n-1}=0} \mathcal{A}_{\mathcal{S}_{2n}}[ X_{i,j}]  \right] \right].
\label{eq:scaffRes}
\end{equation}

Having defined gluon amplitudes via surface integrals, we then study what happens when we apply a single $\W_e$ as well as collections of $\W_e$'s. As we will see, the fact that the surface integrals underlying gluon amplitudes satisfy ``split'' factorizations --- together with the machinery of surfaceology --- allows us to cleanly describe the action of arbitrary sets of $\W_e$'s at the level of the surface integral. Most interestingly, we show that after applying $(n-2)$ $\W_e$'s, the surface integral gives $X_{e_1,e_2}$ (for the two indices missing out of the $\W_e$) times a shifted integral--- which, at field-theory level, magically conspires to give us the Tr$(\phi^3)$ amplitude. So, by applying $(n-2)$ $\W_e$'s to the amplitude, we succeed in turning all the $n$ gluons into scalars! We then conclude by describing the natural generalization of this operator to one-loop surface kinematics. There, we'll see that, by acting now with all $n$ of them, we also can turn the one-loop YM integrand into the one-loop Tr$(\phi^3)$ integrand.

\section{Polarization Choice for $\W_e$}
\label{sec:PolConfig1}

As just explained, from gauge invariance and multi-linearity we know that the action of $\W_e$ is equivalent to fixing the $X$'s to the kinematical locus given in Eq.~\eqref{eq:KinLocusX}. For simplicity of notation, let us take $e=2n$. We now want to translate this locus from scalar-scaffolded language into standard gluon amplitude language, where it will correspond to conditions on dot products involving $\epsilon_n^\mu$. 

Using the scalar-scaffolding map, we can write: 
\begin{equation}
   \mathcal{A}_n^\mu= \frac{\partial}{\partial \epsilon_{n, \mu}}  \mathcal{A}_n = -\frac{\partial}{\partial x_{2n, \mu}}  \mathcal{A}_n, 
\end{equation}
where $x_{2n}^\mu$ is the coordinate associated with vertex $2n$ in the scalar momentum polygon, and the minus sign is just a convention. But, since the scalar-scaffolded amplitude depends on the chords $X_{i,j} = (x_i-x_j)^2$, it is useful to recast this derivative in terms of $X$'s as follows:
\begin{equation}
   \mathcal{A}_n^\mu =  \sum_{j=2}^{2n-2} -2\left( x_{2n}^\mu -x_j^\mu\right) \frac{\partial \mathcal{A}_n}{\partial X_{j,2n}} .
   \label{eq:pol1}
\end{equation}

So, a polarization $\epsilon_{n,\mu} - \alpha_n q_{n,\mu}$ (with its gauge-dependence made explicit) that corresponds to the action of $\W_{2n}$ is simply one that satisfies
\begin{equation}
 2\left[ \epsilon_{n,\mu} - \alpha_n q_{n,\mu}\right]  ( x_{2n}^\mu -x_j^\mu) = -1, \quad \text{ for all } j \in \{2,\ldots,2n-2\}.
  \label{eq:condPol}
\end{equation}
Recall from Eq.~\eqref{eq:GluonScalarMap} that we can write the polarization and momentum of gluon $i$ in terms of scalar momenta as $\epsilon_{i,\mu} - \alpha_i q_{i,\mu} = p_{2i}^\mu - \alpha_i (p_{2i-1}+p_{2i})^\mu$ and $q_i^\mu = (p_{2i-1}+p_{2i})^\mu$. Using this correspondence, we know that
\begin{equation}
    \epsilon_{n}^\mu - \alpha_n q_n^\mu = p_{2n}^\mu - \alpha_n (p_{2n}^\mu+p_{2n-1}^\mu)= (x_1-x_{2n})^\mu -\alpha_n(x_{1}-x_{2n-1})^\mu.
\end{equation}
As a check, plugging this into Eq.~\eqref{eq:condPol} gives
\begin{equation}
    -X_{j,2n} + X_{1,j} +\alpha_n(X_{j,2n-1}-X_{1,j}) = -1 \Leftrightarrow X_{j,2n} = 1 + X_{1,j} + \alpha_n (X_{j,2n-1} - X_{1,j}) ,
\end{equation}
which is precisely the kinematical locus given in Eq.~\eqref{eq:KinLocusX} for $2i=2n$.

Now, purely in terms of scalar momenta, the conditions in Eq.~\eqref{eq:condPol} are equivalent to 
\begin{equation}
\begin{aligned}
    &2[p_{2n} - \alpha_{n} (p_{2n}+ p_{2n-1})] \cdot p_1 = 1, \\
    &2[p_{2n} - \alpha_{n} (p_{2n}+ p_{2n-1})] \cdot p_{2n-2} = -1,
\end{aligned} 
\label{eq:cond1}
\end{equation}
and 
\begin{equation}
[p_{2n} - \alpha_{n} (p_{2n}+ p_{2n-1})] \cdot p_j = 0, \quad \text{ for all } j \in \{2,\ldots,2n-3\}.
\label{eq:cond2}
\end{equation}
We can then trivially translate these constraints into conditions on the dot products of $\epsilon_n^\mu$ to obtain
\begin{equation}
\begin{aligned}
    &q_1 \cdot \epsilon_{n} =1/2, \\
    &q_j \cdot \epsilon_{n} =0, \\
    &q_{n-1} \cdot \epsilon_{n} = -1/2, 
\end{aligned} \quad  \quad
\begin{aligned}
    &\epsilon_1 \cdot \epsilon_{n} = -\alpha_1/2, \\
    &\epsilon_j \cdot \epsilon_{n} =0, \\
    &\epsilon_{n-1} \cdot \epsilon_{n} = - (1 - \alpha_{n-1})/2,
\end{aligned}
\quad \text{with } j \in \{2,\ldots,n-2\},
\label{eq:condPolE}
\end{equation}
where the factor of $1/2$ is a normalization convention. So, we find that the only non-zero dot products with $\epsilon_n^\mu$ are those involving the two adjacent gluons, $1$ and $n-1$, and that the precise value of these dot products is \textit{gauge dependent} -- encoded in their explicit dependence in $\alpha_1$ and $\alpha_{n-1}$.
This means we have a full gauge orbit worth of polarization configurations corresponding to the action of $\W_{2n}$. As suggested by this result, and as we show in \cref{sec:consectivesplits}, $\W_{2n}[\mathcal{A}_n]$ satisfies gauge invariance \eqref{eq:GgInvariance} in all but gluons $1$ and $n-1$, a fact which will become important when we study successive applications of this operator.

We can simplify the conditions in Eqs.~\eqref{eq:cond1} and \eqref{eq:cond2} by selecting the special gauge choices $\alpha_n=0,1$:
\begin{equation}
\begin{aligned}
    &\underline{\alpha_n=0}: \quad 2p_{2n} \cdot p_1 = - 2p_{2n-2}\cdot p_{2n} =1, \quad 2p_{2n} \cdot p_j =0,\\
    &\underline{\alpha_n=1}: \quad 2p_{2n-1} \cdot p_1 = - 2p_{2n-2}\cdot p_{2n-1} =-1, \quad 2p_{2n-1} \cdot p_j =0, 
\end{aligned}  
\label{eq:splitAlpha}
\end{equation}
with  $j \in \{2,\ldots,2n-3\} $. Though obviously not gauge invariant, these constraints are exactly those from which we obtain a simple type of \textit{split factorization} \cite{Zeros,Splits}, which we explore in the next section.

In Ref.~\cite{Cheung:2017ems}, the authors define a transmutation operator
\begin{equation}
    \mathcal{T}_{i,j,k} = \partial_{q_i \cdot \epsilon_j} - \partial_{q_k \cdot \epsilon_j},
\end{equation}
which acts on the gluon amplitude decreasing the spin of particle $j$ and inserting it between particles $i$ and $k$ within a color trace structure. Of course, like we did with $\W_e$, we can also interpret this operator as a polarization configuration for gluon $j$. In particular, if we take $i =1$, $j=n$ and $k=n-1$, then acting with $\mathcal{T}_{1,n,n-1}$ corresponds to choosing a polarization $\epsilon_n^\mu$ such that:
\begin{equation}
\begin{aligned}
    &q_1 \cdot \epsilon_{n} =1, \\
    &q_{n-1} \cdot \epsilon_{n} = -1, 
\end{aligned} \quad  \quad
\begin{aligned}
    &q_j \cdot \epsilon_{n} =0, \quad \text{with } j \in \{2,\ldots,n-2\},\\
    &\epsilon_k \cdot \epsilon_{n} = 0, \quad \text{with } k \in \{1,\ldots,n-1\}, \\
\end{aligned}
\label{eq:TransmOp}
\end{equation}
which, up to a multiplicative factor, matches what we found in Eq.~\eqref{eq:condPolE}, for the particular gauge choice of $\alpha_1 =0$ and $\alpha_{n-1}=1$!

In addition, in Ref.~\cite{Cheung:2017ems}, the authors show that acting with $\mathcal{T}_{i,j,k}$ in $n-2$ adjacent gluons together with the action of $\mathcal{T}_{l,m} = \partial_{\epsilon_l \cdot \epsilon_m} $ on the remaining two, $i.e.$, acting with
\begin{equation}
    \mathcal{T}[\{a_1,a_2,\ldots,a_n\}] = \mathcal{T}_{a_1,a_n} \prod_{i=2}^{n-1} \mathcal{T}_{a_{i-1}, a_i,a_n},
    \label{eq:TransmOp}
\end{equation}
where $\{a_1,a_2,\ldots,a_n\} = \alpha$ is an ordered set of labels, turns the $n$-point gluon amplitude with color-ordering $\alpha$ into a pure Tr$(\phi^3)$ amplitude with the same color-ordering.\footnote{If, instead, we acted with \cref{eq:TransmOp} on a gluon amplitude with ordering $\beta \neq \alpha$, the resulting amplitude would be a doubly color-ordered amplitude of bi-adjoint scalar theory.} This then suggests that, by acting consecutively with $(n-2)$ $\W_e$ operators, we should also be very close to turning the gluon amplitude into a Tr$(\phi^3)$ one. As we explain in the next section, this is exactly the case. 

However, to flesh this out, we first need to have some control on the objects we obtain after acting with $\W_e$. We will achieve this via the special gauge choices $\alpha_n=0,1$, which allow us to recast the action of $\W_e$ as a split factorization, using the constraints given in  Eq.~\eqref{eq:splitAlpha}. Splits hold at the level of the surface integral that defines scalar-scaffolded gluon amplitudes at low energies. As we will see, working directly at string-level will give us a natural way of explaining what happens when we apply $\W$ consecutively on the different gluons. Then, in \cref{sec:polConfigFinal}, we translate the results from splits back to standard polarization/momentum space and compare the action of our operators with that of $\mathcal{T}[a_1,a_2,\cdots,a_n]$ from Eq.~\eqref{eq:TransmOp}.

\section{$\W_{e}$ as a Split Configuration}
\label{sec:Splits}

We will now discuss the two split factorizations that land us on the kinematical loci corresponding to $\alpha_n=0$ and $\alpha_n=1$ in Eq.~\eqref{eq:KinLocusX}, from which we derive a concrete expression --- in the form of a surface integral --- for $\W_{2n}[\mathcal{A}_n]$. Throughout this section, we will deal mostly with the surface integral defining $2n$-point scalar scattering $\mathcal{A}_\mathcal{S}$, and we will take the scaffolding residues as necessary. 

Looking back at the original unshifted Tr($\phi^3$) surface integral \eqref{eq:SurfaceIntTr}, it was pointed out in Ref.~\cite{Splits} that simple features of $u$-variables allow us to derive kinematic loci where the integral ``splits'' into the product of two integrals associated to subsurfaces that overlap on a triangle. To derive these loci --- also referred to as ``split kinematics'' --- we map each curve $X_{i,j}$ on the original surface $\mathcal{S}_{2n}$ into the sum of its images in the subsurfaces $\mathcal{S}^{(1)}$ (with kinematics $z_{i,j}$) and $\mathcal{S}^{(2)}$ (with kinematics $x_{i,j}$). Now, by going on split kinematics, the $\delta$-shifted surface integral in \cref{eq:2nScalarInt} reduces to a \textit{product} of shifted surface integrals associated to $\mathcal{S}_1$ and $\mathcal{S}_2$:
\begin{equation}
    \mathcal{A}_{\mathcal{S}_{2n}}[X_{i,j}^{\delta}] \xrightarrow{\text{Split}}  \mathcal{A}_{\mathcal{S}^{(1)}}[z_{i,j}^\delta] \times  \mathcal{A}_{\mathcal{S}^{(2)}}[x_{i,j}^\delta], 
\end{equation}
where the $\delta$ subscript is included to remind us that the kinematics on both sides will be shifted. While the $X_{i,j}$ are shifted with the standard $\delta =\pm 1$, the shifts for $x_{i,j}$ and $z_{i,j}$ are determined via the split map, as we explain in more detail in the next section.

\subsection{One split}
\begin{figure}[t]
    \centering
    \includegraphics[width=\linewidth]{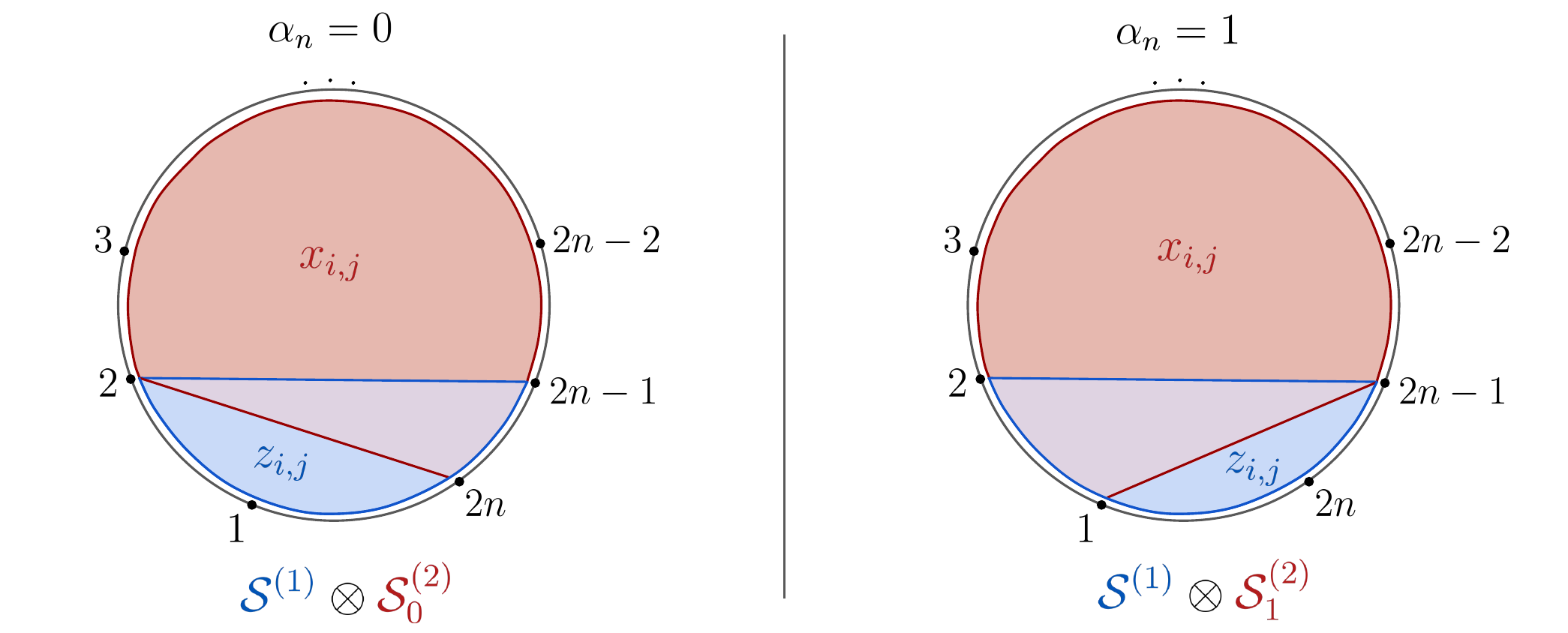}
    \caption{(Left) Two surfaces $\mathcal{S}^{(1)}$ (in blue) and $\mathcal{S}^{(2)}_0$ (in red) associated with the split relevant for the gauge choice $\alpha_n=0$. (Right) Surfaces $\mathcal{S}^{(1)}$ (in blue) and $\mathcal{S}^{(2)}_1$ (in red) entering the split for the gauge choice $\alpha_n=1$. }
    \label{fig:Split}
\end{figure}

In order to describe the action of $\W_e$, the relevant split factorizations are of the simplest type, in which we take $\mathcal{S}^{(1)}$ to be a disk with $4$ marked points and $\mathcal{S}^{(2)}$ a disk with $2n-1$ marked points. Concretely, for the gauge choice corresponding to $\alpha_n=0$, we want the split where $\mathcal{S}^{(1)} = (1,2,2n-1,2n)$ and $\mathcal{S}_0^{(2)} = (2,3,\ldots,2n-1,2n)$, while, for $\alpha_n=1$, we want the same $\mathcal{S}^{(1)}$ but with $\mathcal{S}_1^{(2)} = (1,2,\ldots,2n-2,2n-1)$ (see Fig.~\ref{fig:Split}). Working out the split kinematic configurations associated to these choices, we find
\begin{equation}
\begin{aligned}
&\underline{\text{Split }\alpha_n=0:}\\ 
    &\, \,X_{1,2n-1} = z_{1,2n-1}, \\
    &\,\,X_{2,2n} = z_{2,2n},\\
    &\,\,X_{1,j} =x_{j,2n},  \, \text{ for } j \in \mathcal{S}_0^{(2)}\setminus\{2n-1\},\\
    &\,\,X_{i,j} = x_{i,j},  \, \text{ for }(i,j) \in \mathcal{S}_0^{(2)}\setminus\{2n\}, \\
    &\,\,\underbracket[0.4pt]{X_{j,2n} = z_{2,2n} +x_{j,2n}}\\
     & \hspace{15mm} \downarrow  \\
    & \
    X_{j,2n}=X_{2,2n} + X_{1,j}
\end{aligned} \quad \quad \, \,
\begin{aligned}
&\underline{\text{Split }\alpha_n=1:}\\ 
    &\,\,X_{1,2n-1} = z_{1,2n-1}, \\
    &\,\,X_{2n-2,2n} = z_{2,2n},\\
    &\,\,X_{j,2n-1} =x_{j,2n-1},  \, \text{ for } j \in \mathcal{S}_1^{(2)}\setminus\{1\},\\
    &\,\,X_{i,j} = x_{i,j}, \, \text{ for }(i,j) \in \mathcal{S}_1^{(2)}\setminus\{2n-1\}, \\
    &\,\,\underbracket[0.4pt]{X_{j,2n} = z_{2,2n} +x_{j,2n-1}}\\
    & \hspace{15mm} \downarrow \\
    & \
    X_{j,2n}=X_{2n-2,2n} + X_{j,2n-1},
\end{aligned}
    \label{eq:SplitMaps}
\end{equation}
for $j = 2, 3, \ldots, 2n-2$. If we further set $z_{2,2n} = 1$ in both splits, we are left with precisely the kinematic conditions listed in Eq.~\eqref{eq:splitAlpha}, for the respective gauge-choices.

Let's now see what happens at the level of the surface integral on the support of these two split configurations. Starting with the $\alpha_n=1$ case, we know that the surface integral will factorize as
\begin{equation}
    \mathcal{A}_{\mathcal{S}_{2n}}(X^\delta_{i,j})  \xrightarrow[\alpha_n=1]{\text{Split}}  \mathcal{A}_{4}[z_{1,2n-1}^\delta,z_{2,2n}^\delta] \times  \mathcal{A}_{\mathcal{S}_1^{(2)}}[x_{i,j}^\delta].
\end{equation}
Using the split map \eqref{eq:SplitMaps}, we derive that $z_{1,2n-1}^\delta = X_{1,2n-1}^\delta \Rightarrow z_{1,2n-1}^\delta = z_{1,2n-1}-1$, and similarly, $z_{2,2n}^\delta =X_{2n-2,2n}^\delta \Rightarrow z_{2,2n}^\delta = z_{2,2n}+1$; as for the kinematics $x_{i,j}^\delta$ of the surface $\mathcal{S}_1^{(2)}$, they have exactly the same shift as their respective $X_{i,j}$. Therefore, we find that
\begin{equation}
\begin{aligned}
    \mathcal{A}_{\mathcal{S}_{2n}}  &\xrightarrow[\alpha_n=1]{\text{Split}}  \int_{\mathcal{S}_1^{(2)}} \Omega^{(2)}_{y} \prod u_{i,j}^{x_{i,j}}  \left(\frac{\prod u_{e,e}}{\prod u_{o,o}}\right) \times \int_{\mathcal{S}^{(1)}} \Omega^{(1)}_{y}  u_{1,2n-1}^{z_{1,2n-1}} u_{2,2n}^{z_{2,2n}} \frac{u_{2,2n}}{u_{1,2n-1}}\\
    &= \mathcal{A}_{\mathcal{S}_1^{(2)}}[x_{i,j} = X_{i,j}] \times \frac{\Gamma(z_{1,2n-1}-1) \Gamma(z_{2,2n}+1) }{\Gamma(z_{1,2n-1}+z_{2,2n}) },
    \label{eq:split-alpha-1}
\end{aligned}
\end{equation}
which, upon taking a scaffolding residue at $X_{1,2n-1} = z_{1,2n-1}=0$ (turning the $n^{\rm{th}}$ gluon on-shell), gives us
\begin{equation}
\begin{aligned}
    &\mathop{\mathrm{Res}}_{X_{1,2n-1}=0} \mathcal{A}_{\mathcal{S}_{2n}} \xrightarrow[\alpha_n=1]{\text{Split}} - z_{2,2n} \times \mathcal{A}_{\mathcal{S}_1^{(2)}}[X_{i,j}]  \\
    &\Leftrightarrow \quad \sum_{j}\underbracket[0.4pt]{(X_{2n,j}-X_{2n-1,j})}_{z_{2,2n}} \partial_{X_{2n,j}} \mathcal{A}_{\mathcal{S}_{2n}} = - z_{2,2n} \mathcal{A}_{\mathcal{S}_{2n-1}}(1,2,\ldots,2n-1) \\
    &\Leftrightarrow \quad \mathcal{W}_{2n}\left[\mathcal{A}_{\mathcal{S}_{2n}}\right] = - \mathcal{A}_{\mathcal{S}_{2n-1}}(1,2,\ldots,2n-1),
\end{aligned}
\end{equation}
where we have used the fact that after taking scaffolding residue in $X_{1,2n-1}$ recover the statement of gauge invariance and linearity in the polarization vector of the $n^{\rm{th}}$ gluon. $\mathcal{S}_{2n-1} \equiv \mathcal{S}^{(2)}_1$ stands for surface we are left with -- a disk with $(2n-1)$ marked points on the boundary. By following the same procedure for the $\alpha_{n}=0$ split, we find
\begin{equation}
\begin{aligned}
    \mathcal{A}_{\mathcal{S}_{2n}}  &\xrightarrow[\alpha_n=0]{\text{Split}}  \int_{\mathcal{S}_0^{(2)}} \Omega^{(2)}_{y} \prod u_{i,j}^{x_{i,j}}  \left(\frac{\prod_{e \neq 2n} u_{e,e}}{\prod u_{o,o} u_{2n,o}}\right) \times \int_{\mathcal{S}^{(1)}} \Omega^{(1)}_{y}  u_{1,2n-1}^{z_{1,2n-1}} u_{2,2n}^{z_{2,2n}} \frac{u_{2,2n}}{u_{1,2n-1}}\\
    &= \mathcal{A}_{\mathcal{S}_0^{(2)}}[x_{j,2n}=X_{1,j},x_{i,j} = X_{i,j}] \times \frac{\Gamma(z_{1,2n-1}-1) \Gamma(z_{2,2n}+1) }{\Gamma(z_{1,2n-1}+z_{2,2n}) },
\end{aligned}
\end{equation}
where, as previously noted, $\mathcal{S}_0^{(2)}=(2,3,\ldots,2n-1,2n)$. However, because we have $x_{j,2n}=X_{1,j}$ from split kinematics, the particular surface integral $\mathcal{A}_{\mathcal{S}_0^{(2)}}$ given above is equivalent to $\mathcal{A}_{\mathcal{S}_1^{(2)}}\left[ X_{i,j}\right] \equiv \mathcal{A}_{\mathcal{S}_{2n-1}}\left[ X_{i,j}\right]$, $i.e.$, the same integral appearing in the last line of Eq.~\eqref{eq:split-alpha-1}. And so, by taking the scaffolding residue on $X_{1,2n-1}=z_{1,2n-1}=0$, we find
\begin{equation}
    \mathop{\mathrm{Res}}_{X_{1,2n-1}=0} \mathcal{A}_{\mathcal{S}_{2n}} \xrightarrow[\alpha_n=0]{\text{Split}} -\mathcal{A}_{\mathcal{S}_{2n-1}}[X_{i,j}]  \quad \Leftrightarrow \quad \mathcal{W}_{2n}\left[\mathcal{A}_{\mathcal{S}_{2n}}\right] = - \mathcal{A}_{\mathcal{S}_{2n-1}}(1,2,\ldots,2n-1),
\end{equation}
which precisely agrees with the result from the previous split. Of course, this was expected, since on the support of $X_{1,2n-1} = 0$, the two split configurations are gauge-equivalent.

So, in summary, we find that when we act with the operator $\W_{2i}$ on the full $2n$-point scalar-scaffolded string amplitude $\mathcal{A}_{\mathcal{S}_{2n}}$, we are left with the surface integral with the appropriate even/odd $\delta$-shifts for a surface \textit{without} point $2i$:
\begin{equation}    \mathcal{W}_{2i}\left[\mathcal{A}_{\mathcal{S}_{2n}}\right] = -\int_{\mathcal{S}_{2n-1}= \mathcal{S}_{2n} \setminus\{2i\}} \Omega_{y} \prod_{(i,j)\in \mathcal{S}_{2n-1}} u_{i,j}^{X_{i,j}} \times \left( \frac{\prod u_{e,e}}{\prod u_{o,o}} \right),
\label{eq:ActW2n}
\end{equation}
where $\mathcal{S}_{2n-1}$ is the disk $(1, 2, \ldots, 2i - 1, 2i + 1, \ldots, 2n)$. Of course, by taking the rest of the scaffolding residues $X_{1,3}=X_{3,5}=\cdots=X_{2i - 3, 2i - 1} = X_{2i + 1, 2i + 3}=\cdots=X_{2n-1,1} = 0$ on Eq.~\eqref{eq:ActW2n}, we can extract the action of $\W_{2i}$ on the full string gluon amplitude.

\subsection{Consecutive splits}
\label{sec:consectivesplits}

Having understood the result of applying a single $\W_e$, we can now ask what happens if we act with another $\W_{e^\prime}$ on the integral \eqref{eq:ActW2n}. In particular, we want to understand if we can still describe this new polarization configuration as associated to the same splits identified above, but now applied to a different gluon.

As we will see now, since the integral in \cref{eq:ActW2n} has an \textit{odd} number of points, the ratio of $u_{e,e}/u_{o,o}$ turns out to not simplify in the same way as it does when the integral is \textit{even}. As a result, the gauge invariance statement \eqref{eq:GgInvarianceII} for the gluons adjacent to $2i$ are modified. Nonetheless, we find that for any gluon there is always at least one split pattern that correctly gives the action of $\W$. Of course, we are free to define the action of $\W$ via a particular split choice (or in other words a convenient gauge), and since the final answer is gauge invariant, it won't depend on this choice. 

Let's see what happens explicitly with a simple example for $n=4$ ($2n=8$); it will be clear that the result generalizes to all $n$. Looking at Eq.~\eqref{eq:ActW2n}, we see that, after acting with $\W_8$ on $\mathcal{A}_{\mathcal{S}_{2n}}$ (and taking a scaffolding residue on $X_{1,2n-1} = 0$), the surface integral associated to $\mathcal{S}_7$ becomes
\begin{equation}
  \W_{8} \left[ \,\begin{gathered}
    \begin{tikzpicture}[line width=1,scale=10,baseline={([yshift=0.0ex]current bounding box.center)}]
    \node[regular polygon, regular polygon sides=8,  minimum size=2cm] (p) at (0,0) {$\mathcal{
    S}_8$};
    
    \node[scale=0.8,xshift=5,yshift=5] at (p.corner 8) {$6$};
    \node[scale=0.8,xshift=5,yshift=-5] at (p.corner 7) {$7$};
    \node[scale=0.8,xshift=5,yshift=-5] at (p.corner 6) {$8$};
    \node[scale=0.8,xshift=-5,yshift=-5] at (p.corner 5) {$1$};
    \node[scale=0.8,xshift=-5,yshift=-5] at (p.corner 4) {$2$};
    \node[scale=0.8,xshift=-5,yshift=5] at (p.corner 3) {$3$};
    \node[scale=0.8,xshift=-5,yshift=5] at (p.corner 2) {$4$};
    \node[scale=0.8,xshift=5,yshift=5] at (p.corner 1) {$5$};
    
    \foreach \i in {1,...,8}
    {\draw[fill] (p.corner \i) circle [radius=0.1pt];}

    \draw (0,0) circle [radius=0.102cm];
    \end{tikzpicture}
\end{gathered} \, \right] = \begin{gathered}
    \begin{tikzpicture}[line width=1,scale=10,baseline={([yshift=0.0ex]current bounding box.center)}]
    \node[regular polygon, regular polygon sides=8,  minimum size=2cm] (p) at (0,0) {$\mathcal{
    S}_7$};
        \draw (0,0) circle [radius=0.102cm];
    \node[circle,radius=0.2cm] (c) at (p.corner 5){}; \filldraw[fill=white,draw=white,yshift=-40] (p.corner 7) rectangle (c.south);

    \node[scale=0.8,xshift=5,yshift=5] at (p.corner 8) {$6$};
    \node[scale=0.8,xshift=5,yshift=-5] at (p.corner 7) {$7$};
    \node[scale=0.8,xshift=-5,yshift=-5] at (p.corner 5) {$1$};
    \node[scale=0.8,xshift=-5,yshift=-5] at (p.corner 4) {$2$};
    \node[scale=0.8,xshift=-5,yshift=5] at (p.corner 3) {$3$};
    \node[scale=0.8,xshift=-5,yshift=5] at (p.corner 2) {$4$};
    \node[scale=0.8,xshift=5,yshift=5] at (p.corner 1) {$5$};
    
    \foreach \i in {1,...,5}
    {\draw[fill] (p.corner \i) circle [radius=0.1pt];}
     \foreach \i in {7,8}
    {\draw[fill] (p.corner \i) circle [radius=0.1pt];}

    \draw[Maroon] (p.corner 7) -- (p.corner 5);
    
    \end{tikzpicture}
\label{eq:act-w-8}
\end{gathered} .
\end{equation}

Now, we want to apply the operator $\W_2$, associated with gluon $1$, to the integral associated with $\mathcal{A}_{\mathcal{S}_7}$, as defined in  Eq.~\eqref{eq:act-w-8}. To do this, let's look at the split configuration that would give us the action of $\W_2$. Just like in the $\W_8$ case, we should have naively two options --- $\alpha_1=0$ and $\alpha_1=1$ --- where the first one imposes the locus $X_{2,j}=1+X_{3,j}$ and the second one imposes $X_{2,j}=1+X_{1,j}$. Let's start by looking at the first case, for which we have the following split:
\begin{equation}
\begin{aligned}
&\underline{\text{Gluon 1:}} \\
&(\alpha_1=0)
\end{aligned}\quad \quad 
\begin{gathered}
    \begin{tikzpicture}[line width=1,scale=10,baseline={([yshift=0.0ex]current bounding box.center)}]
    
    \node[regular polygon, regular polygon sides=8,  minimum size=3cm] (p) at (0,0) {};
    
    \node[circle,radius=0.2cm] (c) at (p.corner 5){}; \filldraw[fill=white,draw=white,yshift=-40] (p.corner 7) rectangle (c.south);
    \draw[Maroon,fill=Maroon,fill opacity=0.1] (p.corner 5) -- (p.corner 4)--(p.corner 2) -- (p.corner 1)-- (p.corner 8)  -- (p.corner 7); 

    \draw[Blue,fill=Blue,fill opacity=0.1] (p.corner 5) --(p.corner 4)-- (p.corner 3)--(p.corner 2) -- (p.corner 5);

    \node[scale=0.8,xshift=5,yshift=5] at (p.corner 8) {$6$};
    \node[scale=0.8,xshift=5,yshift=-5] at (p.corner 7) {$7$};
    \node[scale=0.8,xshift=-5,yshift=-5] at (p.corner 5) {$1$};
    \node[scale=0.8,xshift=-5,yshift=-5] at (p.corner 4) {$2$};
    \node[scale=0.8,xshift=-5,yshift=5] at (p.corner 3) {$3$};
    \node[scale=0.8,xshift=-5,yshift=5] at (p.corner 2) {$4$};
    \node[scale=0.8,xshift=5,yshift=5] at (p.corner 1) {$5$};
    
    \foreach \i in {1,...,5}
    {\draw[fill] (p.corner \i) circle [radius=0.1pt];}
     \foreach \i in {7,8}
    {\draw[fill] (p.corner \i) circle [radius=0.1pt];}

    \draw[Maroon] (p.corner 7) -- (p.corner 5);

    \path (p.corner 2) -- (p.corner 4) coordinate[pos=0.5] (anchor1);
     \path (p.corner 5) -- (p.corner 1) coordinate[pos=0.7] (anchor2);
    
    \draw[->, thick, bend left=20] (-0.13,0.13) to (anchor1);
     \node[scale=0.8,xshift=-5,yshift=10] at (-0.13,0.13) {$z_{i,j}$};

     \draw[->, thick, bend right=20] (0.13,0.13) to (anchor2);
     \node[scale=0.8,xshift=13,yshift=0] at (0.13,0.13) {$x_{i,j}$};
    
    \end{tikzpicture}
\end{gathered}  \Rightarrow \quad \begin{cases}
    X_{2,4} = z_{2,4}, \\
    X_{1,3} = z_{1,3},\\
    X_{3,j}=x_{2,j}, \, j\neq 1 \\
     X_{i,j} = x_{i,j}, \, (i,j)\in \mathcal{S}^{(2)}\setminus\{2\}, \\
     X_{2,j} = x_{2,j} + z_{2,4}. \\
\end{cases} 
\end{equation}
We can check whether this split is compatible with the $\delta$-shift by reading off the shifts on the $x_{i,j}$ and $z_{i,j}$. From the first two equalities we have $z_{2,4}^\delta = X_{2,4}^\delta \Rightarrow z_{2,4}^\delta = z_{2,4}+1$, and similarly, $z_{1,3}^\delta = z_{1,3}-1$. From the third equality we find $x_{2,j}^\delta = x_{2,j} -1$ if $j$ is odd, and $x_{2,j}^\delta = x_{2,j}$ otherwise. Finally, for the split to be compatible with the $\delta$-shift, we must have that the last equation can be consistently shifted: on the $l.h.s.$, we see that $X_{2,j}$ is shifted with $+1$ if $j$ is even, and not shifted otherwise. This precisely agrees with the shift we obtain for the $r.h.s.$, which we just derived for $z_{2,4}$ and $x_{2,j}$. Therefore, applying this split and taking the scaffolding residue on $X_{1,3}=z_{1,3}=0$ yields:
\begin{equation}
\begin{aligned}
      &\mathop{\mathrm{Res}}_{X_{1,3}=0} \W_{8}\left[\mathcal{A}_{\mathcal{S}_{8}}\right] \xrightarrow[\alpha_1=0]{\text{Split}_1}  \int_{\mathcal{S}^{(2)}=(1,3,4,5,6,7)} \Omega \prod_{(i,j)\in \mathcal{S}^{(2)}} u_{i,j}^{X_{i,j}}  \left(\frac{\prod u_{e,e}}{\prod u_{o,o}}\right) \\
      &\Leftrightarrow \W_{2}\left[\W_{8}\left[\mathcal{A}_{\mathcal{S}_{8}}\right]\right] = \int_{\mathcal{S}_{2n}\setminus\{2,8\}} \Omega \prod_{(i,j)\in \mathcal{S}_{2n}\setminus\{2,8\}} u_{i,j}^{X_{i,j}}  \left(\frac{\prod u_{e,e}}{\prod u_{o,o}}\right). 
\end{aligned}
\end{equation}
Before proceeding, let's see what we would have obtained if we had instead considered the split corresponding to $\alpha_1=1$, given by:
\begin{equation}
\begin{aligned}
&\underline{\text{Gluon 1:}} \\
&(\alpha_1=1)
\end{aligned}\quad \quad 
\begin{gathered}
    \begin{tikzpicture}[line width=1,scale=10,baseline={([yshift=0.0ex]current bounding box.center)}]
    
    \node[regular polygon, regular polygon sides=8,  minimum size=3cm] (p) at (0,0) {};
    
    \node[circle,radius=0.2cm] (c) at (p.corner 5){}; \filldraw[fill=white,draw=white,yshift=-40] (p.corner 7) rectangle (c.south);
    \draw[Maroon,fill=Maroon,fill opacity=0.1] (p.corner 5) -- (p.corner 3)--(p.corner 2) -- (p.corner 1)-- (p.corner 8)  -- (p.corner 7); 

    \draw[Blue,fill=Blue,fill opacity=0.1] (p.corner 5) --(p.corner 4)-- (p.corner 3)--(p.corner 2) -- (p.corner 5);

    \node[scale=0.8,xshift=5,yshift=5] at (p.corner 8) {$6$};
    \node[scale=0.8,xshift=5,yshift=-5] at (p.corner 7) {$7$};
    \node[scale=0.8,xshift=-5,yshift=-5] at (p.corner 5) {$1$};
    \node[scale=0.8,xshift=-5,yshift=-5] at (p.corner 4) {$2$};
    \node[scale=0.8,xshift=-5,yshift=5] at (p.corner 3) {$3$};
    \node[scale=0.8,xshift=-5,yshift=5] at (p.corner 2) {$4$};
    \node[scale=0.8,xshift=5,yshift=5] at (p.corner 1) {$5$};
    
    \foreach \i in {1,...,5}
    {\draw[fill] (p.corner \i) circle [radius=0.1pt];}
     \foreach \i in {7,8}
    {\draw[fill] (p.corner \i) circle [radius=0.1pt];}

    \draw[Maroon] (p.corner 7) -- (p.corner 5);

    \path (p.corner 2) -- (p.corner 4) coordinate[pos=0.5] (anchor1);
     \path (p.corner 5) -- (p.corner 1) coordinate[pos=0.7] (anchor2);
    
    \draw[->, thick, bend left=20] (-0.13,0.13) to (anchor1);
     \node[scale=0.8,xshift=-5,yshift=10] at (-0.13,0.13) {$z_{i,j}$};

     \draw[->, thick, bend right=20] (0.13,0.13) to (anchor2);
     \node[scale=0.8,xshift=13,yshift=0] at (0.13,0.13) {$x_{i,j}$};
    
    \end{tikzpicture}
\end{gathered}  \Rightarrow \quad \begin{cases}
    X_{2,7} \to z_{2,4}, \\
    X_{1,3} \to z_{1,3},\\
     X_{1,j} \to x_{1,j}, \,j\neq 3 \\
      X_{i,j} \to x_{i,j}, \,(i,j)\in\mathcal{S}^{(2)}\setminus\{1\} \\
     X_{2,j} \to x_{1,j} + z_{2,4}. \\
\end{cases} 
\label{eq:Split1}
\end{equation}
Reading off the resulting $\delta$-shifts, we obtain $z_{2,4}^\delta =z_{2,4}$, $z_{1,3}^\delta =z_{1,3}-1$, $x_{1,j}^\delta=X_{1,j}^\delta$. However, given these shifts, the final equality is not consistently shifted: while on the $l.h.s.$ we have that $X_{2,j}$ is shifted by $+1$ when $j$ is even, and not shifted otherwise, on the $r.h.s.$ for even $j$ we obtain no shift and a shift of $-1$ for odd $j$. Therefore this split kinematics is \textit{not} compatible with the $\delta$-shift, and this split doesn't hold. 

However, there is a simple way to fix this. Let us define $\hat{X}_{2,7} = X_{2,7}-1$. If we write the integral \eqref{eq:ActW2n} (for $2i=8$) in terms of $\hat{X}_{2,7}$, we find that $\hat{X}_{2,7}$ will be shifted with $+1$ (since $X_{2,7}$ has no shift). By doing this and applying the same split \eqref{eq:Split1}, by construction we find that it is compatible with the ``new'' $\delta$-shift, but now this shift happens in the locus
\begin{equation}
    \hat{X}_{2,7}=1 \Leftrightarrow X_{2,7} = 2, \quad X_{2,j} = 1+X_{1,j}, \, \text{for }j\neq7.
    \label{eq:DifGauge1}
\end{equation}
This is simply reflecting the fact that the gauge invariance statement for $2$, when making manifest the dependence in $(X_{2,j}-X_{1,j})$, is different than the standard one for the integral defining $\W_8[\mathcal{A}_{\mathcal{S}_8}] \equiv \mathcal{A}_{\W_8}$. In particular, from \cref{eq:DifGauge1}, we have
\begin{equation}
    \mathcal{A}_{\W_8}= \sum_{j\neq7} (X_{2,j}-X_{1,j})\frac{\partial \mathcal{A}_{\W_8}}{\partial X_{2,j}} + (X_{2,7}-1)\frac{\partial \mathcal{A}_{\W_8}}{\partial X_{2,7}}.
    \label{eq:Gauge21}
\end{equation}
However, since there were no problems with the initial split corresponding to the locus $X_{2,j} \to 1+X_{3,j}$, we still have 
\begin{equation}
    \mathcal{A}_{\W_8}= \sum_{j} (X_{2,j}-X_{3,j})\frac{\partial \mathcal{A}_{\W_8}}{\partial X_{2,j}}.
    \label{eq:Gauge23}
\end{equation}
Therefore, subtracting \cref{eq:Gauge23} and \cref{eq:Gauge21}, we obtain the following identity
\begin{equation}
    \sum_{j\neq7} (X_{3,j}-X_{1,j})\frac{\partial \mathcal{A}_{\W_8}}{\partial X_{2,j}} + (X_{3,7}-1) \frac{\partial \mathcal{A}_{\W_8}}{\partial X_{2,7}} = 0,
\end{equation}
and, using this, we can describe the gauge orbit corresponding to the polarization choice of $\W_2$ on $\mathcal{A}_{\W_8}$ as
\begin{equation}
X_{2,7} = 1 + X_{3,7} + \alpha_1(1-X_{3,7}), \quad  
    X_{2,j} = 1 + X_{3,j} + \alpha_1(X_{1,j}-X_{3,j}), \text{ for }j\neq \{7,8\}.
\end{equation}
For $\alpha_1=1$, this polarization choice is of the same form as what we found for $\W_8$, but in general it is different.

Looking back at the split mappings, it is simple to understand that this modification in the gauge-invariant statement comes from the fact that the non-scaffolding chord inside the four-point subsurface, which should be shifted with $+1$, gets mapped into an $X_{e,o}$ in the big surface, which is not shifted. This only happens because after acting with $\W$, the surface we are dealing has one fewer point and hence has two consecutive odd indices. Therefore, it is straightforward that the only ``gluons'' that will have modifications in their gauge-invariance statements are those adjacent to the ones that have already been acted on with $\W_e$ (where the $e$'s are no longer part of the surface). 

In particular, in the four-point example we are considering, this means that the gauge invariance statement for gluon $2$ after acting with $\W_8$ should be unaffected, but the one for gluon $3$ should have a modification. Indeed, if we repeat the exercise we just did for gluon $1$, but now considering the relevant splits for gluons $2$ and $3$ (giving the action of $\W_4$ and $\W_6$, respectively), we find the following. For gluon $2$, both gauge choices $\alpha_2=0$ and $\alpha_2=1$ lead to a split map consistent with the $\delta$-shift, reflecting standard gauge invariance for gluon $2$. As for gluon $3$, we find that the split corresponding to $\alpha_3 =1$, which is given by the mapping $X_{6,j} = 1 +X_{5,j}$, is compatible with the $\delta$-shift, but the same is not true for the case $\alpha_3 =0$, which gives the mapping $X_{6,j} = 1 +X_{7,j}$, precisely as we expected. Concretely, we find that the gauge invariance statements in $6$ are
\begin{equation}
\begin{aligned}
     \mathcal{A}_{\W_8}&= \sum_{j\neq1} (X_{6,j}-X_{j,7})\frac{\partial \mathcal{A}_{\W_8}}{\partial X_{6,j}} + (X_{1,6}-1)\frac{\partial \mathcal{A}_{\W_8}}{\partial X_{1,6}}\\
     &= \sum_{j} (X_{6,j}-X_{j,5})\frac{\partial \mathcal{A}_{\W_8}}{\partial X_{6,j}}
\end{aligned},
\end{equation}
which is completely analogous to what we found for gluon $1$. Note that the direction in which the modification happens is the one touching the $(1,7)$ edge, in this case associated with the differences $(X_{6,j}-X_{7,j})$. 

In summary, having applied any number of $\W_e$, the gauge-invariance statements for the gluons adjacent to the even indices that have been removed will be modified, in one of the directions. We now proceed to studying what happens when we act with $(n-2)$ $\W_e$'s. To do this, we make use of the fact described above -- that, at each step, there always is a ``good direction'' for which the polarization configuration giving the action of $\W_{e}$ is the standard one, $i.e.$ $X_{e,j} \to 1 + X_{e\pm1,j}$. Using this fact, in \cref{sec:polConfigFinal}, we give a simple description of the polarization choice corresponding to acting with multiple $\W_e$'s. Of course, since the gluon amplitude is gauge invariant, any polarization configuration related to the one found in this simple way will also correspond to applying the same set of $\W_e$. 

Let's for now stick to the two simple examples at four-points, where we have acted with $\W_4$ or $\W_6$ on $\mathcal{A}_{\W_8}$, and give explicit expressions of the integrals we get in both cases. From the action of $\W_4$, which we can induce by applying either one of the splits and taking the residue on $X_{3,5}=0$, we obtain 
\begin{equation}
    \W_4 \left[ \begin{gathered}
    \begin{tikzpicture}[line width=1,scale=10,baseline={([yshift=0.0ex]current bounding box.center)}]
    \node[regular polygon, regular polygon sides=8,  minimum size=1.5cm] (p) at (0,0) {$\mathcal{
    S}_7$};
        \draw (0,0) circle [radius=0.077cm];
    \node[circle,radius=0.2cm] (c) at (p.corner 5){}; \filldraw[fill=white,draw=white,yshift=-40] (p.corner 7) rectangle (c.south);

    \node[scale=0.8,xshift=5,yshift=5] at (p.corner 8) {$6$};
    \node[scale=0.8,xshift=5,yshift=-5] at (p.corner 7) {$7$};
    \node[scale=0.8,xshift=-5,yshift=-5] at (p.corner 5) {$1$};
    \node[scale=0.8,xshift=-5,yshift=-5] at (p.corner 4) {$2$};
    \node[scale=0.8,xshift=-5,yshift=5] at (p.corner 3) {$3$};
    \node[scale=0.8,xshift=-5,yshift=5] at (p.corner 2) {$4$};
    \node[scale=0.8,xshift=5,yshift=5] at (p.corner 1) {$5$};
    
    \foreach \i in {1,...,5}
    {\draw[fill] (p.corner \i) circle [radius=0.1pt];}
     \foreach \i in {7,8}
    {\draw[fill] (p.corner \i) circle [radius=0.1pt];}

    \draw[Maroon] (p.corner 7) -- (p.corner 5);
    
    \end{tikzpicture}
\end{gathered} \right] = \begin{gathered}
    \begin{tikzpicture}[line width=1,scale=10,baseline={([yshift=0.0ex]current bounding box.center)}]
    \node[regular polygon, regular polygon sides=8,  minimum size=1.5cm] (p) at (0,0) {$\mathcal{
    S}_6$};
        \draw (0,0) circle [radius=0.077cm];
    \node[circle,radius=0.2cm] (c) at (p.corner 5){}; 
    \node[circle,radius=0.2cm] (c2) at (p.corner 1){};
    \filldraw[fill=white,draw=white,yshift=-40] (p.corner 7) rectangle (c.south);
    \filldraw[fill=white,draw=white,yshift=-40] (p.corner 3) rectangle (c2.north);

    \node[scale=0.8,xshift=5,yshift=5] at (p.corner 8) {$6$};
    \node[scale=0.8,xshift=5,yshift=-5] at (p.corner 7) {$7$};
    \node[scale=0.8,xshift=-5,yshift=-5] at (p.corner 5) {$1$};
    \node[scale=0.8,xshift=-5,yshift=-5] at (p.corner 4) {$2$};
    \node[scale=0.8,xshift=-5,yshift=5] at (p.corner 3) {$3$};
    \node[scale=0.8,xshift=5,yshift=5] at (p.corner 1) {$5$};
    
    \draw[fill] (p.corner 1) circle [radius=0.1pt];
    \foreach \i in {3,...,5}
    {\draw[fill] (p.corner \i) circle [radius=0.1pt];}
     \foreach \i in {7,8}
    {\draw[fill] (p.corner \i) circle [radius=0.1pt];}

    \draw[Maroon] (p.corner 7) -- (p.corner 5);
    \draw[Maroon] (p.corner 3) -- (p.corner 1);
    
    \end{tikzpicture}
\end{gathered} \quad \equiv \quad \int_{\mathcal{S}_6} \prod \frac{d y}{y} \prod_{(i,j) \in \mathcal{S}_6} u_{i,j}^{X_{i,j}} \left( \frac{u_{2,6}}{u_{1,3}u_{1,5}u_{3,7}u_{5,7}} \right).
\label{eq:ActW4}
\end{equation}
On the other hand, applying the ``good'' split to define $\W_6$ and applying the scaffolidng residue $X_{5,7} = 0$, we find
\begin{equation}
\label{eq:ActW6}
    \W_6 \left[ \begin{gathered}
    \begin{tikzpicture}[line width=1,scale=10,baseline={([yshift=0.0ex]current bounding box.center)}]
    \node[regular polygon, regular polygon sides=8,  minimum size=1.5cm] (p) at (0,0) {$\mathcal{
    S}_7$};
        \draw (0,0) circle [radius=0.077cm];
    \node[circle,radius=0.2cm] (c) at (p.corner 5){}; \filldraw[fill=white,draw=white,yshift=-40] (p.corner 7) rectangle (c.south);

    \node[scale=0.8,xshift=5,yshift=5] at (p.corner 8) {$6$};
    \node[scale=0.8,xshift=5,yshift=-5] at (p.corner 7) {$7$};
    \node[scale=0.8,xshift=-5,yshift=-5] at (p.corner 5) {$1$};
    \node[scale=0.8,xshift=-5,yshift=-5] at (p.corner 4) {$2$};
    \node[scale=0.8,xshift=-5,yshift=5] at (p.corner 3) {$3$};
    \node[scale=0.8,xshift=-5,yshift=5] at (p.corner 2) {$4$};
    \node[scale=0.8,xshift=5,yshift=5] at (p.corner 1) {$5$};
    
    \foreach \i in {1,...,5}
    {\draw[fill] (p.corner \i) circle [radius=0.1pt];}
     \foreach \i in {7,8}
    {\draw[fill] (p.corner \i) circle [radius=0.1pt];}

    \draw[Maroon] (p.corner 7) -- (p.corner 5);
    
    \end{tikzpicture}
\end{gathered} \right] = \begin{gathered}
    \begin{tikzpicture}[line width=1,scale=10,baseline={([yshift=0.0ex]current bounding box.center)}]
    \node[regular polygon, regular polygon sides=8,  minimum size=1.5cm] (p) at (0,0) {$\mathcal{
    S}_6$};
        \draw (0,0) circle [radius=0.077cm];
    \node[circle,radius=0.2cm] (c) at (p.corner 5){}; 
    \node[circle,radius=0.2cm] (c2) at (p.corner 7){};
    \filldraw[fill=white,draw=white,yshift=-40] (p.corner 7) rectangle (c.south);
    \filldraw[fill=white,draw=white,yshift=-40] (p.corner 1) rectangle (c2.east);

    \node[scale=0.8,xshift=5,yshift=-5] at (p.corner 7) {$7$};
    \node[scale=0.8,xshift=-5,yshift=-5] at (p.corner 5) {$1$};
    \node[scale=0.8,xshift=-5,yshift=-5] at (p.corner 4) {$2$};
    \node[scale=0.8,xshift=-5,yshift=5] at (p.corner 3) {$3$};
    \node[scale=0.8,xshift=-5,yshift=5] at (p.corner 2) {$4$};
    \node[scale=0.8,xshift=5,yshift=5] at (p.corner 1) {$5$};
    
    \draw[fill] (p.corner 1) circle [radius=0.1pt];
    \foreach \i in {2,...,5}
    {\draw[fill] (p.corner \i) circle [radius=0.1pt];}
     \foreach \i in {7}
    {\draw[fill] (p.corner \i) circle [radius=0.1pt];}

    \draw[Maroon] (p.corner 7) -- (p.corner 5);
    \draw[Maroon] (p.corner 1) -- (p.corner 7);
    
    \end{tikzpicture}
\end{gathered} \quad \equiv \quad \int_{\mathcal{S}_6} \prod \frac{d y}{y} \prod_{(i,j) \in \mathcal{S}_6} u_{i,j}^{X_{i,j}} \left( \frac{u_{2,4}}{u_{1,3}u_{1,5}u_{3,5}u_{3,7}} \right). 
\end{equation}

Now, since the original four-point YM amplitude has units of $X^2$, after acting twice with $\W$ operators, the resulting units are $X^0$. Therefore, we want to extract the piece with units $X^0$ from the low-energy expansion of \cref{eq:ActW4,eq:ActW6}. Since we further know that the final answer should be linear in indices $(2,6)$ and $(2,4)$, respectively, it can only be of the form $X/X$. In addition, since there can at most be a single $X$ in the numerator, this linearity statement implies that, in the case where we acted with $\W_4$ \eqref{eq:ActW4}, the numerator is $X_{2,6}$, and, for the case of $\W_6$, the $X$ in the numerator must be $X_{2,4}$. 

Let's start with \cref{eq:ActW4} and extract the scaffolding residue corresponding to $X_{1,3}=0$. Since we only care about the piece proportional to $X_{2,6}$, we can write
\begin{equation*}
\begin{aligned}
    \mathop{\mathrm{Res}}_{X_{1,3}=0} \mathcal{A}_{\mathcal{S}_6} =& X_{2,6} \int_0^\infty \frac{d y_{5,7}}{y_{5,7}}\frac{d y_{1,5}}{y_{1,5}} \prod_{(i,j) \in \{1,3,4,5,7\}} u_{i,j}^{X_{i,j}} \frac{1}{u_{1,3}u_{1,5}u_{3,7}u_{5,7}} \times \frac{\partial \log(u_{2,6})}{\partial y_{1,3}} \bigg \vert_{y_{1,3}=0} \\
    &+ (\text{independent of }X_{2,6}).
\end{aligned}
\end{equation*}
where from the $u$-equation for $u_{2,6}$ we have that $\partial_{y_{1,3}} \log(u_{2,6}) \vert_{y_{1,3}} = u_{1,3}u_{1,5}u_{3,7}u_{5,7}$, precisely canceling all the denominator factors in the integral. So, by further taking the scaffolding residue in $X_{5,7}$, we are left with
\begin{equation}
    X_{2,6} \times \int_0^{\infty} \frac{d y_{1,5}}{y_{1,5}} u_{1,5}^{X_{1,5}} u_{3,7}^{X_{3,7}} \xrightarrow[X\ll 1]{\text{low energies}} X_{2,6} \times \underbracket[0.4pt]{\left( \frac{1}{X_{1,5}} + \frac{1}{X_{3,7}} \right)}_{\mathcal{A}_4^{\text{Tr}(\phi^3)}(1,3,5,7)} + \mathcal{O}(X),
\end{equation}
where we find the appearance of the Tr($\phi^3$) amplitude for the gluon inner-gon! Of course, we can make the $X_{2,6}$ factor disappear trivially by acting with either $\W_2$ or $\W_6$. 

So we find that, by acting consecutively with $\W_e$ on all the gluons but one, we turn the gluon amplitude into a purely Tr$(\phi^3)$ amplitude! If instead we only acted with all $\W_e$'s but two, we get the Tr$(\phi^3)$ amplitude multiplied by a single $X$ that ensures linearity in the remaining two gluons. 

We will prove this result for the general $n$-point gluon amplitude in the next section. But, before doing that, let's look at what would have happened in the case described by \cref{eq:ActW6}, where instead of acting with $\W_8$ and $\W_4$, we acted with $\W_8$ and $\W_6$. In this case, keeping the part proportional to $X_{2,4}$ in the $X_{1,3}$ residue, we obtain
\begin{equation*}
\begin{aligned}
    \mathop{\mathrm{Res}}_{X_{1,3}=0} \mathcal{A}_{\mathcal{S}_6} =& X_{2,4} \int_0^\infty \frac{d y_{3,5}}{y_{3,5}}\frac{d y_{1,5}}{y_{1,5}} \prod_{(i,j) \in \{1,3,4,5,7\}} u_{i,j}^{X_{i,j}} \frac{1}{u_{1,3}u_{1,5}u_{3,5}u_{3,7}} \times \frac{\partial \log(u_{2,4})}{\partial y_{1,3}} \bigg \vert_{y_{1,3}=0} \\
    &+ (\text{independent of }X_{2,4}).
\end{aligned}
\end{equation*}
But now, we have that $\partial_{y_{1,3}} \log(u_{2,4}) \vert_{y_{1,3}} = u_{1,3}u_{3,7}u_{3,5}$, which cancels all the $u$'s in the denominator except for $u_{1,5}$. Therefore, after taking the scaffolding residue on $X_{5,7}$, we derive
\begin{equation}
    X_{2,4} \times \int_0^{\infty} \frac{d y_{1,5}}{y_{1,5}} u_{1,5}^{X_{1,5}} u_{3,7}^{X_{3,7}} \times \frac{1}{u_{1,5}} = X_{2,4} \times \underbracket[0.4pt]{\frac{\Gamma(X_{1,5}-1)\Gamma(X_{3,7})}{\Gamma(X_{1,5}+X_{3,7}-1)}}_{\to \mathcal{A}_4^{\text{Tr}(\phi^3)}(1,3,5,7) + \mathcal{O}(X)}.
    \label{eq:stringyX24}
\end{equation}
So, in this case we don't get exactly the surface Tr$(\phi^3)$ integral \eqref{eq:SurfaceIntTr} --- instead, we have an extra factor of $1/u_{1,5}$. Even so, in this case it is trivial to see that the leading low-energy part of this integral is \textit{still} the four-point Tr$(\phi^3)$ amplitude. And, once again, if we act with $\W_4$ or $\W_2$ on this object are left with the purely scalar amplitude just like before. In general, after applying $(n-2)$ $\W$'s, we will find integrals like this one, that are not flat-out the standard ``surface'' Tr$(\phi^3)$ integrals shown in \cref{eq:SurfaceIntTr}. But, we will show that, for this new class of integrals, the leading order in the low-energy expansion is still Tr$(\phi^3)$ theory. We explain this now.

At higher points, the strategy is clear: we start by applying the $(n-2)$ splits that give us the action of $(n-2)$ $\W_e$ operators, for all $e$ except $e_1, e_2$ (see the $l.h.s.$ of \cref{fig:consecutive_splits}). From this, we obtain the $\delta$-shifted integral associated to surface where all even indices except $e_1, e_2$ have been deleted from the original surface (first equation on the $r.h.s.$ of \cref{fig:consecutive_splits}). Then, when we extract the scaffolding residue associated with $e_1$ ($X_{i_1,j_1} = 0$, as depicted in \cref{fig:consecutive_splits}), we want to collect the piece proportional to $X_{e_1,e_2}$ in order to ensure linearity in polarizations $e_1$ and $e_2$. Again, the term proportional to $X_{e_1,e_2}$ brings to the integrand a factor of $\partial_{y_{i_1,j_1}} \log(u_{e_1,e_2}) \vert_{y_{i_1,j_1}=0}$, which we can compute using the $u$-equation for $u_{e_1,e_2}$; the result is $\partial_{y_{i_1,j_1}} \log(u_{e_1,e_2}) \vert_{y_{i_1,j_1}=0} = - \prod_{i\in I, j\in J} u_{i,j}$, where $I$ is the set of \textit{odd} indices between $i_1$ and $i_2$ (including $i_1$ and $i_2$), and $J$ is that between $j_1$ and $j_2$  (including $j_1$ and $j_2$), since these are all the curves that cross $(e_1,e_2)$. This leaves us with the integral in the second line in \cref{fig:consecutive_splits}. There, we see a factor of one over the product of $u$'s corresponding to curves that start and end in $I$, and the same for $J$ --- these are represented in blue on the $l.h.s.$ of \cref{fig:consecutive_splits}. Finally, extracting the last residue in $X_{i_2,j_2}$, we obtain the bottom line in  \cref{fig:consecutive_splits}, which is now a surface integral defined over the inner \textit{odd} n-gon that we get after taking all scaffolding residue. 

\begin{figure}
\begin{minipage}{.4\textwidth}
    \begin{tikzpicture}[line width=1,scale=10]
    \node[regular polygon, regular polygon sides=22,  minimum size=5cm] (p) at (0,0) {$\mathcal{
    S}_{2n}$};
     \draw (0,0) circle [radius=0.251cm];
     
     \node[circle,radius=1cm] (c1) at (p.corner 22){}; 
     \node[circle,radius=0.5cm] (c2) at (p.corner 8){};
     \filldraw[fill=white,draw=white,yshift=-40] (c1.north) rectangle (c2.west);
     \node[circle,radius=0.8cm,fill=white,draw=white] (cp) at (p.corner 2){};
    \filldraw[fill=white,draw=white,yshift=-40] (cp.north) rectangle (p.corner 22);   
    \filldraw[fill=white,draw=white,yshift=-40] (cp.north) rectangle (p.corner 4);

     \node[circle,radius=1.5cm] (c3) at (p.corner 20){}; 
     \node[circle,radius=0.5cm] (c4) at (p.corner 12){};
     \node[circle,radius=0.5cm] (c5) at (p.corner 18){};
     \filldraw[fill=white,draw=white,yshift=-40] (c3.east) rectangle (c4.south);
     \filldraw[fill=white,draw=white,yshift=-40] (c3.center) rectangle (c5.east);
     \filldraw[fill=white,draw=white,yshift=-40] (p.corner 16) rectangle (c5.east);
     \filldraw[fill=white,draw=white,yshift=-40] (p.corner 10) rectangle (c4.south);

    \node[scale=0.8,xshift=-5,yshift=-5] at (p.corner 8) {$i_1$};
    \node[scale=0.8,xshift=-5,yshift=-5] at (p.corner 9) {$e_1$};
    \node[scale=0.8,xshift=-5,yshift=-5] at (p.corner 10) {$j_1$};
    
    \node[scale=0.8,xshift=7,yshift=7] at (p.corner 22) {$i_2$};
    \node[scale=0.8,xshift=5,yshift=5] at (p.corner 21) {$e_2$};
    \node[scale=0.8,xshift=7,yshift=7] at (p.corner 20) {$j_2$};
    
    \foreach \i in {20,...,22}
    {\draw[fill] (p.corner \i) circle [radius=0.1pt];}
    \foreach \i in {8,...,10}
    {\draw[fill] (p.corner \i) circle [radius=0.1pt];}

     \draw[Gray,dashed,opacity=0.5] (p.corner 8) -- (p.corner 7)-- (p.corner 6) --(p.corner 5)--(p.corner 4)--(p.corner 3)--(p.corner 2)--(p.corner 1)--(p.corner 22);
     \draw[Gray,dashed,opacity=0.5] (p.corner 10) -- (p.corner 11)-- (p.corner 12) --(p.corner 13)--(p.corner 14)--(p.corner 15)--(p.corner 16)--(p.corner 17)--(p.corner 18)--(p.corner 19)--(p.corner 20);

    \draw[fill] (p.corner 6) circle [radius=0.1pt];
    \draw[fill] (p.corner 4) circle [radius=0.1pt];
    \draw[fill] (p.corner 2) circle [radius=0.1pt];

    \draw[fill] (p.corner 18) circle [radius=0.1pt];
    \draw[fill] (p.corner 16) circle [radius=0.1pt];
    \draw[fill] (p.corner 14) circle [radius=0.1pt];
    \draw[fill] (p.corner 12) circle [radius=0.1pt];

    \draw[Maroon] (p.corner 8) -- (p.corner 6)-- (p.corner 4)-- (p.corner 2)-- (p.corner 22);
    \draw[Maroon] (p.corner 10) -- (p.corner 12)-- (p.corner 14)-- (p.corner 16)-- (p.corner 18)-- (p.corner 20);

    \path[Maroon,bend right,dashed] (p.corner 22) edge (p.corner 20);
    \path[Maroon,bend left,dashed] (p.corner 8) edge (p.corner 10);

    \path[Blue,opacity=0.4] (p.corner 8) edge (p.corner 22);
    \path[Blue,opacity=0.4] (p.corner 8) edge (p.corner 2);
    \path[Blue,opacity=0.4] (p.corner 8) edge (p.corner 4);
    \path[Blue,opacity=0.4] (p.corner 22) edge (p.corner 4);
    \path[Blue,opacity=0.4] (p.corner 22) edge (p.corner 6);
     \path[Blue,opacity=0.4] (p.corner 6) edge (p.corner 2);

    \path[Blue,opacity=0.4] (p.corner 10) edge (p.corner 20);
    \path[Blue,opacity=0.4] (p.corner 10) edge (p.corner 18);
    \path[Blue,opacity=0.4] (p.corner 10) edge (p.corner 16);
     \path[Blue,opacity=0.4] (p.corner 10) edge (p.corner 14);
     \path[Blue,opacity=0.4] (p.corner 14) edge (p.corner 20);
     \path[Blue,opacity=0.4] (p.corner 12) edge (p.corner 20);
     \path[Blue,opacity=0.4] (p.corner 16) edge (p.corner 20);
     \path[Blue,opacity=0.4] (p.corner 12) edge (p.corner 18);
     \path[Blue,opacity=0.4] (p.corner 12) edge (p.corner 16);
     \path[Blue,opacity=0.4] (p.corner 18) edge (p.corner 14);


  
    \end{tikzpicture} 
  \end{minipage}
  \begin{minipage}{.6\textwidth}

      \begin{align}
    &\int \frac{d y_{i_1,j_1}}{y_{i_1,j_1}} \frac{d y_{i_2,j_2}}{y_{i_2,j_2}} \prod \frac{dy_{o,o}}{y_{o,o}} \prod_{X}u_X^X \frac{u_{e_1,e_2}}{\prod\limits_{i\in I,j \in J} u_{i,j} \prod u_{i,i^\star} \prod u_{j,j^\star}} \nonumber\\
    & \Bigg \downarrow  \mathop{\mathrm{Res}}_{X_{i_1,j_1}=0} \nonumber\\
    &=X_{e_1,e_2} \int \frac{d y_{i_2,j_2}}{y_{i_2,j_2}} \prod \frac{dy_{o,o}}{y_{o,o}}\prod_{X}u_X^X \frac{1}{ \prod\limits_{i,i^*\in I} u_{i,i^*} \prod\limits_{j,j^*\in J} u_{j,j^\star}} + \cdots\nonumber\\
    & \Bigg \downarrow  \mathop{\mathrm{Res}}_{X_{i_2,j_2}=0} \nonumber\\
     &=X_{e_1,e_2} \int_{\mathcal{S}_n}\prod\frac{dy_{o,o}}{y_{o,o}} \prod_{X}u_X^X \frac{1}{ \prod\limits_{i,i^*\in I} u_{i,i^\star} \prod\limits_{j,j^*\in J} u_{j,j^\star}} \nonumber
\end{align}

  \end{minipage}%
  \caption{(Left) In red, we draw the leftover surface obtained after acting with $\W_e$ on all even indices apart from $e_1$ and $e_2$. The set of indices between $i_1$ and $i_2$ we call $I$, and those between $j_1$ and $j_2$ we call $J$. The curves in blue are the curves starting and ending in either $I$ or $J$. (Right) On the first line is the surface integral associated to the red surface. After taking the two remaining scaffolding residue in $X_{i_1,j_1}$ and $X_{i_2,j_2}$, we find the surface Tr$(\phi^3)$ integral with an extra factor of one over the product of the $u$'s associated with the curves highlighted in blue on the left.}
  \label{fig:consecutive_splits}
\end{figure}

So, in summary, after applying $(n-2)$ consecutive splits, the result is simply a pre-factor of $X_{e_1,e_2}$ times a particular surface integral over the disk with $n$-marked points on the boundary. All of the points on the disk are labeled by odd indices, corresponding to the inner $n$-gon, and, in addition to the standard logarithmic form and Koba-Nielsen factor, we have a factor of one over the product of all the $u$'s living in the two smaller subsurfaces determined by the chords $(i_1,i_2)$ and $(j_1,j_2)$:
\begin{equation}
\begin{gathered}
    \begin{tikzpicture}[line width=1,scale=10]

    \node[regular polygon, regular polygon sides=22,  minimum size=2cm] (p) at (0,0) {};
     \draw (0,0) circle [radius=0.101cm];

    \node[scale=0.8,xshift=-5,yshift=-5] at (p.corner 8) {$i_1$};
    \node[scale=0.8,xshift=-5,yshift=-5] at (p.corner 10) {$j_1$};    
    \node[scale=0.8,xshift=7,yshift=7] at (p.corner 22) {$i_2$};
    \node[scale=0.8,xshift=7,yshift=7] at (p.corner 20) {$j_2$};
    
    \foreach \i in {20}
    {\draw[fill] (p.corner \i) circle [radius=0.1pt];}
    \foreach \i in {8}
    {\draw[fill] (p.corner \i) circle [radius=0.1pt];}
    \draw[fill] (p.corner 10) circle [radius=0.1pt];
     \draw[fill] (p.corner 22) circle [radius=0.1pt];

     \fill[Blue,opacity=0.2]  (p.corner 8) -- (p.corner 7) -- (p.corner 6) -- (p.corner 5) -- (p.corner 4) -- (p.corner 3) -- (p.corner 2) -- (p.corner 1) -- (p.corner 22)-- cycle;

     \fill[Blue,opacity=0.2]  (p.corner 10) -- (p.corner 11) -- (p.corner 12) -- (p.corner 13) -- (p.corner 14) -- (p.corner 15) -- (p.corner 16) -- (p.corner 17) -- (p.corner 18)-- (p.corner 19)-- (p.corner 20)-- cycle;

     \node[scale=0.8,xshift=18,yshift=30] at (p.corner 8) {$\mathcal{S}_L$};
     \node[scale=0.8,xshift=37,yshift=13] at (p.corner 10) {$\mathcal{S}_R$};

    \end{tikzpicture}     
\end{gathered}
\equiv X_{e_1,e_2} \times \int_{\mathcal{S}_n} \prod \frac{d y}{y}  \prod_{(i,j) \in \mathcal{S}_n} u_{i,j}^{X_{i,j}} \frac{1}{\prod_{(i,j)\in \mathcal{S}_L} u_{i,j} \times \prod_{(i,j)\in \mathcal{S}_R} u_{i,j} },
\label{eq:case1}
\end{equation}
where $\mathcal{S}_L$ and $\mathcal{S}_R$ include curves $(i_1,i_2)$ and $(j_1,j_2)$, respectively. If instead we took $e_2 = e_1 +2$ (this is $i_1 = i_2$), we obtain the extremal case where $\mathcal{S}_L$ disappears, and instead we get the following integral:
\begin{equation}
\begin{gathered}
    \begin{tikzpicture}[line width=1,scale=10]
    \node[regular polygon, regular polygon sides=22,  minimum size=2cm] (p) at (0,0) {};
     \draw (0,0) circle [radius=0.101cm];

    \node[scale=0.8,xshift=-10,yshift=0] at (p.corner 7) {$k_2$};
    \node[scale=0.8,xshift=-7,yshift=-5] at (p.corner 10) {$k_1$};    
    \node[scale=0.8,xshift=-7,yshift=7] at (p.corner 4) {$k_3$};
    
    \foreach \i in {7}
    {\draw[fill] (p.corner \i) circle [radius=0.1pt];}
    \foreach \i in {4}
    {\draw[fill] (p.corner \i) circle [radius=0.1pt];}
    \draw[fill] (p.corner 10) circle [radius=0.1pt];


     \fill[Blue,opacity=0.2]   (p.corner 10) -- (p.corner 11) -- (p.corner 12) -- (p.corner 13) -- (p.corner 14) -- (p.corner 15) -- (p.corner 16) -- (p.corner 17) -- (p.corner 18)-- (p.corner 19)-- (p.corner 20) -- (p.corner 21) -- (p.corner 22) -- (p.corner 1) -- (p.corner 2) -- (p.corner 3) -- (p.corner 4) -- cycle;

     \node[scale=0.8,xshift=25,yshift=30] at (p.corner 10) {$\mathcal{S}_R$};

    \end{tikzpicture}     
\end{gathered} \, \, 
\equiv X_{e_1,e_2} \times \int_{\mathcal{S}_n} \prod \frac{d y}{y}  \prod_{(i,j) \in \mathcal{S}_n} u_{i,j}^{X_{i,j}} \frac{1}{ \prod_{(i,j)\in \mathcal{S}_R} u_{i,j} },
\label{eq:extremalcase}
\end{equation}
where again $\mathcal{S}_R$ includes curve $(k_1,k_3)$.

So, in both cases, the integral we obtain is not \textit{precisely} the ``surface'' Tr$(\phi^3)$ integral --- it is instead one where certain kinematics $X_{i,j}$ are shifted by $-1$. Yet, as we will now demonstrate, all of these integrals remarkably give the $n$-point Tr$(\phi^3)$ amplitude at leading order in the low-energy expansion! With this result, we find that, by acting with $(n-2)$ $\W_e$'s on the $n$-point gluon amplitude ($2n$ scalars), we produce the corresponding $n$-point Tr($\phi^3$) amplitude, multiplied by the factor $X_{e_1,e_2}$ for the two even indices left out of the set of $\W_e$. If we further act with either $\W_{e_1}$ or $\W_{e_2}$, we kill the $X_{e_1,e_2}$ pre-factor, and we are left with the pure scalar amplitude --- succeeding in converting \textit{all} the external gluons into colored scalars!

\subsection{Shifted stringy integrals and Tr($\phi^3$) at low energies}
\label{sec:shiftsTrPhi3}

Let's now prove that the integrals found in \cref{eq:case1,eq:extremalcase} indeed give Tr($\phi^3$) amplitudes at low energies. To do this, we want to show that the residue in any full triangulation/cubic diagram is equal to $1$ (at leading order). We will demonstrate this recursively, by taking one residue at a time in the triangulation chords, and understanding how the result of each residue always lands us on integrals of the form of \cref{eq:case1,eq:extremalcase}. 

We begin by looking at integrals of the form of \cref{eq:case1}. Pick a triangulation of the surface $\mathcal{S}_n$. There are then two possibilities: 
\begin{enumerate}
    \itemsep-0em
    \item the triangulation contains a chord $(k,m)$ crossing both curves $(i_1,i_2)$ and $(j_1,j_2)$ (the boundaries of $\mathcal{S}_L/\mathcal{S}_R$);
    \item the triangulation contains either chord $(i_1,j_2)$ or $(i_2,j_1)$.
\end{enumerate}
Let's start by looking at case 1. In this scenario, since the exponent of curve $(k,m)$ is \textit{not} shifted, when we take the residue at $X_{k,m} \to 0$ we simply find the product of the two lower surface integrals that result from cutting $\mathcal{S}_n$ along $(k,m)$:

\begin{equation}
\begin{gathered}
    \begin{tikzpicture}[line width=1,scale=10]
    
    \node[regular polygon, regular polygon sides=22,  minimum size=2cm] (p) at (0,0) {};
     \draw (0,0) circle [radius=0.101cm];

    \node[scale=0.8,xshift=-5,yshift=-5] at (p.corner 8) {$i_1$};
    \node[scale=0.8,xshift=-5,yshift=-5] at (p.corner 10) {$j_1$};    
    \node[scale=0.8,xshift=7,yshift=7] at (p.corner 22) {$i_2$};
    \node[scale=0.8,xshift=7,yshift=7] at (p.corner 20) {$j_2$};
    
    \foreach \i in {20}
    {\draw[fill] (p.corner \i) circle [radius=0.1pt];}
    \foreach \i in {8}
    {\draw[fill] (p.corner \i) circle [radius=0.1pt];}
    \draw[fill] (p.corner 10) circle [radius=0.1pt];
     \draw[fill] (p.corner 22) circle [radius=0.1pt];

     \fill[Blue,opacity=0.2]  (p.corner 8) -- (p.corner 7) -- (p.corner 6) -- (p.corner 5) -- (p.corner 4) -- (p.corner 3) -- (p.corner 2) -- (p.corner 1) -- (p.corner 22)-- cycle;

     \fill[Blue,opacity=0.2]  (p.corner 10) -- (p.corner 11) -- (p.corner 12) -- (p.corner 13) -- (p.corner 14) -- (p.corner 15) -- (p.corner 16) -- (p.corner 17) -- (p.corner 18)-- (p.corner 19)-- (p.corner 20)-- cycle;


     \draw[Maroon] (p.corner 4) -- (p.corner 15);
    \fill[Maroon] (p.corner 4) circle [radius=0.15pt];
     \fill[Maroon] (p.corner 15) circle [radius=0.15pt];
     \node[scale=0.8,xshift=-7,yshift=7,Maroon] at (p.corner 4) {$k$};
     \node[scale=0.8,xshift=7,yshift=-7,Maroon] at (p.corner 15) {$m$};
    \end{tikzpicture}     
\end{gathered} \quad \xrightarrow{\mathop{\mathrm{Res}}_{X_{k,m}=0}} \quad 
\begin{gathered}
    \begin{tikzpicture}[line width=1,scale=10]
    
    \node[regular polygon, regular polygon sides=22,  minimum size=2cm] (p) at (0,0) {};
     \draw (0,0) circle [radius=0.101cm];

    \node[scale=0.8,xshift=-5,yshift=-5,Maroon] at (p.corner 8) {$k$};
    \node[scale=0.8,xshift=-5,yshift=-5,Maroon] at (p.corner 10) {$m$};    
    \node[scale=0.8,xshift=7,yshift=7] at (p.corner 22) {$i_2$};
    \node[scale=0.8,xshift=7,yshift=7] at (p.corner 20) {$j_2$};
    
    \foreach \i in {20}
    {\draw[fill] (p.corner \i) circle [radius=0.1pt];}
    \foreach \i in {8}
    {\draw[fill] (p.corner \i) circle [radius=0.1pt];}
    \draw[fill] (p.corner 10) circle [radius=0.1pt];
     \draw[fill] (p.corner 22) circle [radius=0.1pt];

     \fill[Blue,opacity=0.2]  (p.corner 8) -- (p.corner 7) -- (p.corner 6) -- (p.corner 5) -- (p.corner 4) -- (p.corner 3) -- (p.corner 2) -- (p.corner 1) -- (p.corner 22)-- cycle;

     \fill[Blue,opacity=0.2]  (p.corner 10) -- (p.corner 11) -- (p.corner 12) -- (p.corner 13) -- (p.corner 14) -- (p.corner 15) -- (p.corner 16) -- (p.corner 17) -- (p.corner 18)-- (p.corner 19)-- (p.corner 20)-- cycle;

     \node[scale=0.8,xshift=18,yshift=30] at (p.corner 8) {$\mathcal{S}_L^{(1)}$};
     \node[scale=0.8,xshift=37,yshift=13] at (p.corner 10) {$\mathcal{S}_R^{(1)}$};

    \fill[Maroon] (p.corner 8) circle [radius=0.15pt];
     \fill[Maroon] (p.corner 10) circle [radius=0.15pt];
    
    \end{tikzpicture}     
\end{gathered} \times \begin{gathered}
    \begin{tikzpicture}[line width=1,scale=10]
    
    \node[regular polygon, regular polygon sides=22,  minimum size=2cm] (p) at (0,0) {};
     \draw (0,0) circle [radius=0.101cm];

    \node[scale=0.8,xshift=-5,yshift=-5] at (p.corner 8) {$i_1$};
    \node[scale=0.8,xshift=-5,yshift=-5] at (p.corner 10) {$j_1$};    
    \node[scale=0.8,xshift=7,yshift=7,Maroon] at (p.corner 22) {$k$};
    \node[scale=0.8,xshift=7,yshift=7,Maroon] at (p.corner 20) {$m$};
    
    \foreach \i in {20}
    {\draw[fill] (p.corner \i) circle [radius=0.1pt];}
    \foreach \i in {8}
    {\draw[fill] (p.corner \i) circle [radius=0.1pt];}
    \draw[fill] (p.corner 10) circle [radius=0.1pt];
     \draw[fill] (p.corner 22) circle [radius=0.1pt];

     \fill[Blue,opacity=0.2]  (p.corner 8) -- (p.corner 7) -- (p.corner 6) -- (p.corner 5) -- (p.corner 4) -- (p.corner 3) -- (p.corner 2) -- (p.corner 1) -- (p.corner 22)-- cycle;

     \fill[Blue,opacity=0.2]  (p.corner 10) -- (p.corner 11) -- (p.corner 12) -- (p.corner 13) -- (p.corner 14) -- (p.corner 15) -- (p.corner 16) -- (p.corner 17) -- (p.corner 18)-- (p.corner 19)-- (p.corner 20)-- cycle;

     \node[scale=0.8,xshift=18,yshift=30] at (p.corner 8) {$\mathcal{S}_L^{(2)}$};
     \node[scale=0.8,xshift=37,yshift=13] at (p.corner 10) {$\mathcal{S}_R^{(2)}$};

    \fill[Maroon] (p.corner 22) circle [radius=0.15pt];
     \fill[Maroon] (p.corner 20) circle [radius=0.15pt];
    
    \end{tikzpicture}     
\end{gathered} ,
\label{eq:Cut1_km}
\end{equation}
where in both cases we have precisely a kinematic shift of the same type of that of the original surface, $i.e.$:
\begin{equation*}
     \int_{\mathcal{S}^{(1)}} \Omega^{(1)} \prod_{ \mathcal{S}^{(1)}} u_{i,j}^{X_{i,j}}\frac{1}{\prod_{\mathcal{S}_L^{(1)}} u_{i,j} \prod_{\mathcal{S}_R^{(1)}} u_{i,j} }  \times \int_{\mathcal{S}^{(2)}} \Omega^{(2)} \prod_{ \mathcal{S}^{(2)}} u_{i,j}^{X_{i,j}}\frac{1}{\prod_{\mathcal{S}_L^{(2)}} u_{i,j}  \prod_{\mathcal{S}_R^{(2)}} u_{i,j} } .
\end{equation*}
Here, we have lost the contribution coming from all the $u_{i,j}$ associated to curves that cross $(k,m)$ because these go to one when $u_{k,m} \to 0$. 

Now, let's consider case 2, and say our triangulation contains curve $(i_2,j_1)$. Then, once more, since $X_{i_2,j_1}$ is not shifted, extracting the residue at $X_{i_2,j_1}=0$ simply gives us the product of the two lower surface integrals: 
\begin{equation}
    \begin{gathered}
    \begin{tikzpicture}[line width=1,scale=10]
    
    \node[regular polygon, regular polygon sides=22,  minimum size=2cm] (p) at (0,0) {};
     \draw (0,0) circle [radius=0.101cm];

    \node[scale=0.8,xshift=-5,yshift=-5] at (p.corner 8) {$i_1$};
    \node[scale=0.8,xshift=-5,yshift=-5,Maroon] at (p.corner 10) {$j_1$};    
    \node[scale=0.8,xshift=7,yshift=7,Maroon] at (p.corner 22) {$i_2$};
    \node[scale=0.8,xshift=7,yshift=7] at (p.corner 20) {$j_2$};
    
    \foreach \i in {20}
    {\draw[fill] (p.corner \i) circle [radius=0.1pt];}
    \foreach \i in {8}
    {\draw[fill] (p.corner \i) circle [radius=0.1pt];}
    \draw[fill] (p.corner 10) circle [radius=0.1pt];
     \draw[fill] (p.corner 22) circle [radius=0.1pt];

     \fill[Blue,opacity=0.2]  (p.corner 8) -- (p.corner 7) -- (p.corner 6) -- (p.corner 5) -- (p.corner 4) -- (p.corner 3) -- (p.corner 2) -- (p.corner 1) -- (p.corner 22)-- cycle;

     \fill[Blue,opacity=0.2]  (p.corner 10) -- (p.corner 11) -- (p.corner 12) -- (p.corner 13) -- (p.corner 14) -- (p.corner 15) -- (p.corner 16) -- (p.corner 17) -- (p.corner 18)-- (p.corner 19)-- (p.corner 20)-- cycle;


     \draw[Maroon] (p.corner 10) -- (p.corner 22);
    \fill[Maroon] (p.corner 10) circle [radius=0.15pt];
     \fill[Maroon] (p.corner 22) circle [radius=0.15pt];
    \end{tikzpicture}     
\end{gathered} 
\quad \xrightarrow{\mathop{\mathrm{Res}}_{X_{i_2,j_1}=0}} \quad \begin{gathered}
    \begin{tikzpicture}[line width=1,scale=10]
    \node[regular polygon, regular polygon sides=22,  minimum size=2cm] (p) at (0,0) {};
     \draw (0,0) circle [radius=0.101cm];

    \node[scale=0.8,xshift=-10,yshift=0,Maroon] at (p.corner 7) {$j_1$};
    \node[scale=0.8,xshift=-7,yshift=-5,Maroon] at (p.corner 10) {$i_2$};    
    \node[scale=0.8,xshift=-7,yshift=7] at (p.corner 4) {$i_1$};
    
    \foreach \i in {7}
    {\draw[fill] (p.corner \i) circle [radius=0.1pt];}
    \foreach \i in {4}
    {\draw[fill] (p.corner \i) circle [radius=0.1pt];}
    \draw[fill] (p.corner 10) circle [radius=0.1pt];


     \fill[Blue,opacity=0.2]   (p.corner 10) -- (p.corner 11) -- (p.corner 12) -- (p.corner 13) -- (p.corner 14) -- (p.corner 15) -- (p.corner 16) -- (p.corner 17) -- (p.corner 18)-- (p.corner 19)-- (p.corner 20) -- (p.corner 21) -- (p.corner 22) -- (p.corner 1) -- (p.corner 2) -- (p.corner 3) -- (p.corner 4) -- cycle;

     \node[scale=0.8,xshift=25,yshift=30] at (p.corner 10) {$\mathcal{S}^{(1)}_R$};
      \fill[Maroon] (p.corner 10) circle [radius=0.15pt];
       \fill[Maroon] (p.corner 7) circle [radius=0.15pt];

    \end{tikzpicture}     
\end{gathered} \, \,   \times 
 \begin{gathered}
    \begin{tikzpicture}[line width=1,scale=10]
    \node[regular polygon, regular polygon sides=22,  minimum size=2cm] (p) at (0,0) {};
     \draw (0,0) circle [radius=0.101cm];

    \node[scale=0.8,xshift=-10,yshift=0,Maroon] at (p.corner 7) {$i_2$};
    \node[scale=0.8,xshift=-7,yshift=-5,Maroon] at (p.corner 10) {$j_1$};    
    \node[scale=0.8,xshift=-7,yshift=7] at (p.corner 4) {$j_2$};
    
    \foreach \i in {7}
    {\draw[fill] (p.corner \i) circle [radius=0.1pt];}
    \foreach \i in {4}
    {\draw[fill] (p.corner \i) circle [radius=0.1pt];}
    \draw[fill] (p.corner 10) circle [radius=0.1pt];


     \fill[Blue,opacity=0.2]   (p.corner 10) -- (p.corner 11) -- (p.corner 12) -- (p.corner 13) -- (p.corner 14) -- (p.corner 15) -- (p.corner 16) -- (p.corner 17) -- (p.corner 18)-- (p.corner 19)-- (p.corner 20) -- (p.corner 21) -- (p.corner 22) -- (p.corner 1) -- (p.corner 2) -- (p.corner 3) -- (p.corner 4) -- cycle;

     \node[scale=0.8,xshift=25,yshift=30] at (p.corner 10) {$\mathcal{S}^{(2)}_R$};
      \fill[Maroon] (p.corner 10) circle [radius=0.15pt];
       \fill[Maroon] (p.corner 7) circle [radius=0.15pt];

    \end{tikzpicture}     
\end{gathered} ,
\end{equation}
where now we see the appearance of two integrals of the extremal type given in \cref{eq:extremalcase}. 

Let us analyze what happens when we take a residue in this extremal case, where there is only a factor of products of $u$'s associated with a single subsurface. Take the integral given in \cref{eq:extremalcase} as an example. Whatever the triangulation we are considering for $\mathcal{S}_n$ is, it either contains a chord \textit{crossing} $(k_1,k_3)$, or it \textit{contains} chord $(k_1,k_3)$. In the first case, just like previously, it is trivial to check that under the residue of a curve that crosses $(k_1,k_3)$ (say $(k_2,m)$ for some $m\in \{k_3+1,k_3+2,\cdots,k_1-1\}$) the integral factorizes into the product of two lower integrals of the same type as the original one, $i.e.$ both associated to a similar extremal case \eqref{eq:extremalcase}. 

If instead the triangulation contains the chord $(k_1,k_3)$, it is not so clear whether the residue on $X_{k_1,k_3}=0$ will still land us on an integral of the form of \cref{eq:case1,eq:extremalcase}, since $X_{k_1,k_3}$ gets shifted. Let's now compute this residue explicitly. To simplify the notation, we will take (without loss of generality) $k_1=1,k_2=2,k_3=3$. We start by defining the object that we are computing via the shift of the standard ``stringy'' Tr($\phi^3$) integral as follows
\begin{equation}
 \mathcal{A}_{\mathcal{S}_n}(\hat{X}_{1,3}, \hat{X}_{i,j}), \text{ where } \hat{X}_{1,3} = X_{1,3} -1, \text{ and } \hat{X}_{i,j} = \begin{cases}
      X_{i,j} -1, \text{ if }i,j \neq 2,\\
      X_{i,j}, \text{ otherwise.}
 \end{cases} 
\end{equation}
Now, due to the kinematic shift, taking the residue at $X_{1,3} =0$ corresponds to extracting the first residue of the unshifted amplitude at $\hat{X}_{1,3}=-1$. To compute this residue, we can use the formula given in Ref.~\cite{StringAmps}: 
\begin{equation}    
\mathop{\mathrm{Res}}_{\hat{X}_{1,3}=-1}  \mathcal{A}_{\mathcal{S}_n}(\hat{X}_{1,3}, \hat{X}_{i,j}) = \sum_{j=3}^{n-1} (-\hat{c}_{1,j})\times \mathcal{A}_{\mathcal{S}_R}\left[\hat{X}_{1,m} \to \hat{X}_{1,m} + \Theta(j-m)\right],
\end{equation}
where $\hat{c}_{1,j} = \hat{X}_{1,j} +\hat{X}_{2,j+1} - \hat{X}_{2,j} - \hat{X}_{1,j+1}$, and $\Theta(j-m)=1$, if $j \geq m$, and $0$ otherwise. But crucially, we only want to keep the piece that has units of $X^0$ from this residue, which comes from the shifts in $\hat{X}$. In particular, the shifts in $\hat{X}$, imply that some of the $\hat{c}_{1,j}$ are also shifted by negative integers. It is easy to check that the \textit{only} $\hat{c}_{1,j}$ that gets shifted is that of $j=n-1$, for which we have $\hat{c}_{1,n-1} = c_{1,n-1} -1$. Therefore, the piece of order $X^0$ of this residue is simply 
\begin{equation}
    \mathop{\mathrm{Res}}_{\hat{X}_{1,3}=-1}  \mathcal{A}_{\mathcal{S}_n}(\hat{X}_{1,3}, \hat{X}_{i,j}) = \mathcal{A}_{\mathcal{S}_R}\left[\hat{X}_{1,m} \to \hat{X}_{1,m}+1\right] + \mathcal{O}(X).
\end{equation}
So, we obtain the lower-point amplitude associated to the subsurface bounded by $(1,3)$, but where now the kinematics $\hat{X}_{1,m} \to \hat{X}_{1,m} +1$ --- in terms of the original $X$'s, this means that $X_{1,m}$ is \textit{unshifted}! Therefore, we have that the residue in $X_{1,3}$ is again an integral of the form of \cref{eq:extremalcase}, where now $\mathcal{S}_R$ is bounded by curve $(3,n)$. 

So, putting everything together, by computing the residue in $X_{k_1,k_3}=0$ on \cref{eq:extremalcase}, we get
\begin{equation}
   \mathop{\mathrm{Res}}_{X_{k_1,k_3}=0} \left[  \begin{gathered}
    \begin{tikzpicture}[line width=1,scale=10]
    \node[regular polygon, regular polygon sides=22,  minimum size=2cm] (p) at (0,0) {};
     \draw (0,0) circle [radius=0.101cm];

    \node[scale=0.8,xshift=-10,yshift=0] at (p.corner 7) {$k_2$};
    \node[scale=0.8,xshift=-7,yshift=-5,Maroon] at (p.corner 10) {$k_1$};    
    \node[scale=0.8,xshift=-7,yshift=7,Maroon] at (p.corner 4) {$k_3$};

    \foreach \i in {7}
    {\draw[fill] (p.corner \i) circle [radius=0.1pt];}
    \foreach \i in {4}
    {\draw[fill] (p.corner \i) circle [radius=0.1pt];}
    \draw[fill] (p.corner 10) circle [radius=0.1pt];


     \fill[Blue,opacity=0.2]   (p.corner 10) -- (p.corner 11) -- (p.corner 12) -- (p.corner 13) -- (p.corner 14) -- (p.corner 15) -- (p.corner 16) -- (p.corner 17) -- (p.corner 18)-- (p.corner 19)-- (p.corner 20) -- (p.corner 21) -- (p.corner 22) -- (p.corner 1) -- (p.corner 2) -- (p.corner 3) -- (p.corner 4) -- cycle;

     \node[scale=0.8,xshift=25,yshift=30] at (p.corner 10) {$\mathcal{S}^{(1)}$};
      \draw[Maroon] (p.corner 4) -- (p.corner 10);
    \fill[Maroon] (p.corner 10) circle [radius=0.15pt];
     \fill[Maroon] (p.corner 4) circle [radius=0.15pt];

    \end{tikzpicture}     
\end{gathered} \, \, \right] = \,  \begin{gathered}
    \begin{tikzpicture}[line width=1,scale=10]
    \node[regular polygon, regular polygon sides=22,  minimum size=2cm] (p) at (0,0) {};
     \draw (0,0) circle [radius=0.101cm];

    \node[scale=0.8,xshift=-10,yshift=0] at (p.corner 7) {$k_1$};
    \node[scale=0.8,xshift=-7,yshift=-10] at (p.corner 10) {$k_1 -1$};    
    \node[scale=0.8,xshift=-7,yshift=7] at (p.corner 4) {$k_3$};

    \foreach \i in {7}
    {\draw[fill] (p.corner \i) circle [radius=0.1pt];}
    \foreach \i in {4}
    {\draw[fill] (p.corner \i) circle [radius=0.1pt];}
    \draw[fill] (p.corner 10) circle [radius=0.1pt];


     \fill[Blue,opacity=0.2]   (p.corner 10) -- (p.corner 11) -- (p.corner 12) -- (p.corner 13) -- (p.corner 14) -- (p.corner 15) -- (p.corner 16) -- (p.corner 17) -- (p.corner 18)-- (p.corner 19)-- (p.corner 20) -- (p.corner 21) -- (p.corner 22) -- (p.corner 1) -- (p.corner 2) -- (p.corner 3) -- (p.corner 4) -- cycle;

     \node[scale=0.8,xshift=25,yshift=30] at (p.corner 10) {$\mathcal{S}_R^{(1)}$};

    \end{tikzpicture}    
\end{gathered} \, \equiv  \int_{\mathcal{S}^{(1)}} \Omega^{(1)} \prod_{\mathcal{S}^{(1)}} u_{i,j}^{X_{i,j}} \frac{1}{\prod_{\mathcal{S}_R^{(1)}}u_{i,j}}.
\end{equation}
In all the residues, we obtain a recursive result that always lands us on integrals of the form of either \cref{eq:case1} or \cref{eq:extremalcase}. This means that we will eventually hit a situation where one of the lower-point problems is in the form of the previously-considered example at four-points. Since there we proved we get the $\text{Tr}(\phi^3$)amplitude at low energies, we have finished the proof.  

This proof is largely pictorial and conceptual, but it is amusing to note its extensive use of much of the basic technology of $u$-variables and surfaceology. The final object we land on --- an (integer) kinematically-shifted stringy integral which nonetheless also reduces to Tr$(\phi^3)$ at low energies --- is especially interesting. This suggests that the world of integer kinematically-shifted integrals is even richer than we have realized so far.

\section{Polarization Configuration for Consecutive Splits}
\label{sec:polConfigFinal}

Having understood that, after applying $(n-2)$ $\W_e$ on the $n$-point gluon amplitude, we get the $n$-point Tr$(\phi^3)$ amplitude (times $X_{e_1,e_2}$), we now turn to discussing the polarization configuration corresponding to this limit. 

To do this we use the simple split kinematic configurations that define the action of $\W_e$ at each step, starting from the gluon amplitude all the way down to Tr$(\phi^3)$. Note that, in this way of describing the kinematical locus, the order in which we apply the different $\W_e$ will now be important. This is simply because, after applying a given $\W_e$, index $e$ disappears, and so the map for the next operator will be with respect to the kinematics that do not depend on $e$. However, this ordering is just a practicality that makes it especially straightforward to derive a particular polarization configuration for the action of these $\W_e$'s. After obtaining one such configuration, gauge invariance of the gluon amplitude tells us that any gauge-equivalent configuration will similarly describe the action of the $\W_e$'s. 

Let's consider a set of $(n-2)$ operators that includes all the even indices in $I_{2n}=\{1,2,3,\cdots,2n-1,2n\}$ except for $e_1$ and $e_2$, and let's for the moment assume that $e_1$ and $e_2$ are not adjacent to each other (which would correspond to case \cref{eq:case1}). So we have two subsets of even indices adjacent to each other: 
\begin{equation}
    E_{L} = \{ e_1^L, e_2^L, \cdots, e_{n_L}^L\}, \quad  E_{R} = \{ e_1^R, e_2^R, \cdots, e_{n_R}^R\}, 
\end{equation}
with $E_L$ being the subset to the left of chord $X_{e_1,e_2}$, and $E_{R}$ that to the right of $X_{e_1,e_2}$, with $n_R+n_L = n-2$. One possible ordering in applying these operators is to start from the one closest to $e_1$ in $E_R$ and end with the one closest to $e_2$, and then proceed with the one closest to $e_2$ in $E_L$ and end with the one closest to $e_1$. In this way, we are going around the disk counter-clock wise --- once more, this is just one possibility. Doing it in this way, since we are going counter-clockwise, it is simpler to consider the kinematic configuration corresponding to $X_{e,j} \to X_{e-1,j}+1$\footnote{Strictly speaking, we don't need to make this choice for the first ones we apply in either set, as for these both gauge invariance statements are unaltered.}. This yields:
\begin{equation}
\begin{aligned}
    \underline{E_R}:& \quad X_{e_1^R,j}=1 +X_{e_1^R-1,j}, \text{ for }j\in I_{2n} ,\\
    & \quad X_{e_2^R,j}=1 +X_{e_2^R-1,j}, \text{ for }j\in I_{2n}\setminus\{ e_1^R\}, \\
    & \quad \quad \vdots \\
    & \quad X_{e_{n_R}^R,j}=1 +X_{e_{n_R}^R-1,j}, \text{ for }j\in I_{2n}\setminus\{ e_1^R, e_2^R, \cdots, e_{n_R-1}^R\}, \\ 
    &\\
    \underline{E_L}:& \quad X_{e_1^L,j}=1 +X_{e_1^L-1,j}, \text{ for }j\in I_{2n}\setminus E_R ,\\
    & \quad X_{e_2^L,j}=1 +X_{e_2^L-1,j}, \text{ for }j\in I_{2n}\setminus\{ E_R \cup e_1^L\}, \\
    & \quad \quad \vdots \\
    & \quad X_{e_{n_L}^L,j}=1 +X_{e_{n_L}^L-1,j}, \text{ for }j\in I_{2n}\setminus\{ E_R \cup e_1^L, e_2^L, \cdots, e_{n_L-1}^L\}. \\ 
\end{aligned}
\label{eq:finalCond}
\end{equation}
This configuration defines one representative in the gauge orbit that gives the action of $\W_e$ with $e\in E_R \cup E_L$. Due to gauge invariance, we can reach all the remaining configurations by performing the standard gauge transformations $X_{2i,j} \to X_{2i,j} + \alpha_{i}(X_{2i+1,j}-X_{2i-1,j})$, for any even index $2i$; as the gluon amplitude is invariant under these, they will all lead to the same final result. Similarly, if instead we put $e_1$ and $e_2$ next to each other, we would only have a single subset $E_R$ (where $n_R=n-2$), but the polarization configuration above would still hold. 

We can easily translate \cref{eq:finalCond} into collections of non-adjacent dot products of scalar momenta $p_i\cdot p_j$ being set to zero (just like we saw for the action of a single $\W_e$ in \cref{sec:PolConfig1}). Using this we can translate condition \eqref{eq:finalCond} into a statement directly on the dots products involving the polarizations of the $(n-2)$ gluons in $E_L$ and $E_R$. We will now do explicitly this in the four-point example, and use it to compare our operator with the one proposed in Ref.~\cite{Cheung:2017ems}. 

\subsection{Four-point example}

Let us now go back to the four-point example studied in \cref{sec:consectivesplits}, and read off the polarization configuration corresponding to the action of 2 $\W$'s. In particular, we'll consider $\W_4[\W_8[\mathcal{A}_4]]$, which corresponds to case described in \cref{eq:ActW4}, and $\W_6[\W_8[\mathcal{A}_4]]$, which corresponds to that of \cref{eq:ActW6}.

In both cases, we apply $\W_8$ first, and so we can pick either split factorization, as for the first one there are no subtleties. Without loss of generality, let's choose the one corresponding the map $X_{8,j} = 1 + X_{1,j}$, which is equivalent to the following locus in terms of the scalar dot products $p_i\cdot p_j$:
\begin{equation}
2p_8 \cdot p_1 = 1, \, 2p_8 \cdot p_6 =-1, \, 2p_8\cdot p_j =0, \text{ for }j\in\{2,3,4,5\},
\end{equation}
where of course we always have $p_8\cdot p_7=0$ since this is the same as asking for $q_4^2=0$. Picking $\epsilon_4^\mu = p_8^\mu$ (which is consistent with $\alpha_4=0$), but letting the remaining polarizations be generic, $i.e.$ $\epsilon_i^\mu = p_{2i}^\mu - \alpha_i(p_{2i} + p_{2i-1})^\mu $, we obtain 
\begin{equation}
\begin{aligned}
    &\epsilon_4 \cdot \epsilon_1 =- \alpha_1/2, \\
    &\epsilon_4 \cdot \epsilon_2 =0, \\
    &\epsilon_4 \cdot \epsilon_3 =-(1-\alpha_3)/2, \\
\end{aligned} \quad \quad 
\begin{aligned}
    &\epsilon_4 \cdot q_1 =1/2, \\
    &\epsilon_4 \cdot q_2 =0, \\
    &\epsilon_4 \cdot q_3 =-1/2. \\
\end{aligned}
\label{eq:ConstW8}
\end{equation}
This is precisely the four-point analog of the locus derived in \cref{eq:condPolE}, for the $2n$ case. 

Now let's consider the case in which we further act with $\W_4$. In this scenario, we are allowed to choose either split mapping, as both gauge invariance statements are unaffected for gluon $2$. Let's say we pick the map $X_{4,j}= 1+X_{5,j}$, which at the level of the scalar dot products corresponds to
\begin{equation}
\begin{aligned}
2p_4 \cdot p_5 = 1, &\quad  2p_4 \cdot p_2 =-1, \, 2p_4\cdot p_j =0, \text{ for }j\in\{1,6\}, \\
 &2p_4\cdot (p_7+p_8)=0 \Leftrightarrow 2p_4 \cdot p_7=0,
\end{aligned}
\end{equation}
where in the last constraint we have used the fact that $p_4\cdot p_8=0$ from the action of $\W_8$. Therefore, picking $\epsilon_2^\mu = p_4^\mu$, these turn into 
\begin{equation}
\begin{aligned}
    &\epsilon_2 \cdot \epsilon_1 =- (1-\alpha_1)/2, \\
    &\epsilon_2 \cdot \epsilon_3 =-\alpha_3/2, \\
    &(\epsilon_2 \cdot \epsilon_4 =0), \\
\end{aligned} \quad \quad 
\begin{aligned}
    &\epsilon_2 \cdot q_1 =-1/2, \\
    &\epsilon_2 \cdot q_3 =1/2, \\
    &\epsilon_2 \cdot q_4 =0, \\
\end{aligned}
\label{eq:ConstW4}
\end{equation}
where we put the constraint $\epsilon_2 \cdot \epsilon_4 =0$ in parenthesis since it was imposed by the action of $\W_8$, and by the time we act with $\W_4$ the answer no longer depends on $\epsilon_4^\mu$. Therefore, the kinematical locus describing $\W_4[\W_8[\mathcal{A}_4]]$ is given by \cref{eq:ConstW4}, as well as any gauge-equivalent configurations. In this locus, we therefore obtain
\begin{equation}
    \W_4[\W_8[\mathcal{A}_4]] = X_{2,6} \times \left( \frac{1}{X_{1,5}} + \frac{1}{X_{3,7}}\right) = X_{2,6} \times \mathcal{A}_4^{\text{Tr}(\phi^3)}(X_{1,5},X_{3,7}).
\end{equation}
If, instead of acting with $\W_4$, we had acted with $\W_6$, we would have had to be careful to use the correct split mapping, which in this case is $X_{6,j} = 1 - X_{5,j}$. Translating this into the scalar dot products yields
\begin{equation}
\begin{aligned}
2p_4 \cdot p_5 = 1, &\quad  2p_5 \cdot (p_7+p_8) =-1, \Leftrightarrow 2p_5 \cdot p_7 =-1, \\
 &2p_5\cdot p_j =0, \text{ for }j\in\{1,2,3\},
\end{aligned}
\end{equation}
where we have used the fact that $p_5\cdot p_8 = 0$, from $\W_8$. Note that if we had chosen the other split, which imposes similar conditions but now on the dot products involving $p_6$, we would have obtained $2 p_6 \cdot(p_7+p_8) =1 $. Since $2p_6\cdot p_8 =-1 $ from $\W_8$, this would give us $2p_6\cdot p_7 =2$, which precisely agrees with what we found in \cref{sec:consectivesplits}. In any case, let's stick to this simpler choice. We can easily translate it into conditions on $\epsilon_3^\mu$ by picking $\alpha_3=1$, for which $\epsilon_3^\mu= -p_5^\mu$. We thus find
\begin{equation}
\begin{aligned}
    &\epsilon_3 \cdot \epsilon_1 =0, \\
    &\epsilon_3 \cdot \epsilon_2 =-(1-\alpha_2)/2, \\
    &(\epsilon_3 \cdot \epsilon_4 =0), \\
\end{aligned} \quad \quad 
\begin{aligned}
    &\epsilon_3 \cdot q_1 =0, \\
    &\epsilon_3 \cdot q_2 =-1/2, \\
    &\epsilon_3 \cdot q_4 =1/2, \\
\end{aligned}
\label{eq:ConstW6}
\end{equation}
where, once again, we put the constraint $\epsilon_3 \cdot \epsilon_4 =0$ in parenthesis since it was already imposed by $\W_8$ provided we set $\alpha_3=1$ in \cref{eq:ConstW8}. 
To summarize, for $\W_6[\W_8[\mathcal{A}_4]]$, we derive the following locus
\begin{equation}
\begin{aligned}
    &\epsilon_4 \cdot \epsilon_1 =- \alpha_1/2, \\
    &\epsilon_4 \cdot \epsilon_2 =0, \\
    &\epsilon_4 \cdot \epsilon_3 =0, \\
\end{aligned} \quad \quad 
\begin{aligned}
    &\epsilon_4 \cdot q_1 =1/2, \\
    &\epsilon_4 \cdot q_2 =0, \\
    &\epsilon_4 \cdot q_3 =-1/2, \\
\end{aligned} \quad \quad 
\begin{aligned}
    &\epsilon_3 \cdot \epsilon_1 =0, \\
    &\epsilon_3 \cdot \epsilon_2 =-(1-\alpha_2)/2, \\
\end{aligned} \quad \quad 
\begin{aligned}
    &\epsilon_3 \cdot q_1 =0, \\
    &\epsilon_3 \cdot q_2 =-1/2, \\
    &\epsilon_3 \cdot q_4 =1/2, \\
\end{aligned}
\label{eq:ConstW8}
\end{equation}
which, at the level of the amplitude, gives us
\begin{equation}
    \W_6[\W_8[\mathcal{A}_4]] = X_{2,4} \times \left( \frac{1}{X_{1,5}} + \frac{1}{X_{3,7}}\right) = X_{2,4} \times \mathcal{A}_4^{\text{Tr}(\phi^3)}(X_{1,5},X_{3,7}).
\label{eq:W6W8}    
\end{equation}

\paragraph{Comparison with transmutation operators from Ref.~\cite{Cheung:2017ems}} At this stage, it is natural to go back to the operator introduced in Ref.~\cite{Cheung:2017ems} (also used in Ref.~\cite{Wei:2024ynm} to derive higher order soft statements), and ask whether it matches any sequence of actions with the $\W_e$ operators. According to Ref.~\cite{Cheung:2017ems}, one should get the Tr$(\phi^3)$ amplitude from the gluon one by acting with
\begin{equation}
    \mathcal{T}[a_1,a_2,\cdots,a_n] = \mathcal{T}_{a_1,a_n} \prod_{i=2}^{n-1} \mathcal{T}_{a_{i-1}, a_i,a_n},
    \label{eq:TransmOpP}
\end{equation}
where $\{a_1,a_2,\cdots,a_n\}$ is the ordered set of labels, and $\mathcal{T}_{i,j,k} = \partial_{q_i \cdot \epsilon_j} - \partial_{q_k \cdot \epsilon_j}$ and $\mathcal{T}_{l,m} = \partial_{\epsilon_l \cdot \epsilon_m} $. From its definition in \cref{eq:TransmOpP}, it is clear that this operator acts on $(n-2)$ gluons with $\mathcal{T}_{a_{i-1}, a_i,a_n}$, and then kills off the dependence on the remaining two polarizations with $\mathcal{T}_{a_1,a_n}$. This seems very similar to our case, with the exception that in the operator $\mathcal{T}[a_1,a_2,\cdots,a_n]$, the last two polarizations are always associated with two adjacent gluons, $a_1$ and $a_n$, while in our case, depending on the set of $\W_e$'s we choose to apply, the final two gluons can also be separated from each other. 

Nonetheless, to compare explicitly with Ref.~\cite{Cheung:2017ems}, let us consider a case in which the last two gluons (say gluons $1$ and $2$) are adjacent to each other at four-points. From our perspective, this corresponds to applying $\W_6$ and $\W_8$, just like we described previously. The closest limit to this in Ref.~\cite{Cheung:2017ems} would be to consider the ordered set in \eqref{eq:TransmOpP} to be $\{2,3,4,1\}$, in which case we have
\begin{equation}
     \mathcal{T}[\{2,3,4,1\}] = \mathcal{T}_{2,1} \prod_{i=\{3,4\}} \mathcal{T}_{i-1,i,1} = \partial_{\epsilon_1\cdot \epsilon_2} \left(\partial_{q_2\cdot \epsilon_3} - \partial_{q_1\cdot \epsilon_3}\right)\left(\partial_{q_3\cdot \epsilon_4} - \partial_{q_1\cdot \epsilon_4}\right). 
\end{equation}
So, starting with $\mathcal{T}_{3,4,1} =\left(\partial_{q_3\cdot \epsilon_4} - \partial_{q_1\cdot \epsilon_4}\right)$, as explained in \cref{sec:PolConfig1} this operator is picking a polarization configuration for $\epsilon_4^\mu$ which agrees with what we get from acting with $\W_8$, provided we make the gauge choice $\alpha_1=0$ and $\alpha_3=1$; this is so that, in \cref{eq:ConstW8}, we ensure that all the dot products $\epsilon_4\cdot \epsilon_i =0$. Note that it is convenient to have $\alpha_3=1$, since this is also the gauge we pick to read off the kinematic locus giving $\W_6$ in \cref{eq:ConstW6}.

Now, after acting with $\W_6$ and $\W_8$, from \cref{eq:W6W8} we get a prefactor $X_{2,4}$ times the four-point Tr$(\phi^3)$ amplitude, and quite remarkably, for the gauge choice $\alpha_1=0,\alpha_3=1$, we find that $X_{2,4} = \epsilon_1 \cdot \epsilon_2$! This means that we can directly compare the action of $\mathcal{T}_{2,3,1}\mathcal{T}_{3,4,1}$ with that of $\W_6\W_8$. For this gauge choice, $\W_8$ is equivalent to $\mathcal{T}_{3,4,1}$, so let's now look at $\W_6$ and $\mathcal{T}_{2,3,1}$. From the definition of $\mathcal{T}_{2,3,1}$, we see that it is equivalent to the polarization configuration
\begin{equation}
\epsilon_3\cdot \epsilon_i =0, \quad \epsilon_3 \cdot q_1 =-1, \quad \epsilon_3 \cdot q_2 =1, \quad \epsilon_3 \cdot q_4 =0.
\label{eq:T6}
\end{equation}

Comparing with \cref{eq:ConstW6}, by picking $\alpha_2=1$ we can have $\epsilon_3\cdot \epsilon_i=0$, but the conditions on $\epsilon_3\cdot q_i$ from $\W_6$ don't match those of $\mathcal{T}_{2,3,1}$ --- they are also not related by any gauge transformation. Therefore, even though the two operators are closely related, already at four-points they are not the same. 

Another way one can conclude this is by looking at the action of the two operators directly at the level of the string amplitudes. In Ref.~\cite{Cheung:2017ems}, the authors claim that by starting with the the $n$-point tree-level gluon bosonic string amplitude~\cite{Green:1987sp}
\begin{equation}	\mathcal{A}^{\text{YM}}_n(1,2,\dots,n) = \int_{\mathcal{D}} \frac{\diff^{n} z_i}{\text{SL(2,}\mathbb{R})} \,  \prod_{i<j} z_{i,j}^{2 q_i \cdot q_j}  \, \left.\exp{ \sum_{i\neq j} 2 \frac{\epsilon_i \cdot \epsilon_j }{z_{i,j}^2} - \frac{\epsilon_i \cdot q_j}{z_{i,j}}}\right\vert_{\text{linear in  all }\epsilon_i}
\label{eq:BosStringTree}
\end{equation}
with $\mathcal{D}=z_1<z_2<\cdots<z_n$ and $z_{i,j} = z_j -z_i$, and acting on it with the operator in \cref{eq:TransmOpP}, one obtains directly the stringy Tr$(\phi^3)$ shown in \cref{eq:SurfaceIntTr}. We can write the latter integral in terms of the $z_{i,j}$'s, reproducing the $Z$-theory integral \cite{Huang:2016tag}
\begin{equation}	
    \mathcal{A}^{\text{Tr}(\phi^3)}_n(1,2,\dots,n) = \int_{\mathcal{D}} \frac{\diff^{n} z_i}{\text{SL(2,}\mathbb{R})} \, \frac{1}{z_{1,2}z_{2,3} \cdots z_{n,1}}  \prod_{i<j} z_{i,j}^{2 q_i \cdot q_j} . 
\label{eq:Ztheory}
\end{equation}

Indeed, already at four-points, by going on the kinematic locus specified by $\mathcal{T}_{2,3,1}\mathcal{T}_{3,4,1}$ we find
\begin{equation}
\begin{aligned}
    \mathcal{A}^{\text{YM}}_n(1,2,\dots,n) &= \int_{\mathcal{D}} \frac{\diff^{n} z_i}{\text{SL(2,}\mathbb{R})} \,  \prod_{i>j} z_{i,j}^{2\alpha^{\prime} p_i \cdot p_j}  \, \frac{2\epsilon_1 \cdot \epsilon_2}{z_{1,2}^2} \times \left(\frac{1}{z_{3,1}} - \frac{1}{z_{3,2}} \right)\times \left(\frac{1}{z_{4,1}}-\frac{1}{z_{4,3}} \right)\\
    &= \int_{\mathcal{D}} \frac{\diff^{n} z_i}{\text{SL(2,}\mathbb{R})} \,  \prod_{i>j} z_{i,j}^{2\alpha^{\prime} p_i \cdot p_j}  \, \frac{2\epsilon_1 \cdot \epsilon_2}{z_{1,2}z_{2,3}z_{3,4}z_{4,1}} ,
\end{aligned}
\label{eq:stringTau}
\end{equation}
which is precisely stringy Tr$(\phi^3)$. If instead we apply the polarization configuration corresponding to $\W_6\W_8$ (with $\alpha_1=0,\alpha_3=1,\alpha_2=1$) we arrive at
\begin{equation}
\begin{aligned}
    \mathcal{A}^{\text{YM}}_n(1,2,\dots,n) &=\frac{1}{4} \int_{\mathcal{D}} \frac{\diff^{n} z_i}{\text{SL(2,}\mathbb{R})} \,  \prod_{i>j} z_{i,j}^{2\alpha^{\prime} p_i \cdot p_j}  \, \frac{2\epsilon_1 \cdot \epsilon_2}{z_{1,2}^2} \times \left(\frac{1}{z_{3,2}} - \frac{1}{z_{3,4}} \right)\times \left(\frac{1}{z_{4,3}}-\frac{1}{z_{4,1}} \right)\\
    &= \frac{1}{4}\int_{\mathcal{D}} \frac{\diff^{n} z_i}{\text{SL(2,}\mathbb{R})} \,  \prod_{i>j} z_{i,j}^{2\alpha^{\prime} p_i \cdot p_j}  \, \frac{2\epsilon_1 \cdot \epsilon_2}{z_{1,2}z_{2,3}z_{3,4}z_{4,1}} \times \frac{z_{2,4} z_{1,3}}{z_{4,3}z_{1,2}} ,
\end{aligned}
\end{equation}
which is manifestly different than \cref{eq:stringTau}. In addition, 
recalling that we can write the $u_{i,j}$ in terms of the $z_{i,j}$ as $u_{i,j} = z_{i,j-1}z_{i-1,j}/(z_{i,j}z_{i-1,j-1})$, we can recognize the extra factor to be $1/u_{1,3}$, which precisely agrees with what we found at the level of the integral in Eq.~\eqref{eq:stringTau}!

\section{Transmutation at One-Loop}
\label{sec:TrasmLoop}

Having understood the action of $\W_e$ on tree-level gluon amplitudes, it is natural to ask how this extends to one-loop. Recall that, in Eq.~\eqref{eq:defOp}, we took $\W_e$ to be a sum of derivatives of $\mathcal{A}_n$ with respect to all $X$ variables depending on the index $e$. In this way, the resulting object was independent of $e$. Thus, the simplest way to generalize to loop-level is to define $\W^{(l)}_e$ in exactly the same way:
\begin{equation}
    \W^{(l)}_e = \sum_{j\neq e} \frac{\partial}{\partial X_{e,j}} +\sum_{j\neq e} \frac{\partial}{\partial X_{j,e}} +  \frac{\partial}{\partial X_{e,p}},
\label{eq:loop-W}
\end{equation}
where we are dealing with surface kinematics, and so $X_{e,j} \neq X_{j,e}$. 

Using the explicit expressions for the residues $R_{2n-3,1}$ and $R_{2n-1,3}$, one can show that we precisely recover the lower-point integrands by acting with this operator:
\begin{equation}
    \W_e^{(l)}[\mathcal{I}_n] = \frac{\mathcal{I}_{n-1}(1, 2, \ldots, 2n-2)}{X_{2n-3,1}} + \frac{\mathcal{I}_{n-1}(2n-1,2,\ldots, 2n-2)}{X_{2n-1,3}} + \cdots,
\end{equation}
where the ellipses contain the remaining terms with poles in neither $X_{2n-3,1}$ nor $X_{2n-1,3}$. This formula is strikingly similar to that found in \cref{eq:W-on-amp} for the tree-level case. As a result, we may hope that, using our simple generalization $\W_e^{(l)}$, we can transmutate gluons into scalars not only at tree-level but also at loop-integrand-level.

Note that, in the first appearance of transmutation operators, the authors introduce $\mathcal{T}_{i,j,k}$ (which is equivalent to $\W_e$) as an operator that \textit{preserves} tree-level gauge invariance~\cite{Cheung:2017ems}. Since gauge invariance only holds post-loop integration, it is hard to see how, in the standard language, such an operator might also exist for the one-loop integrand. However, once again working in surface kinematics, we find that the simplest, most natural generalization of $\W_e$ given in \cref{eq:loop-W} precisely gives us what we want!

Let us start by consider the simplest case, the one-point integrand given in \cref{eq:one-pt-exam}. Applying to it the operator $\W_2^{(l)}$, one can check simply that
\begin{equation}
    \W_2^{(l)}[\mathcal{I}_1] = \frac{\partial \mathcal{I}_1}{\partial X_{1,2}} + \frac{\partial \mathcal{I}_1}{\partial X_{2,1}} + \frac{\partial \mathcal{I}_1}{\partial X_{2,p}} = \frac{1 - \Delta}{X_{1,p}}.
\end{equation}
As anticipated, this is exactly the one-point integrand in Tr$(\phi^3)$ theory multiplied by $1 - \Delta = d$. Note that, already in this simplest case, it is crucial to keep the boundary curves $X_{1,2}$ and $X_{2,1}$, as otherwise, we would not get the Tr$(\phi^3)$ integrand. So, once more we see clearly that surface kinematics plays a fundamental role in allowing us to extend tree-level statements to loop integrands.  

We can go one step further, and see what happens when we apply $\W_2^{(l)}$ and $\W_4^{(l)}$ sequentially to the two-point integrand shown in \cref{eq:two-pt-exam}. This gives us
\begin{equation}
    \W_2^{(l)}[\W_4^{(l)}[\mathcal{I}_2]] = \W_4^{(l)}[\W_2^{(l)}[\mathcal{I}_2]] = \frac{d}{X_{1,p} X_{3,p}} + \frac{d}{X_{1,p} X_{1,1}} + \frac{d}{X_{3,p} X_{3,3}},
\end{equation}
where, again, we find the integrand for Tr$(\phi^3)$ theory times $d$, the dimension of spacetime. We have checked that this behavior continues up to five-points, and therefore we make the conjecture that applying all $n$ $\W_e^{(l)}$ operators to $\mathcal{I}_n$ always returns the $n$-point integrand in Tr$(\phi^3)$ theory multiplied by $d$. We suspect that, with an argument similar to the one presented at tree-level ($i.e.$, using splits of the one-loop surface integral), one should be able to prove the result above, but we leave it to future work. For other instances of transmutation at one-loop, see Refs.~\cite{Tao:2022nqc,Chen:2023bji,Cao:2024olg}.

Finally, let us end by making some remarks on the generalization of transmutation operators to higher loops. As it turns out, already at two-loops it is clear that the naive extension doesn't hold purely by a units argument. At $l$-loops, the integrand has units of $X^{2-2l}$, and so at two-loops it has units of $X^{-2}$. Already in the simplest case of $n = 1, l = 2$, the part of the integrand that has all the Tr$(\phi^3)$ singularities has five $X$'s in the denominator and three in the numerator. So, to hope to turn the gluon integrand into a Tr$(\phi^3)$ one, we would need to act with at least three differential operators. However, since we are at one-point, we only have a single even index $e = 2$. For this reason, it is not obvious how or if any sort of transmutation holds for $l \geq 2$. On the other hand, the form of the $\W_e$ operators were suggested by gauge invariance, and the statement of gauge invariance at higher loops has not yet been fully fleshed out. So, it is conceivable that understanding this could offer a way out of the units dilemma.

%% file: Sections_v2/Outlook_v2.tex
\section{Outlook}
\label{sec:outlook}

The scalar-scaffolded formalism offers us a new window into the study of gluon amplitudes. In this paper, we leveraged the advantages of this representation --- the trivialization of on-shell kinematics using scalar $X$ variables, a locked form for the amplitude, and the use of surface kinematics to give a well-defined surface integrand at loop-level --- to explore two aspects of gluon amplitudes, tied together by their close relationship to the scaffolding representation of surface gauge invariance.

In \cref{part:1}, we used the scalar variables to give a precise kinematical definition for the soft limit, as a well-defined Laurent expansion in a set of soft factors. At tree-level, we derived universal leading and subleading terms in the soft expansion, which followed directly from the combination of factorization on the collinear poles adjacent to the soft gluon and the requirement of gauge invariance. A general feature of the scaffolding formalism is that the terms in the amplitude Laurent expansion are well-defined but not individually gauge invariant, so gauge invariance can link different terms in the expansion. This led us to propose a simple ``gauge-invariantifying'' operation that produced gauge-invariant soft terms at both leading and subleading levels.  At one-loop, working in the expanded surface-kinematics basis gave us a clear way to generalize our tree-level soft limit and apply it to the YM surface integrand. The use of surface kinematics makes it possible to define a canonical Yang-Mills integrand, enjoying both gauge invariance as well as correct cuts. These two properties led to well-defined soft factors at the level of the surface loop integrand. We obtained a universal leading term in the one-loop soft expansion, which has the expected Weinberg factor plus specific corrections proportional to the tadpole variable $X_{s,s}$; of course, these corrections vanish upon loop integration. It is striking to find a well-defined soft term at integrand level, whose existence is made possible only by the use of surface kinematics. 

In \cref{part:2}, we switched gears and investigated a differential operator $\W_e$ acting on external gluon labels, suggested naturally from the requirements of gauge invariance. After taking some hints about its structure from the soft expansion in \cref{part:1}, we demonstrated that the action of this operator is equivalent to imposing some of the simplest ``split'' kinematic configurations on the amplitude. Working at the level of the full surface integral, we used these split kinematics to find the surface integral that results from applying any number of $\W_e$ operators. In particular, after applying $(n - 2)$ of them in any order, we obtained a class of ``shifted'' integrals, which remarkably reduce to Tr$(\phi^3)$ amplitudes at low energies!  For a single operation, our $\W_e$ turns out to correspond precisely to the insertion operator $\mathcal{T}_{i,j,k}$ discussed in Ref.~\cite{Cheung:2017ems}, whereas, for multiple applications, differences appear. We then showed that, using surface kinematics, the $\W_e$ operator can be generalized to the one-loop integrand. We conjectured (and verified up to $n=5$ points) that applying the $\W_e$ to all $n$ gluons of the YM integrand returns the Tr$(\phi^3)$ integrand. We suspect that this statement can be proven by an extension of our tree-level split kinematics argument to loop-level. 

We close by briefly discussing a set of possible next steps suggested by our investigations. 

When we defined the soft limit in \cref{part:1}, we chose a certain ``minimal'' limit, where we changed only the momenta of the soft gluon and its two adjacent gluons. We also ensured that the on-shell conditions of all gluons remained true during the limit, while momentum conservation was handled \textit{ipso facto} by the dual variables of the momentum polygon. In this limit, we derived a universal subleading soft theorem. It would be interesting to investigate the relationship between this subleading soft theorem and the ones that have been intensively studied in the literature on celestial holography over the past decade . In particular, while our ``minimal'' soft limit can be realized for general polarizations in high enough $(d \geq 6)$ spacetime dimensions, it cannot be realized for all helicities in $d=4$ dimensions. It is obvious that a ``less minimal'' soft limit, where more $X$'s are varied, can be defined in any number of dimensions, and it would be interesting to compute subleading theorems for these extended soft limits.  

As we discussed in \cref{part:2}, the transmutation operator $\W_e$ applied to the full surface integral has a beautiful interpretation from the perspective of split kinematics: it returns the surface integral for the surface with marked point $e$ removed \eqref{eq:ActW2n}. One might therefore hope that $\W_e[ \mathcal{A}_n ]$ has some interpretation as an ``amplitude'' itself, where one (or a few) gluons are turned into scalars. Indeed, at field-theory level, we see from \cref{eq:W-on-amp} that $\W_e[ \mathcal{A}_n ]$ factorizes on the poles $X_{1,2n-3}$ and $X_{3,2n-1}$ as on a scalar exchange, giving some possible hints at its structure. Could there be some diagrammatic way to understand the field theory limit of the surface integral \eqref{eq:ActW2n}? Even if not an amplitude, is there a set of rules that determines its structure in $X$ language? 

It would be even more fascinating to find an algorithmic way to build YM theory from Tr$(\phi^3)$ theory directly at field-theory level, where one would hope to obtain a prescription to ``undo'' each $\W_e$ operation. Since transmutation also exists at one-loop, one could hope to do this for both the tree amplitude and the one-loop integrand. Of course, to find such a procedure for the integrand, we would likely first need to understand the proof of transmutation at one-loop; in particular, is there a surface integral similar to \cref{eq:ActW2n} for the punctured disk after applying $\W_e^{(l)}$? And finally, as noted in \cref{sec:TrasmLoop}, it it natural to seek out a transmutation operator valid for more than one-loop --- a challenge we can only sensibly undertake after finding a systematic understanding of surface gauge invariance at all loop orders.

%% file: Sections_v2/AppHigherOrderTree_v2.tex
\section{Higher Orders in the Tree-Level Soft Expansion}
\label{app:HigherOrdSoft}

In this Appendix, we will finish the discussion started in \cref{sec:1.1}; namely, we will derive the subleading term in the soft expansion and see what constraints we can place at sub-subleading and higher.

To do this, we will start with the residue in $X_{1,2n-3}$, which, as discussed in the main text, takes the form
\begin{equation}
    R_{1,2n-3}= \sum_{j \in L, \ J\in \{2n-2,2n-1,2n\}} (X_{j,J}-X_{1,j} -X_{2n-3,J}) \frac{\partial M_L}{\partial X_{j,x_L}}  \times \frac{\partial M_R}{\partial X_{J,x_R}},
    \label{eq:factX12}
\end{equation}
where $L = \{ 2, 3, \ldots, 2n -4 \}$, $M_L \equiv M_L(1,2,\ldots,2n-4,2n-3,x_L)$, which is an $(n-1)$-point amplitude, and $M_R \equiv M_R(1,x_R,2n-3,2n-2,2n-1,2n)$, which is a $3$-point amplitude. Using the explicit form of $3$-point amplitude for $M_R$, as given in Eq.~\eqref{eq:3ptGluons}, we can write out each term in the $J$-sum above as follows:
\begin{equation}
\begin{aligned}
    &\underline{J=2n-2}: \quad X_{2n-3,2n}\times \sum_j(X_{j,2n-2} - X_{1,j})\frac{\partial M_L}{\partial X_{j,2n-2}},\\
     &\underline{J=2n-1}: \quad (X_{2n-2,2n}-X_{2n-3,2n} - \underbracket[0.4pt]{X_{1,2n-2}}_{\delta_{2n-2}^{(1)}}) \times \sum_j\underbracket[0.4pt]{(X_{j,2n-1} - X_{1,j})}_{\delta_j^{(2n-1)} - \delta_j^{(1)}}\frac{\partial M_L}{\partial X_{j,2n-2}},\\
     &\underline{J=2n}: \quad \quad \quad \underbracket[0.4pt]{X_{1,2n-2}}_{\delta_{2n-2}^{(1)}} \times \sum_j(X_{j,2n}- \underbracket[0.4pt]{X_{1,j}}_{\delta_j^{(1)} + X_{s,j}} -X_{2n-3,2n})\frac{\partial M_L}{\partial X_{j,2n-2}},\\    
\end{aligned}
\label{eq:J_X1}
\end{equation}
where we have relabeled $x_L = 2n-2$, and under brackets we show the soft map for the $X$'s on the prefactors. 

Let us look at each of these contributions one-by-one. Using gauge invariance of $M_L$ in gluon $n - 1$, the $J = 2n-2$ contribution becomes
\begin{equation}
    X_{2n-3,2n}\times \sum_{j\in L}(X_{j,2n-2} - X_{1,j})\frac{\partial M_L}{\partial X_{j,x_L}} = X_{2n-3,2n} \times M_L(1, 2, \ldots, 2n-2).
\end{equation}
Of course, this is exactly the term that gave us half of the leading Weinberg term in the main text. However, if we want to look at higher orders, we have to take into account that $M_L$ depends on the soft factors, via $X_{1,j} = X_{s,j} +\delta_j^{(1)} $. Taylor expanding $M_L$ yields:
\begin{equation}
    M_L(1,2,\ldots,2n-2) = \mathcal{A}_{n-1} + \sum_{i = 4}^{2n-4} \delta_i^{(1)} \frac{\partial \mathcal{A}_{n-1}}{\partial X_{s,i}} + \frac{1}{2} \sum_{i,k} \delta_i^{(1)} \delta_k^{(1)} \frac{\partial^2 \mathcal{A}_{n-1}}{\partial X_{s,i} \partial X_{s,k}} + \cdots,
\end{equation}
with $\mathcal{A}_{n-1} \equiv \mathcal{A}_{n-1}(s,2,3,\cdots,2n-2)$.
This then gives the all-order 
 soft expansion of the $J = 2n-2$ contribution. Quite nicely, the $J = 2n-1$ and $J = 2n$ terms are already organized as expansions in soft factors, so all we must do is plug in the expansion of $M_L$. In summary, using the residue formula together with the expansion of $M_L$, we can derive the all-order soft expansion of the pole in $X_{1,2n-3}$. In particular, at subleading order $\mathcal{O}(\delta)$, we find:
\begin{equation}
\begin{aligned}
    &\underline{R_{1,2n-3}^{(1)}}: X_{2n-3,2n}\sum_{i=4}^{2n-4} \delta_i^{(1)} \frac{\partial \mathcal{A}_{n-1}}{\partial X_{i,s}} + (X_{2n-2,2n} - X_{2n-3,2n}) \sum_{j=2}^{2n-4} (\delta_j^{(2n-1)} - \delta_j^{(1)} ) \frac{\partial \mathcal{A}_{n-1}}{\partial X_{j,2n-2}} \\
     & \quad \quad \quad \quad + \delta_{2n-2}^{(1)}  \sum_{j=2}^{2n-4}(X_{j,2n}- X_{s,j} -X_{2n-3,2n})\frac{\partial  \mathcal{A}_{n-1}}{\partial X_{j,2n-2}}.
\label{eq:Res1sub}
\end{aligned}
\end{equation}

We can repeat this analysis for the other residue $R_{3,2n-1}$, where now the two lower-point amplitudes that enter the factorization formula
\begin{equation}
     R_{3,2n-1}= \sum_{j \in R, \ J\in \{2,1,2n\}} (X_{j,J}-X_{2n-1,j} -X_{3,J}) \frac{\partial A_L}{\partial X_{J,x_L}}  \times \frac{\partial A_R}{\partial X_{j,x_R}}
\end{equation}
are $A_L\equiv A_L(1,2,3,x_L,2n-1,2n)$, and $A_R \equiv A_R(2n-1, x_R, 3,4,\ldots,2n-2)$, which are, respectively, a $3$- and an $(n-1)$-point amplitudes; and $R = \{ 4, 5, \ldots, 2n-2\}$. Using again the explicit form of the $3$-point amplitude given in Eq.~\eqref{eq:3ptGluons}, the different terms in the $J$-sum take values
\begin{equation}
\begin{aligned}
    &\underline{J=2}: \quad X_{3,2n}\times \sum_j(X_{2,j} - X_{j,2n-1})\frac{\partial A_R}{\partial X_{j,2}},\\
     &\underline{J=1}: \quad (X_{2,2n}-X_{3,2n} - \underbracket[0.4pt]{X_{2,2n-1}}_{\delta_{2}^{(2n-1)}}) \times \sum_j\underbracket[0.4pt]{(X_{1,j}-X_{j,2n-1} )}_{\delta_j^{(1)} - \delta_j^{(2n-1)}}\frac{\partial A_R}{\partial X_{j,2}},\\
     &\underline{J=2n}:  \quad \underbracket[0.4pt]{X_{2,2n-1}}_{\delta_{2}^{(2n-1)}} \times \sum_j(X_{j,2n}- \underbracket[0.4pt]{X_{j,2n-1}}_{X_{j,s}+\delta_j^{(2n-1)}} -X_{3,2n})\frac{\partial A_R}{\partial X_{j,2}},\\    
\label{eq:Js3}
\end{aligned}
\end{equation}
where now we have relabeled $x_R = 2$. As before, we can use the gauge invariance of $A_R$ to rewrite the $J = 2$ contribution as
\begin{equation}
    X_{3,2n}\times \sum_j(X_{2,j} - X_{j,2n-1})\frac{\partial A_R}{\partial X_{j,2}} = X_{3,2n} \times A_R(2n-1,2,\ldots, 2n-2),
\end{equation}
where now $A_R$ depends on the soft factors via $X_{j,2n-1} = X_{j,s} + \delta^{(2n-1)}_j$. Taylor expanding $A_R$ in the soft factors we obtain
\begin{equation}
    A_R(2n-1,2,\ldots,2n-2) = \mathcal{A}_{n-1} + \sum_{i=4}^{2n-4} \delta_i^{(2n-1)} \frac{\partial \mathcal{A}_{n-1}}{\partial X_{i,s}} + \cdots.
\end{equation}
Of course, as with $R_{1,2n-3}$, this nails all orders in the soft expansion of $R_{3,2n-1}$; in particular, the term at subleading order takes the form
\begin{equation}
\begin{aligned}
    &\underline{R_{3,2n-1}^{(1)}}: X_{3,2n} \sum_{i=4}^{2n-4} \delta_i^{(2n-1)} \frac{\partial \mathcal{A}_{n-1}}{\partial X_{i,s}} + (X_{2,2n} - X_{3,2n}) \sum_{j=4}^{2n-2} (\delta_j^{(1)} - \delta_j^{(2n-1)}) \frac{\partial \mathcal{A}_{n-1}}{\partial X_{j,2}} \\ 
    & \quad \quad \quad \quad + \delta_2^{(2n-1)} \sum_{j=4}^{2n-2} (X_{j,2n} - X_{j,s} - X_{3,2n}) \frac{\partial \mathcal{A}_{n-1}}{\partial X_{j,2}}.
\label{eq:Res3sub}
\end{aligned}
\end{equation}

To fully understand the subleading term in the soft expansion of the amplitude, we also need control over the soft expansion of $\mathcal{R}$ in Eq.~\eqref{eq:LaurentSeries}. As alluded to in the main text, we can use the constraint coming from gauge invariance in the $n^{\text{th}}$ gluon~\eqref{eq:constContact} to determine the leading-order piece of $\mathcal{R}$. 

Let us start by rewriting \cref{eq:constraintContact2}, but now making manifest the different terms on the $l.h.s.$:
\begin{equation}
      \sum_{k=2}^3 \delta_k^{(2n-1)} C_k -\sum_{k=2n-3}^{2n-2}  \delta_{k}^{(1)} C_{k} +  \sum_{j=4}^{2n-4} (\delta_j^{(2n-1)} -\delta_j^{(1)}) C_j= \frac{\partial R_{1,2n-3}}{\partial X_{2n-3,2n}}  -\frac{\partial R_{3,2n-1}}{\partial X_{3,2n}}.
      \label{eq:ConstraintApp}
\end{equation}
Now, using \cref{eq:J_X1,eq:Js3}, we can easily extract the derivatives of $R_{1,2n-3}$ and $R_{3,2n-1}$, and write the difference entering on the $r.h.s.$ in terms of lower-point amplitudes $M_L(1,2,\ldots,2n-2)$ and $A_R(2n-1,2,\ldots,2n-2)$:
\begin{equation}
\begin{aligned}
    \frac{\partial R_{1,2n-3}}{\partial X_{2n-3,2n}} - \frac{\partial R_{3,2n-1}}{\partial X_{3,2n}} = &M_L - A_R - \sum_{k=2}^3\delta_k^{(2n-1)} \frac{\partial M_L}{\partial X_{k,2n-2}}+\sum_{k=2n-3}^{2n-2}\delta_k^{(1)} \frac{\partial A_R}{\partial X_{2,k}}\\ 
    &- \sum_{i = 4}^{2n-4} \left(\delta_i^{(2n-1)} - \delta_i^{(1)}\right) \left[ \frac{\partial M_L}{\partial X_{i,2n-2}} +\frac{\partial A_R}{\partial X_{2,i}}\right] \\
    &-\delta_{2n-2}^{(1)} \sum_{j = 2}^{2n-4}\frac{\partial M_L}{\partial X_{j,2n-2}} + \delta_{2}^{(2n-1)}\sum_{j = 4}^{2n-2} \frac{\partial A_R}{\partial X_{2,j}}.
\end{aligned}
\label{eq:Constraint2}
\end{equation}
Expanding $M_L$ and $A_R$ in soft factors, we see that the difference $(M_L - A_R)$ starts at linear order as 
\begin{equation}
    M_L - A_R = \sum_{i=4}^{2n-4} (\delta_i^{(1)} - \delta_i^{(2n-1)})  \frac{\partial \mathcal{A}_{n-1}}{\partial X_{i,s}} + \mathcal{O}(\delta^2).
\end{equation}
So, plugging everything into Eq.~\eqref{eq:Constraint2}, we can read off the leading $\mathcal{O}(\delta^0)$ part of each $C_j$:
\begin{equation}
\begin{aligned}
&C_j^{(0)} = -\frac{\partial  \AA_{n-1}}{\partial X_{s,j}} - \frac{\partial  \AA_{n-1}}{\partial X_{j,2n-2}} - \frac{\partial  \AA_{n-1}}{\partial X_{2,j}}, \quad j=4,5, \ldots, 2n-4, \\
&\begin{aligned}
&C_2^{(0)}  =   \sum_{k=4}^{2n-3} \frac{\partial  \AA_{n-1}}{\partial X_{2,k}}, \\
&C_3^{(0)}  =  -\frac{\partial  \AA_{n-1}}{\partial X_{3,2n-2}},
\end{aligned} \quad \quad 
\begin{aligned}
&C_{2n-3}^{(0)} = -\frac{\partial  \AA_{n-1}}{\partial X_{2,2n-3}} , \\ 
&C_{2n-2}^{(0)}  =  \sum_{j=3}^{2n-4} \frac{\partial  \AA_{n-1}}{\partial X_{j,2n-2}}.
\end{aligned} 
\end{aligned}
\label{eq:C-terms}
\end{equation}
Finally, since at leading order in the soft factors we have $\partial_{X_{2n-3,2n}} R_{1,2n-3}, \partial_{X_{3,2n}} R_{3,2n-1} \to \mathcal{A}_{n-1}$, we can use \cref{eq:rep1,eq:rep2} to write the first two terms in the soft expansion as
\begin{equation*}
    \mathcal{A}_{n} \to \underbracket[0.4pt]{\left( \frac{X_{2n-3,2n}}{\delta_{2n-3}^{(1)}} + \frac{X_{3,2n}}{\delta_{3}^{(2n-1)}} \right) \mathcal{A}_{n-1}}_{\text{Leading, } \mathcal{S}^{-1}} -  \underbracket[0.4pt]{\mathcal{A}_{n-1} + \frac{R_{1,2n-3}^{(1)}}{\delta_{2n-3}^{(1)}} + \frac{R_{3,2n-1}^{(1)}}{\delta_{3}^{(2n-1)}} + \sum_{j=2}^{2n-2} (X_{j,2n}-X_{s,j})C_j^{(0)}}_{\text{Sub-leading, }\mathcal{S}^{0}} + \mathcal{O}(\delta).
\end{equation*}

Now, just like at leading order it is possible to gauge-invariantify the subleading term in the $n^{\rm{th}}$ gluon. To do this, let us first note that, by gauge-invariantifying the leading term, we have subsumed into it the subleading factor of $-\mathcal{A}_{n-1}$. So, to continue with our gauge-invariantifying, we will treat this factor as part of the leading term and focus our attention only on $\overline{\mathcal{S}}^0 \equiv \mathcal{S}^0 + \mathcal{A}_{n-1}$.

Remarkably, we find that the subleading term $\overline{\mathcal{S}}^0$, also satisfies, $ \mathcal{G}_{2n,1}[ \overline{\mathcal{S}}^0] = \mathcal{G}_{2n,2n-1}[\overline{\mathcal{S}}^0]$, concretely we have that 
\begin{equation}
    \mathcal{G}_{2n,1}[ \overline{\mathcal{S}}^0] = \mathcal{G}_{2n,2n-1}[\overline{\mathcal{S}}^0] = \frac{G_{1,2n-3}}{\delta_{2n-3}^{(1)}} + \frac{G_{3,2n-1}}{\delta_{3}^{(2n-1)}} + G_0,
\end{equation}
where one can work out
\begin{equation}
\begin{aligned}
    &G_{1,2n-3} = -\delta_{2n-2}^{(1)} \sum_{j = 2}^{2n-4} \delta_j^{(2n-1)}  \frac{\partial \mathcal{A}_{n-1}}{\partial X_{j,2n-2}}, \\
    &G_{3,2n-1} = -\delta_2^{(2n-1)} \sum_{j = 4}^{2n-2} \delta_j^{(1)} \frac{\partial \mathcal{A}_{n-1}}{\partial X_{2,j}}, \\
    &G_0 = \sum_{j = 4}^{2n-2} \delta_j^{(1)} \frac{\partial \mathcal{A}_{n-1}}{\partial X_{2,j}} + \sum_{j = 2}^{2n-4} \delta_j^{(2n-1)} \frac{\partial \mathcal{A}_{n-1}}{\partial X_{j,2n-2}}.
\end{aligned}
\end{equation} 
Since we find that $\mathcal{G}_{2n,1}[ \overline{\mathcal{S}}^0] = \mathcal{G}_{2n,2n-1}[\overline{\mathcal{S}}^0]$, we can then use the same simple trick \eqref{eq:gaug-inv-form} we used at leading order to gauge-invariantify the subleading soft term in gluon $n$, which is then simply $\overline{\mathcal{S}}^0 + \mathcal{G}_{2n,1}[ \overline{\mathcal{S}}^0]$.

Let us conclude this appendix with some remarks about what happens at higher orders in the soft expansion. Of course, since we have formulae for the residues $R_{1,2n-3}$ and $R_{3,2n-1}$ we can continue to write their soft expansions indefinitely. But, the only constraint we have for the rest of the amplitude comes from \cref{eq:constraintContact2}, which can only be used to write down the leading part of the $C_j$ in the soft expansion. However, we can ask if there is \textit{anything} interesting that can be said about the behavior of the higher-order terms in the $C_j$ expansion using just \cref{eq:constraintContact2}. Indeed, one can see a hint of some structure by noticing that, if one sums the terms in Eq.~\eqref{eq:C-terms}, many terms cancel, and the result is
\begin{equation}
    \sum_{j = 2}^{2n-2} C_j^{(0)} = \sum_{j = 4}^{2n-4} \frac{\partial \mathcal{A}_{n-1}}{\partial X_{j,s}}.
\label{eq:C-sum-rule-1}
\end{equation}
It turns out that this behavior continues to all orders in the soft expansion. To state it most succinctly, let us consider the soft limit where we take $x_s^\mu = x_1^\mu$. In this case, our two sets of soft factors $\delta_j^{(1)}, \delta_j^{(2n-1)}$ become degenerate, and we thus only have one set $\delta_j = X_{j,2n-1} - X_{1,j}$ to deal with. Let us consider $\mathcal{C} = \sum_{j} C_j$ as an expansion in these parameters:
\begin{equation}
    \mathcal{C} = \mathcal{C}_0 + \sum_{i = 2}^{2n-2} \mathcal{C}_i \delta_i + \frac{1}{2!} \sum_{i,j = 2}^{2n-2} \mathcal{C}_{i,j} \delta_i \delta_j + \cdots.
\end{equation}
One can then work out easily from \cref{eq:constraintContact2} that the property in \cref{eq:C-sum-rule-1} continues to all orders in the above expansion:
\begin{equation}
\begin{aligned}
    \mathcal{C}_0 = &\sum_{j = 4}^{2n-4} \frac{\partial \mathcal{A}^{(1)}_{n-1}}{\partial X_{1,j}}, \quad
    \sum_{i = 2}^{2n-2}\mathcal{C}_i = \frac{1}{2}\sum_{i,j = 4}^{2n-4} \frac{\partial \mathcal{A}_{n-1}^{(1)}}{\partial X_{1,i}}\frac{\partial \mathcal{A}_{n-1}^{(1)}}{\partial X_{1,j}}, \\
    &\sum_{i,j = 2}^{2n-2} \mathcal{C}_{i,j} = \frac{1}{3}\sum_{i,j,k = 4}^{2n-4} \frac{\partial \mathcal{A}_{n-1}^{(1)}}{\partial X_{1,i}}\frac{\partial \mathcal{A}_{n-1}^{(1)}}{\partial X_{1,j}}\frac{\partial \mathcal{A}_{n-1}^{(1)}}{\partial X_{1,k}}, \quad \ldots
\end{aligned}
\end{equation}
where $\mathcal{A}_{n-1}^{(1)} = \mathcal{A}_{n-1}(1,2,\ldots,2n-2)$. The object $\mathcal{C}$, though not appearing in the amplitude itself, is interesting in its own right, since it precisely agrees with the piece of $\W_{2n}[\mathcal{A}_n]$ that has no poles in $X_{1,2n-3}$ and $X_{3,2n-1}$. (See \cref{eq:W-on-amp}.) In the main text, we observed that the numerators of those poles simply gave lower-point amplitudes under the action of $\W_{2n}$, and now, after a bit more work, we find that the remainder also behaves nicely --- satisfying the interesting sum rules described above.

%% file: Sections_v2/AppHigherOrderLoop_v2.tex
\section{Higher Orders in the One-Loop Soft Expansion}
\label{app:highorderLoop}

In this Appendix, we discuss the specifics of the soft expansion of the one-loop integrand.

Let us begin by analyzing the residue in $X_{2n-1,3}$. Using Eq.~\eqref{eq:tree-loop}, this corresponds to $i = 2n-1$, $j = 3$, $L = \{2n, 1, 2\} $, and $R = \{ 3, 4, \ldots, 2n-1\}$. In this case, we have that $\mathcal{A}_L = \mathcal{A}_3(1,2,3,x_L,2n-1,2n)$ and $\mathcal{I}_R=\mathcal{I}_{n-1}(2n-1, x_R, 3,4,\ldots,2n-2)$. Just as we did at tree-level, we will consider each contribution of the known three-point amplitude to the gluing rule individually. Starting with $k = 1$, we find that
\begin{equation}
\begin{aligned}
    \underline{k=1}: &\left[X_{2n,2}-X_{2n,3} -X_{2n-1,2}\right] \times \left[ \sum_{m\in R} (X_{1,m}-X_{2n-1,m}) \frac{\partial \mathcal{I}_R}{\partial X_{2,m}} +(X_{1,p}-X_{2n-1,p} )\frac{\partial \mathcal{I}_R}{\partial X_{2,p}}\right.\\
    &\left. +\sum_{m\in R} (X_{m,1}-X_{m,2n-1}\hat{\theta}_{m,2n-2}) \frac{\partial \mathcal{I}_R}{\partial X_{m,2}} + X_{2n-2,2n-1} \frac{\partial \mathcal{I}_R}{\partial X_{2n-2,2n-1}} \bigg \vert_{2\to 2n-1}\right],
\end{aligned}
\label{eq:k1}
\end{equation}
where, like at tree-level, we have relabeled the dummy index $x_R \to 2$. Also, recall that $\hat{\theta}_{a,b}$ equals $1$ if $a\neq b$ and $0$ otherwise. Expanding the prefactors of the above equation in soft factors, and keeping in mind that $X_{2n-2,2n-1} \to \delta_{2n-2,2n-1}$ is itself a soft factor, we derive
\begin{equation}
\begin{aligned}
    \underline{k=1}: &\left[X_{2n,2}-X_{2n,3} -\delta_{2n-1,2}\right] \times \left[ \sum_{m\in R} (\delta_{1,m}-\delta_{2n-1,m}) \frac{\partial \mathcal{I}_R}{\partial X_{2,m}} +(\delta_{1,p}-\delta_{2n-1,p} )\frac{\partial \mathcal{I}_R}{\partial X_{2,p}}\right.\\
    &\left.  - X_{s,s} \frac{\partial \mathcal{I}_R}{\partial X_{2n-1,2}} +\sum_{m\in R} (\delta_{m,1}-\delta_{m,2n-1}\hat{\theta}_{m,2n-2}) \frac{\partial \mathcal{I}_R}{\partial X_{m,2}} \right.\\
    &\left. + \delta_{2n-2,2n-1} \frac{\partial \mathcal{I}_R}{\partial X_{2n-2,2n-1}} \bigg \vert_{2\to 2n-1}\right].
\end{aligned}
\label{eq:k11}
\end{equation}
Now, the lower-point integrand $\mathcal{I}_R$ depends on the soft factors via $X_{m,2n-1}=X_{m,s} + \delta_{m,2n-1}$ (and similarly for $X_{2n-1,m}$) for $m \in R$. So, expanding $\mathcal{I}_R$ in $\delta_{m,2n-1}$ and $\delta_{2n-1,m}$, we derive the all-soft-order expansion of Eq.~\eqref{eq:k1}; at leading order $\mathcal{O}(\delta^0)$ in particular, we get
\begin{equation}
    \underline{k=1}: \quad(X_{2n,3} - X_{2n,2})X_{s,s} \frac{\partial \mathcal{I}_{n-1}}{\partial X_{s,2}}\bigg \vert_{\mathcal{S}} + \mathcal{O}(\delta),
\end{equation}
where $\mathcal{I}_{n-1} \equiv \mathcal{I}_{n-1}(s, 2, \ldots, 2n-2)$. The notation ``$\vert _\mathcal{S}$'' instructs us to evaluate on the locus of $X_{s,2},X_{2n-2,s}=0$, as required by our definition of the soft limit. In this case, gauge invariance in gluon $2$ means that the derivative $\partial_{X_{s,2}} \mathcal{I}_{n-1}$ is independent of $X_{s,2}$, but we still need to manually send $X_{2n-2,s} \to 0$ to ensure consistency of the limit.

As for $k=2$, we find
\begin{equation}
\begin{aligned}
    \underline{k=2}: \quad &X_{2n,3}  \times \left[ \sum_{m\in R} \left( (X_{2,m}-X_{2n-1,m}) \frac{\partial \mathcal{I}_R}{\partial X_{2,m}} + (X_{m,2}-X_{m,2n-1} \hat{\theta}_{m,2n-2}) \frac{\partial \mathcal{I}_R}{\partial X_{m,2}} \right)\right. \\
    &\left. +(X_{2,p} - X_{2n-1,p})\frac{\partial \mathcal{I}_R}{\partial X_{2,p}} + X_{2n-2,2n-1} \frac{\partial \mathcal{I}_R}{\partial X_{2n-2,2n-1}} \bigg \vert_{2\to 2n-1} \right].
\end{aligned}
\label{eq:k2}
\end{equation}
Quite nicely, just like at tree-level, the large sum inside the brackets simply becomes $\mathcal{I}_R$ due to the surface gauge invariance of $\mathcal{I}_R$! As such, the soft expansion of the $k = 2$ term comes exclusively from the Taylor expansion of $\mathcal{I}_R$, which at leading order yields
\begin{equation}
    X_{2n,3} \mathcal{I}_{n-1}(s, 2, \ldots, 2n-2) \vert_{\mathcal{S}} + \mathcal{O}(\delta).
\end{equation}
Finally, we examine the case of $j = 2n$. Here, since $\partial_{X_{2n,x_L}} \mathcal{A}_L = \delta_{2n-1,2}$, the entire term is subleading in the soft expansion, and will thus not contribute to the leading theorem.

Now, we can undergo exactly the same procedure for the residue $R_{2n-3,1}$, where we take $i = 1$, $j = 2n-3$, $L = \{ 1, 2, \ldots, 2n-3\}$, $R = \{ 2n-2, 2n-1, 2n\}$, $\mathcal{I}_L = \mathcal{I}_{n-1}(1, 2, \ldots, 2n-3, x_L)$ and $\mathcal{A}_R = \mathcal{A}_3(1, x_R, 2n-3, 2n-2, 2n-1, 2n)$ in \cref{eq:tree-loop}. In this case, $m = 2n-1$ gives the analogous expression for \cref{eq:k11}, $m = 2n-2$ for \cref{eq:k2}, and we again find something manifestly subleading for $m = 2n$. So, by further Taylor expanding $\mathcal{I}_L$ now in factors of $\delta_{1,k}$ and $\delta_{k,1}$, we can straightforwardly obtain the all-soft-order expansion of $R_{2n-3,1}$. In doing so, we have also proved the leading soft theorem (pre-gauge-invariantification), displayed in \cref{eq:one-loop-lead}.